\documentclass{article}

\usepackage[utf8]{inputenc}
\usepackage{amssymb}
\usepackage{amsmath}
\usepackage{amsfonts}
\usepackage{graphicx}
\usepackage{setspace}
\usepackage{xcolor}
\usepackage{bm}
\usepackage[margin=1in]{geometry}
\usepackage{url}
\usepackage[sectionbib,numbers]{natbib}
\usepackage{bibunits}
\defaultbibliographystyle{ieeetr}
\defaultbibliography{references}
\usepackage{booktabs}
\usepackage{calc}
\usepackage{caption}
\usepackage{subcaption}
\usepackage{array}
\usepackage{longtable}
\usepackage[figuresright]{rotating}
\usepackage{fancyhdr}
\newcommand{\bsL}{\boldsymbol{\Lambda}}
\newcommand{\bsPH}{\boldsymbol{\Phi}}
\newcommand{\bsP}{\boldsymbol{\Psi}}
\newcommand{\bsS}{\boldsymbol{\Sigma}}

\newcommand{\bsT}{\boldsymbol{\Theta}}

\newcommand{\bsG}{\boldsymbol{\Gamma}}

\newcommand{\mbQ}{\mathbf{Q}}

\begin{document}

\title{Multi-study factor-based Gaussian graphical models with applications to the metabolomics of gestational diabetes mellitus}

\author{Katherine H. Shutta$^{1,2,3,+}$, Denise M. Scholtens$^4$, William L. Lowe, Jr.$^5$, \\ Raji Balasubramanian$^{1}$, Roberta De Vito$^{6,7*}$}
\date{}

\label{firstpage}

\maketitle

\begin{center}
{\small
\noindent$^{1}$ Department of Biostatistics and Epidemiology, University of Massachusetts - Amherst, Amherst, MA, U.S.A. \\
$^{2}$ Department of Biostatistics, Harvard School of Public Health, Boston, MA, U.S.A. \\
$^{3}$ Channing Division of Network Medicine, Department of Medicine, \\ Brigham and Women's Hospital, Boston, MA, U.S.A. \\
$^{4}$ Division of Biostatistics and Informatics, Department of Preventive Medicine, \\ Northwestern University Feinberg School of Medicine, Chicago, IL, U.S.A. \\
$^5$ Division of Endocrinology, Metabolism and Molecular Medicine, Department of Medicine, \\ Northwestern University Feinberg School of Medicine, Chicago, IL, U.S.A.\\
$^{6}$ Department of Statistical Science, Sapienza University of Rome, Rome, Italy \\
$^{7}$ Department of Biostatistics and Data Science Initiative, Brown University, Providence, RI, U.S.A.
\\
\vspace{0.1in}
$^*$Correspondence: roberta.devito@uniroma1.it
\\
$^+$Present address: Harvard School of Public Health}
\end{center}

\begin{abstract}
Gestational diabetes mellitus (GDM) is a metabolic disorder of pregnancy that is associated with a range of adverse maternal and newborn outcomes. Metabolomic profiling plays a key role in investigating the mechanisms underlying GDM; GDM-related metabolites could serve as early biomarkers for intervention during pregnancy. Metabolites typically function in groups and exhibit informative correlation patterns, making latent variable and network-based approaches a promising strategy for understanding metabolic disorders like GDM. This context motivates an integrated approach combining the strengths of latent variable methods with the power of network analysis. To this end, we introduce extended Multi-Study Factor Analysis (MSFA-X), a novel approach leveraging MSFA together with network modeling to facilitate the identification of shared and divergent network patterns based on latent variable representations across multiple conditions. We demonstrate the accuracy of MSFA-X through extensive simulations. Applying MSFA-X to metabolomic data from the Hyperglycemia and Adverse Pregnancy Outcomes (HAPO) study, we identify shared and condition-specific metabolomic networks that reveal important differential connections between key amino acids of glucose metabolism and acylcarnitines in mothers with and without GDM. Our results characterize the divergent metabolomic milieux of GDM and normal gestational glucose metabolism. MSFA-X is freely available as an R package at \url{https://github.com/katehoffshutta/MSFA}.
\end{abstract}

\begin{bibunit}



\section{Introduction}

Gestational diabetes mellitus (GDM) is a metabolic disorder of pregnancy leading to maternal hyperglycemia and is associated with a range of adverse maternal and neonatal outcomes \cite{mcintyre2019gestational}. Metabolomic profiling has been successfully used to characterize GDM at the molecular level, linking metabolite levels to GDM status and related clinical characteristics, including maternal traits such as fasting plasma glucose, BMI, and insulin sensitivity as well as neonatal characteristics such as cord C-peptide levels, birth size, and adiposity \cite{jacob2017targeted,sandler2017associations,kadakia2019maternal,kadakia2019cord}. Such studies have uncovered evidence suggesting that GDM is not a condition of isolated dysfunction in single metabolites; rather, coordinated changes in groups of metabolites form patterns or signatures that are associated with GDM and its various phenotypic characteristics \cite{jacob2017targeted,sandler2017associations}. These groups of metabolites can be viewed as latent variables; thus, statistical approaches that identify latent variables across multiple conditions are a promising lens for understanding differences in glucose metabolism during gestation in women with and without GDM. Moreover, such latent variables may be predictive of consequent adverse maternal and newborn phenotypes. 

Factor analysis is a dimension reduction method that identifies latent variables (``factors”) in multivariate data based on correlation structures of the underlying data \cite{kerr2000analysis, carvalho2008high}. It is a useful framework for two reasons: (i) dimension reduction is generally helpful for wrangling and analyzing high-dimensional data, and (ii) the factors themselves represent summary measures of coordinated underlying features that are often biologically interpretable. Because of the correlated and high-dimensional nature of metabolomics data, factor analysis is an attractive framework for finding latent metabolomic variables related to GDM.  In particular, multi-study factor analysis (MSFA) \cite{de2019multi} extends standard factor analysis to jointly estimate shared and study-specific factors across multiple studies, groups, or conditions. The MSFA approach is a natural fit for latent variable analysis of metabolomic data in women with and without GDM.

Many important previous insights in GDM metabolism have come from the application of network models to metabolomics data, identifying metabolites associated with maternal and newborn traits \cite{scholtens2016metabolic,sandler2017associations,kadakia2019maternal,lee2024metabolomic}. Network models are models in which nodes correspond to biomolecules (e.g., metabolites, genes) and edges encode relationships such as correlations or biochemical interactions; they have been shown to provide a powerful framework for representing multivariate dependencies in a wide range of biological data \cite{ideker2017network,loscalzo2017network,huang2018systematic,altenbuchinger2020gaussian}. Several existing methods aim to integrate network structures across conditions, including approaches for defining network distances \cite{tantardini2019comparing}, estimating differential networks \cite{zhao2014direct}, and jointly modeling networks across multiple conditions to improve estimation before comparison \cite{guo2011joint,danaher2014joint,shojaie2021differential}. However, none of these approaches use factor analysis; existing methods are unable to capture network structures between the types of latent variables we know to be relevant in metabolomics data.  This gap motivates the development of an integrated approach combining the power of network analysis with the strengths of factor analysis to study latent variables in a network framework. To this end, we introduce MSFA-X, an extension of MSFA for constructing latent variable-based Gaussian graphical models (GGMs). MSFA-X jointly estimates a shared GGM, capturing relationships common across studies or conditions, and study-specific GGMs, identifying condition-specific dependencies.  By estimating shared and divergent GGM networks through a latent factor approach, MSFA-X offers a powerful framework for uncovering the molecular basis of disease subtypes and for analyzing related conditions such as metabolic disorders. Although MSFA-X was developed with gestational diabetes mellitus as the motivating application, it is broadly applicable for the analysis of multivariate data exhibiting latent variable structures in multi-condition settings. 

Our specific study is motivated by data from the Hyperglycemia and Adverse Pregnancy Outcomes (HAPO) Study \cite{hapo2008hyperglycemia}, a large prospective observational study that has played a central role in  GDM diagnosis and its association with pregnancy outcomes and neonatal conditions \cite{metzger2019hyperglycemia}. The HAPO study aims to identify key factors influencing GDM and its long-term consequences for both mothers and offspring \cite{hapo2002hyperglycemia}. One of the primary objectives of the study is to explore how maternal metabolomic profiles are associated with insulin resistance \cite{sandler2017associations, liu2020metabolomic}. By applying MSFA-X to the HAPO metabolomic dataset, we identify shared and condition-specific metabolic network structures distinguishing women with and without GDM and demonstrate that the latent factors underlying these network structures are associated with newborn adiposity.

The structure of the paper is as follows. Section 2 describes the HAPO study and its key characteristics that motivated our model. Section 3 presents the statistical framework of MSFA-X, including distributional assumptions, parameter estimation, and identifiability. Section 4 evaluates MSFA-X through extensive simulations, comparing its performance with a graphical lasso-based benchmark. Section 5 applies MSFA-X to the HAPO study, estimating shared and condition-specific metabolomic networks and revealing key network structures related to GDM. Finally, Section 6 contains a discussion highlighting strengths and limitations of our approach and directions for future research.

\section{The Hyperglycemia and Adverse Pregnancy Outcomes (HAPO) Study}

The Hyperglycemia and Adverse Pregnancy Outcomes (HAPO) study is an international prospective cohort study investigating the relationship between maternal glucose metabolism and newborn outcomes in 25,505 mother-newborn pairs \cite{hapo2008hyperglycemia}. The study design and methodology have been previously described \cite{hapo2008hyperglycemia,hapo2002hyperglycemia}. Briefly, fasting plasma metabolomic profiles were collected from mothers prior to an oral glucose tolerance test (OGTT) at approximately 28 weeks gestation, with additional measurements taken one hour after glucose intake \cite{kadakia2019maternal, kadakia2019cord}. Both targeted and untargeted metabolomic profiling were conducted following the methodology described by \cite{kadakia2019maternal}. In this paper, we focus on 60 targeted metabolites collected from a subset of 3463 of these participants, including 2,887 women without GDM and 576 women with GDM. Table \ref{tab:hapo} summarizes the basic characteristics of this sample. Women with GDM were, on average, older than women without GDM, and exhibited higher mean arterial pressure at OGTT,  BMI, and fasting plasma glucose. Additionally, GDM was more prevalent among women of Mexican American and Thai ancestry: 26.4 \% of the GDM group were of Mexican American ancestry compared to 18.3\% in the overall sample, while 35.6\% of the GDM group were of Thai ancestry compared to 30.2\% overall. Despite these metabolic and demographic differences, gestational age at delivery was similar between women with GDM (mean: 39.6 weeks) and those without GDM (mean: 39.8 weeks).

\clearpage

\begin{longtable}[]{@{}
  >{\raggedright\arraybackslash}p{(\columnwidth - 6\tabcolsep) * \real{0.3298}}
  >{\raggedright\arraybackslash}p{(\columnwidth - 6\tabcolsep) * \real{0.2234}}
  >{\raggedright\arraybackslash}p{(\columnwidth - 6\tabcolsep) * \real{0.2553}}
  >{\raggedright\arraybackslash}p{(\columnwidth - 6\tabcolsep) * \real{0.2515}}@{}}
\caption{Basic characteristics of the HAPO metabolomics study participants.} \\
\label{tab:hapo}\\
\toprule\noalign{}
\begin{minipage}[b]{\linewidth}\raggedright
\end{minipage} & \begin{minipage}[b]{\linewidth}\raggedright
Gestational Diabetes
\end{minipage} & \begin{minipage}[b]{\linewidth}\raggedright
No Gestational Diabetes
\end{minipage} & \begin{minipage}[b]{\linewidth}\raggedright
Overall
\end{minipage} \\
\midrule\noalign{}
\endhead
\bottomrule\noalign{}
\endlastfoot
& (N=576) & (N=2887) & (N=3463) \\
\textit{Age at OGTT} & & & \\
Mean (SD) & 31.1 (5.68) & 28.6 (5.69) & 29.0 (5.76) \\
Median {[}Min, Max{]} & 31.2 {[}18.2, 45.6{]} & 28.9 {[}18.0, 47.1{]} &
29.2 {[}18.0, 47.1{]} \\ 
&&&\\
\textit{Mean arterial pressure at OGTT} & & & \\
Mean (SD) & 83.4 (7.74) & 80.2 (7.69) & 80.7 (7.79) \\
Median {[}Min, Max{]} & 83.3 {[}62.8, 108{]} & 80.0 {[}55.3, 110{]} &
80.7 {[}55.3, 110{]} \\
&&&\\
\textit{Body mass index at OGTT} & & & \\
Mean (SD) & 29.6 (5.57) & 27.5 (5.07) & 27.9 (5.21) \\
Median {[}Min, Max{]} & 28.7 {[}19.3, 57.4{]} & 26.6 {[}16.2, 55.2{]} &
27.0 {[}16.2, 57.4{]} \\
&&&\\
\textit{Maternal height at OGTT} & & & \\
Mean (SD) & 159 (7.36) & 161 (6.96) & 161 (7.06) \\
Median {[}Min, Max{]} & 160 {[}141, 180{]} & 161 {[}132, 185{]} & 161
{[}132, 185{]} \\
&&&\\
\textit{Fasting plasma glucose at OGTT} & & & \\
Mean (SD) & 89.2 (7.55) & 79.7 (5.35) & 81.3 (6.77) \\
Median {[}Min, Max{]} & 91.8 {[}68.4, 115{]} & 79.2 {[}61.2, 90.0{]} &
81.0 {[}61.2, 115{]} \\
&&&\\
\textit{Gestational age at delivery} & & & \\
Mean (SD) & 39.6 (1.25) & 39.8 (1.20) & 39.8 (1.21) \\
Median {[}Min, Max{]} & 39.6 {[}37.0, 43.0{]} & 39.9 {[}37.0, 44.7{]} &
39.7 {[}37.0, 44.7{]} \\
&&&\\
\textit{Store time of maternal sample} & & & \\
Mean (SD) & 13.6 (2.64) & 13.9 (2.73) & 13.9 (2.72) \\
Median {[}Min, Max{]} & 13.7 {[}7.84, 18.9{]} & 14.1 {[}7.70, 19.2{]} &
14.0 {[}7.70, 19.2{]} \\
&&&\\
\textit{Ancestry group} & & & \\
Afro Caribbean & 91 (15.8\%) & 743 (25.7\%) & 834 (24.1\%) \\
European & 125 (21.7\%) & 815 (28.2\%) & 940 (27.1\%) \\
Mexican American & 152 (26.4\%) & 482 (16.7\%) & 634 (18.3\%) \\
Thai & 205 (35.6\%) & 842 (29.2\%) & 1047 (30.2\%) \\
Missing & 3 (0.5\%) & 5 (0.2\%) & 8 (0.2\%) \\
\end{longtable}

We apply MSFA-X to model shared and disease-specific partial correlations in glucose response based on changes from fasting metabolite levels to 1-hr post-glucose intake levels. Analyses are adjusted for maternal age at OGTT, mean arterial pressure at OGTT, body mass index (BMI) at OGTT, maternal height at OGTT, fasting plasma glucose, ancestry group, and sample storage time. Further details of metabolite pre-processing, model design, and covariate adjustment are provided in the supplement.

\section{Methods}

Traditional factor analysis (FA) represents the $i^{th}$ observation of $p-$dimensional random variable $\mathbf{x}$ as $\mathbf{x}_i = \bsPH \mathbf{f}_i + \mathbf{e}_i$, where $\bsPH$ is a $p \times k$ matrix with $k \leq p$, the columns of which represent the $k$ factor loadings; $\mathbf{f}_i$ is a $k \times 1$ column vector containing the $k$ latent variables (``factors"); and $\mathbf{e}_i$ is the residual noise, typically assumed to be  independent and identically distributed (i.i.d.) as $\mathbf{e}_i \sim N_p(0,\bsP)$ where $\bsP$ is a diagonal matrix (e.g., \cite{joreskog1967some, joreskog1971simultaneous}). An attractive feature of FA is that under the assumption that $\mathbf{f}_i \sim N_k(\mathbf{0},\mathbf{I}_k)$, the variance-covariance matrix of the data can be decomposed as $\bsS = \bsPH\bsPH^\top + \bsP$. In this way, the observed variation of multivariate data of interest can be decomposed into a component attributable to the factor loadings ($\bsPH\bsPH^\top$), and a component attributable to noise ($\bsP$).

Multi-study factor analysis (MSFA) is an extension of FA to the setting where the same multivariate random variable has been measured across multiple studies or conditions \cite{de2019multi}.  MSFA models two types of  latent variables: shared factors common to all studies and study-specific factors unique to each study. Let $\mathbf{x}_{is}$ be the $p$-dimensional vector of observations for the $i^{th}$ subject in study $s$. MSFA represents the data $\mathbf{x}_{is}$ as:
\begin{equation}
 \mathbf{x}_{is} = \bsPH \mathbf{f}_{is} + \bsL_s \mathbf{l}_{is} + \mathbf{e}_{is} 
\label{eq:msfa}
\end{equation}

where $\bsPH$ is a $p \times k$ matrix of shared loadings, $\mathbf{f}_{is}$ is its corresponding $k \times 1$ vector of shared latent factors, $\bsL_s$ is a $p \times j_s$ matrix of study-specific loadings, $\mathbf{l}_{is}$ is its corresponding $j_s \times 1$ vector of study-specific latent factors, and $\mathbf{e}_{is}$ is the independent residual noise, again assumed to be distributed as i.i.d. $\mathbf{e}_{is} \sim N_p(\bm{0},\bsP_s)$ where $\bsP_s$ is a diagonal matrix specific to study $s$. Under the assumption that $\mathbf{f}_{is} \sim N_k(\bm{0}, \mathbf{I}_k)$ and $\mathbf{l}_{is} \sim N_k(\bm{0}, \mathbf{I}_{j_s})$, the covariance matrix for each study is given by
\begin{equation}
\bsS_s = \bsPH \bsPH^\top + \bsL_s \bsL_s^\top + \bsP_s
\label{eq:msfacovdec}
\end{equation}

Thus, the covariance matrix can be decomposed into three components: i) the part attributable to the shared factors $\bsPH \bsPH^\top$, ii) the part attributable to the study-specific factors $\bsL_s \bsL_s^\top$, and iii) $\bsP$  the residual noise component \cite{de2019multi}.   In \cite{de2021bayesian}, shared correlation networks are derived from $\Phi\Phi^\top$, while study-specific correlation networks from $\Lambda_s\Lambda_s^\top$.

 Correlation networks are typically  dense, making interpretation challenging, as edges represent both direct and indirect associations. Gaussian graphical models (GGMs) differ from correlation networks in that GGMs define edges based on conditional (partial) correlations, whereas correlation networks define edges via marginal correlations. In contrast to edges in correlation networks, edges in a GGM reflect direct associations, resulting in sparser and more interpretable network structures. 

Assuming the data follow a multivariate normal distribution, it is well known that pairwise partial correlations can be derived from the elements of the inverse covariance matrix, also known as the precision matrix \cite{uhler2017gaussian,shutta2022gaussian}. Specifically, let $\mathbf{x} \sim N_p(0, \bsS)$ with $\bsT = \bsS^{-1}$, and let $\theta_{ij}$ denote the $i,j$ entry of $\bsT$. Then the partial correlation between $x_i$ and $x_j$ is given by
\begin{equation}
\rho_{x_i,x_j | x_{-ij}} = - \frac{\theta_{ij}}{\sqrt{\theta_{ii}\theta_{jj}}} 
\label{eq:parkour}
\end{equation}

Therefore, estimating a GGM is equivalent to estimating the precision matrix. However, MSFA does not directly enable precision matrix estimation. To address this limitation, we developed MSFA-X, a new tool based on MSFA that is specifically designed to directly estimate both shared and study-specific GGMs.

When the covariance matrix is invertible, precision matrix estimation is straightforward: the maximum likelihood estimation (MLE) for the precision matrix can be obtained by simply inverting the MLE of the covariance matrix \cite{stewart1969continuity,casella2002statistical}. 
However, in the MSFA framework, the shared and study-specific components in Equation \ref{eq:msfacovdec} are singular.  Specifically, inverting $\bsPH\bsPH^\top$ is not possible, since $\bsPH\bsPH^\top$ is not full rank with $k < p$, and a similar issue arises with $\bsL_s\bsL_s^\top$. 

When the covariance matrix is not invertible, a common approach is to apply the graphical lasso (``glasso") method, which employs a penalized likelihood approach to estimate a sparse precision matrix \cite{glasso2008}. However, applying the graphical lasso to these matrices is not straightforward as it requires tuning parameter selection via cross-validation, which in turn requires estimation of the latent factors $\mathbf{f}_{is}$ and $\mathbf{l}_{is}$. Moreover, tuning parameter selection involves partitioning the data into training and test sets, which introduces an additional layer of variability and computing demands. 

MSFA-X addresses these challenges by providing a tuning-free approach to estimate the shared and the study-specific GGMs. To ensure invertibility,  MSFA-X decomposes the residual covariance $\bsP_s$ into shared and study-specific components, respectively  $\bsG$ and $\mathbf{H}_s$, where $\bsG$ and $\mathbf{H}_s$ where both are full-rank diagonal matrices. This enables an alternative formulation of the covariance decomposition in which both the shared component, $\bsPH\bsPH^\top + \bsG$, and the  study-specific component, $\bsL_s\bsL_s^\top + \mathbf{H}_s$, are invertible, facilitating precision matrix estimation and GGM construction.

To estimate $\bsG$ and $\mathbf{H}_s$, we assume the observed residual term $\mathbf{e}_{is}$ is partitioned into two distinct and independent sources of noise: an overall noise $\mathbf{g}_{is}$ and a study-specific noise $\mathbf{h}_{is}$.  The resulting model is:

\begin{equation}
    \mathbf{x}_{is} = \mathbf{\Phi} \mathbf{f}_{is} + \mathbf{\Lambda}_{s} \mathbf{l}_{is} + \mathbf{g}_{is} + \mathbf{h}_{is} 
    \label{eq:MSFAX}
\end{equation}
    
where the noise components are distributed as $\mathbf{g}_{is} \sim N_P(\bm{0}, \bsG)$ with $\bsG = diag(\gamma_{1}, \dots, \gamma_{P})$ and $\mathbf{h}_{is} \sim N_p(\bm{0}, \mathbf{H}_s)$ with $\mathbf{H}_{s} = diag(\eta_{1s}, \dots, \eta_{Ps})$. Under this model formulation, the covariance matrix for study 
$s$ decomposes as:
\begin{equation}
    \bsS_s = \bsPH \bsPH^\top + \bsL_s \bsL_s^\top  + \bsG + \mathbf{H}_s
    \label{eq:msfax_covdecomp}
\end{equation}

\begin{figure}[ht]
\centering
\caption{An illustration of the covariance decomposition implied by the MSFA-X model formulation for $S=2$ studies on $p=12$ predictors.}
\begin{subfigure}[b]{\textwidth}
\centering
\includegraphics[width=0.8\textwidth]{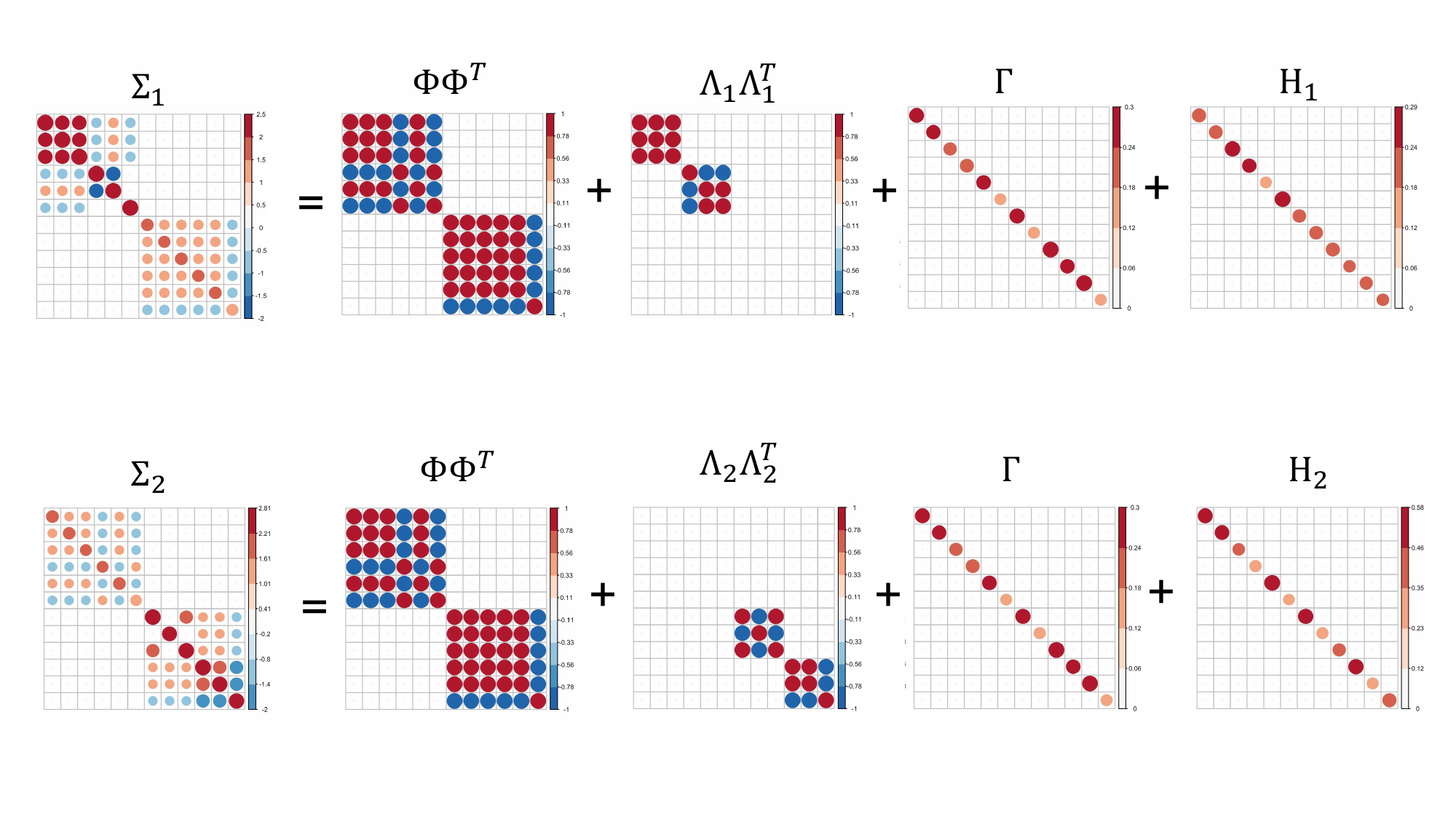}
\caption{Under the MSFA-X model formulation, the covariance matrix of the data in the $s^{th}$ study can be decomposed as $\bsS_s = \bsPH \bsPH^\top + \bsL_s \bsL_s^\top  + \bsG + \mathbf{H}_s$.  Here, we show the decomposition of simulated data. Our goal is to recover this decomposition for a given input dataset.}
\vspace{0.1in}
\label{fig:decomp}
\end{subfigure}
\begin{subfigure}[b]{\textwidth}
    \centering
    \includegraphics[width=0.78\textwidth]{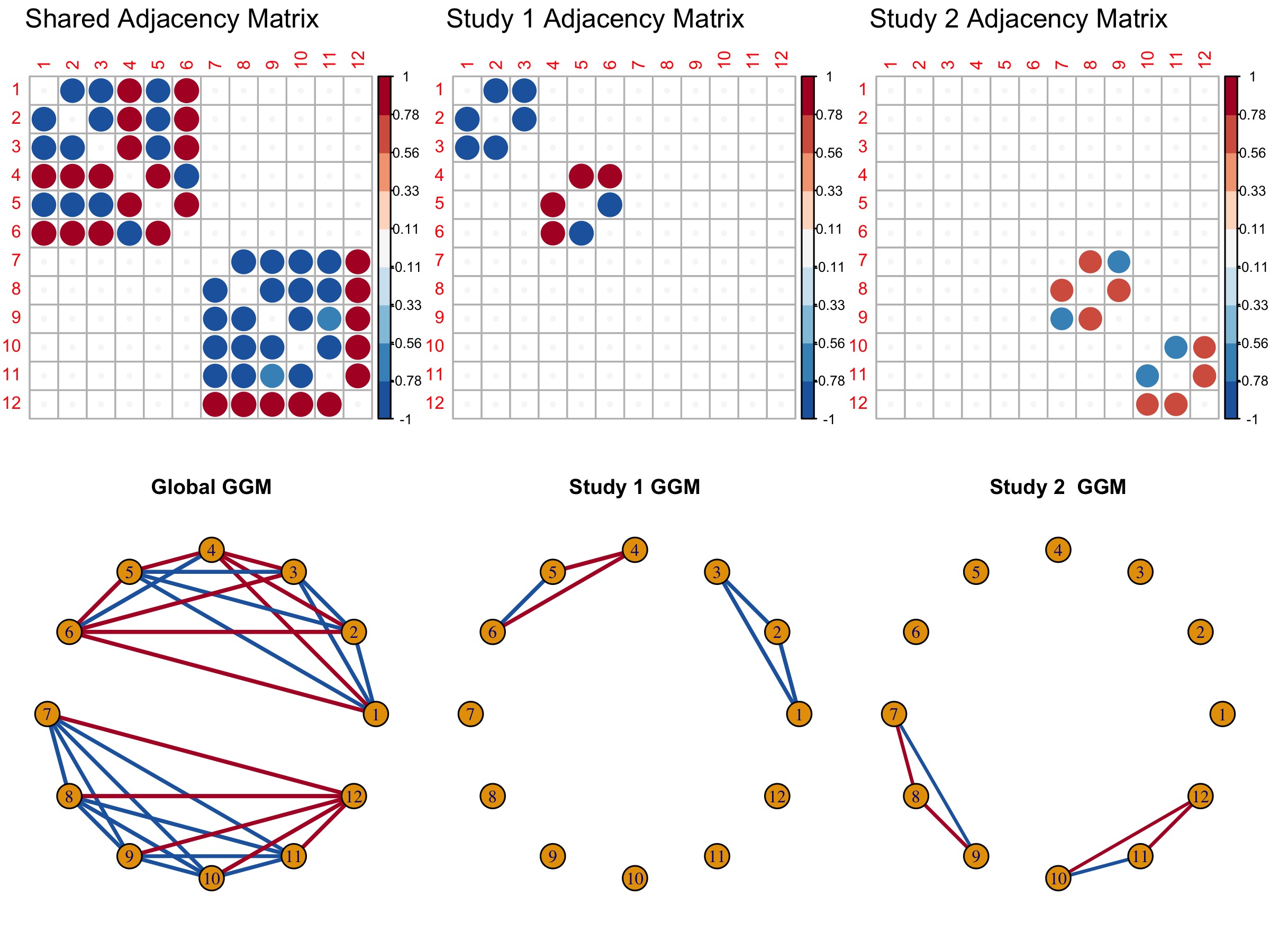}
        \caption{The shared and study-specific precision matrices are shown here. Left, ($\bsPH \bsPH^\top + \bsG)^{-1}$; center, $(\bsL_1\bsL_1^\top + \mathbf{H}_{1})^{-1}$; right, $(\bsL_2\bsL_2^\top + \mathbf{H}_{2})^{-1}$.  Estimating these matrices allows us to estimate the corresponding shared and study-specific GGMs. Red indicates positive values; blue, negative.  Edge width in graphs is proportional to magnitude of partial correlation between two variables.}
        \vspace{0.1in}
\label{fig:precmatSample}
\end{subfigure}
\end{figure}

Figure \ref{fig:decomp} illustrates this decomposition for $S=2$ studies and $P=12$ variables. From Equation \ref{eq:MSFAX}, the conditional distribution of $\mathbf{x}_{is}$ given the study-specific factors $\mathbf{l}_{is}$ and the study-specific noise $\mathbf{g}_{is}$ is:
\begin{equation}
    \mathbf{x}_{is} | \mathbf{l}_{is}, \mathbf{h}_{is} \sim N_p(\mathbf{\Lambda}_{s}\mathbf{l}_{is}, \mathbf{\Phi}\mathbf{\Phi}^\top + \mathbf{\Gamma})
    \label{eq:sharedprec}
\end{equation}

We interpret $\mathbf{\Phi}\mathbf{\Phi}^\top +  \mathbf{\Gamma}$ as the shared covariance matrix, and its inverse $(\mathbf{\Phi}\mathbf{\Phi}^\top +  \mathbf{\Gamma})^{-1}$ defining the shared precision matrix. The adjacency matrix of the shared GGM is then constructed using Equation \ref{eq:parkour}. Similarly,  the study-specific covariance matrix 
 is given by $\mathbf{\Lambda}_{s}\mathbf{\Lambda}_{s}^\top + \mathbf{H}_s$ and its inverse $(\mathbf{\Lambda}_{s}\mathbf{\Lambda}_{s}^\top + \mathbf{H}_s)^{-1}$ defines the study-specific precision matrix.  The adjacency matrices and the corresponding GGMs as implied by the covariance decomposition in Figure \ref{fig:decomp}, are illustrated in Figure  \ref{fig:precmatSample}.

To estimate $\mathbf{\Gamma}$ and $\mathbf{H}_s$, we adopt the expectation-conditional maximization (ECM) algorithm developed by \cite{de2019multi}. We denote the full parameter vector by $\boldsymbol{\theta} = (\mathbf{\Phi}, \mathbf{\Lambda}_1, \dots, \mathbf{\Lambda}_S, \mathbf{\Gamma}, \mathbf{H}_1, \dots, \mathbf{H}_S)$. 

The objective function $Q(\boldsymbol\theta)$ is the expected complete log likelihood of $\boldsymbol\theta$, conditioned on the observed data and taken with respect to the latent variables $\mathbf{f}_{is}$ and $\mathbf{l}_{is}$:
\begin{eqnarray*}
Q(\boldsymbol\theta) &=&  \mathbb{E}[\ell_c(\boldsymbol\theta)|\mathbf{x}_{is}, \boldsymbol\theta^{(t)}] \\
 &\propto& -\frac{1}{2}\log |\bsG + \mathbf{H}_s| \\ &-&\frac{1}{2}tr\{(\bsG + \mathbf{H}_s)(\mathbf{x}_{is}\mathbf{x}_{is}^\top + \bsPH\mathbf{f}_{is}\mathbf{f}_{is}^\top\bsPH^T + \bsL_{s} \mathbf{l}_{is}\mathbf{l}_{is}^\top\mathbf{\Lambda}_{s}^\top  \\
&-& 2\mathbf{x}_{is}\mathbf{f}_{is}^\top\bsPH^\top - 2\mathbf{x}_{is}\mathbf{l}_{is}^\top\bsL_s^\top+2\bsPH\mathbf{f}_{is}\mathbf{l}_{is}^\top\mathbf{\Lambda}_{is}^\top )\}
\label{eq:nonid}
\end{eqnarray*}

\clearpage
\subsection{Assessing edge significance with empirical p-values}
We assign significance of the GGM edge weights using an empirical null approach that allows us to estimate the null distribution of the edge weights for a model following the form of Equation \ref{eq:MSFAX}. Specifically, for $S$ studies on $P$ predictors, with $K$ shared factors and $\{J_s\}$ study-specific factors, we define the null parameters:

\begin{eqnarray*}
\bsPH^{null} &=& \left(\begin{matrix}
diag(K) \\
\vspace{0.02in} \\
\mathbf{0}_{(P-K) \times K}
\end{matrix}\right) \\
\vspace{0.5in}&& \\
\bsL_s^{null} &=& \left(\begin{matrix}
\mathbf{0}_{(K + \sum_{s < S}J_s) \times J_s} \\
\vspace{0.02in} \\
diag(J_s) \\
\vspace{0.02in} \\
\mathbf{0}_{(P-(K + \sum_{s \leq S}J_s)) \times J_s}
\end{matrix}\right), s = 1, \dots, S \\
\vspace{0.5in}&& \\
\mathbf{\Gamma}^{null} &=& 0.5*diag(P) \\
\mathbf{H}_s^{null} &=& 0.5*diag(P), s = 1, \dots, S \\
\end{eqnarray*}

We then sample $N$ multivariate normal factor scores according to the model assumptions as $\mathbf{f}_{is} \sim N_k(\bm{0}, \mathbf{I}_k)$ and $\mathbf{l}_{is} \sim N_k(\bm{0}, \mathbf{I}_{j_s})$. We further sample a noise term:

\begin{equation}
    \mathbf{e}_{is} \sim N_P(\bm{0}, \mathbf{\Gamma}^{null} + \mathbf{H}_s^{null} = diag(P))
\end{equation}

This yields the null data:

\begin{equation}
    \mathbf{X}_{s}^{null} = \left\{\Phi^{null}\mathbf{f}_{is} + \Lambda_s^{null}\mathbf{l}_{is} + \mathbf{e}_{is};  i = 1, \dots, N \right\}
\end{equation}

Applying MSFA-X to $\mathbf{X}_{s}^{null}$ yields the estimated shared and study-specific GGMs under a null case of $K$ shared latent factors and $J_s$ study-specific latent factors exhibiting no conditional associations between any of the $P$ variables. 

Let $\mathcal{E}_{shared}$ represent the set of shared GGM edge weights obtained from this null model. Two-tailed p-values for each estimated shared edge $\hat{E}$ are then calculated as:

\begin{equation}
P(\hat{E}_{ab}) = \frac{1}{|\mathcal{E}_{shared}|}\sum_{E \in \mathcal{E}_{shared}}\mathbb{I}(|E|>|\hat{E}|)
\end{equation}

Similarly, the distribution of $\mathcal{E}_{study,s}$ is used to calculate p-values for the estimated study-specific GGM edges.

\subsection{Identifiability Considerations}

As in the standard factor analysis, a number of identifiability concerns arise in MSFA-X and several conditions are necessary to specify an identifiable model.

The first identifiability issue is the orthogonal rotation indeterminacy in the factor loadings $\bsPH$ and $\bsL_s$.  Specifically, for any $k \times k$ orthogonal matrix $ \mbQ^*$ and $j_s \times j_s$ orthogonal matrices $ \mbQ_s^*, s=1, \dots,S$, we have:

\begin{eqnarray*}
\bsPH Q^*(\bsPH \mbQ^*)^\top = \bsPH  \mbQ^* ( \mbQ^*)^\top \bsPH^\top = \bsPH\bsPH^\top \\
\bsL_s  \mbQ_s^*(\bsL_s  \mbQ_s^*)^\top = \bsL_s  \mbQ_s^* ( \mbQ_s^*)^\top \bsL_s^\top = \bsL_s\bsL_s^\top
\end{eqnarray*}

To resolve this issue, we impose a block lower triangular constraint on each matrix $\Phi,\Lambda_1, \dots,\Lambda_s$ \cite{de2019multi,lopes2004bayesian}.

The second identifiability issue is label switching, which arises from the permutation of the columns of $\bsPH$ and $\bsL_s$ according to permutations of the corresponding rows of $\mathbf{f}_{is}$ and $\mathbf{l}_{is}$. In practice, the lower triangular constraint also addresses this issue. 

A third identifiability issue, unique to multi-study factor analysis, is the need for sufficient constraints to separate shared and study-specific components. The MSFA framework results in $S$ equations of the form $\bsS_s - \bsP_s = \bsPH\bsP^\top + \bsL_s\bsL_s^\top$, but there are $S+1$ unknown matrices ($\bsPH$ and $\bsL_s, s=1, \dots, S$). To uniquely identify the solution, we impose the constraint that the matrix $\Omega = [\bsPH, \bsL_1, \dots, \bsL_s]$ has full column rank $Rk(\Omega) = K + J_1 + \dots + J_s$ and that $K + J_1 + \dots + J_s \leq P$, ensuring that the total number of factors does not exceed the number of observed variables; these constraints are described in detail in \cite{de2019multi}.  

A fourth identifiability issue relates to the number of free parameters in the covariance matrix $\bsS_s$. These parameters include those in the factor loading matrices, $\bsPH,\bsL_1, \dots, \bsL_s$, as well as in the error covariance matrices $\bsG$ and $\mathbf{H}_s$. Taking into account the lower triangular constraint, the total number of free parameters is:
\begin{equation}
PK - K(K-1)/2 + \sum_{s=1}^S\{PJ_s - J_s(J_s-1)/2\} + (S+1)P.
\label{eq:nparams}
\end{equation}

To ensure that $\bsS_s$is estimable, the total number of parameters must be less than the number of unique elements in the sample covariance matrices across all studies. This results in the following constraint:
\begin{equation*}
PK - K(K-1)/2 + \sum_{s=1}^S\{PJ_s - J_s(J_s-1)/2\} + (S+1)P \leq S\left(\frac{P(P+1)}{2}\right).
\end{equation*}

A sufficient, but more restrictive, condition to meet this constraint is that both $K + \sum_{s=1}^S J_s \leq P$ and $K + \sum_{s=1}^S J_s > S$ hold \cite{de2019multi}. 

A fifth identifiability issue is sign indeterminacy in the factor loadings.   Changing the sign of both the factor loadings and the factor scores results in an equivalent model. The sign of factor loadings should not be interpreted as indicating a positive or negative association. This issue does not affect estimation but should be considered when interpreting results.

The final identifiability issue is specific to our extension to graphical modeling, MSFA-X, and is a property of the decomposition of $\bsP_s$ into $\mathbf{\Gamma}$ and $\mathbf{H}_s$. Specifically, for given values of the loadings $\bsPH, \bsL_1, \dots, \bsL_s$, any values of $\mathbf{\Gamma}$ and $\mathbf{H}_s$ that add to the same sum will result in an equivalent expected conditional log likelihood ($Q(\boldsymbol\theta)$) and, consequently, an equivalent model.

We address this identifiability issue as follows. From equations (2) and (5), we have the constraint that: 

\begin{eqnarray*}
\Psi_s &=& \Gamma + \mathbf{H}_s
\end{eqnarray*}

where $\Psi_s, \Gamma,$ and $\mathbf{H}_s$ are all diagonal matrices. For every study $s$, the $p$th diagonal element in $\Psi_s$,  can be expressed as $\psi_{ps} = \gamma_p + \eta_{ps}$ or equivalently $\gamma_p =\psi_{ps} - \eta_{ps}$. 

Given the constraints that $\gamma_p>0, \eta_{ps} > 0 \space \space \forall s$, our model provides a lower and upper bound on $\gamma_p$ that define the region of non-identifiability:

\begin{equation}
    \gamma_p \in (0, min_{s}\psi_{ps})
\end{equation}

\begin{figure}[h!]
    \centering
    \includegraphics[width=0.9\textwidth]{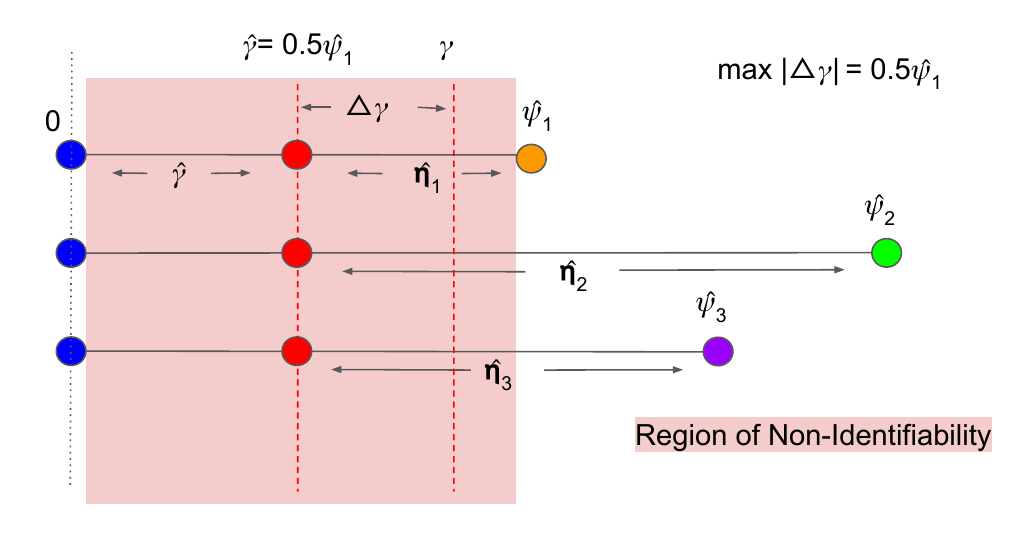}
    \caption{A visual of the region of non-identifiability for one predictor in the case of three studies ($S=3$). $\gamma$: true, unknown shared noise for this predictor; $\hat{\psi}_s$: estimated overall noise; $\hat{\gamma}$: estimated shared noise; $\hat{\eta}_s$: estimated study-specific noise; $\Delta\gamma$: error in estimating $\gamma$.}
    \label{fig:nonid}
\end{figure}

 Figure \ref{fig:nonid} illustrates this situation for a single parameter when $S=3$ studies.  Noting that, because of the properties of the ECM algorithm, $\hat{\psi}_{ps}$ is a consistent estimator of $\psi_{ps}$ \cite{de2019multi}. Therefore, the magnitude of the deviation of $\hat\gamma_p$ from the true value $\gamma_p$ is bounded: $\Delta\gamma_{p} = | \hat\gamma_p - \gamma_p| < 0.5 \ min_{s} \hat{\psi}_{ps}$. Because this error is bounded, and the bounds of the error are consistently estimable, we propose a strategy of estimating $\gamma_p$ at the center of this non-identifiability region. Specifically, the MSFA-X estimate is taken to be: $\hat{\gamma}_p = 0.5*min_{s} \hat{\psi}_{ps}$. Then, the study-specific covariance matrix of the error term can be estimated as $\hat{\eta}_{ps} = \hat{\psi}_{ps} - \hat{\gamma}_p$. 

We propose that this approach yields a practically meaningful model, despite its lack of statistical consistency. We validate this claim through extensive simulation studies, where errors can be quantitatively assessed against a known gold standard. 




\subsection{Determining the number of factors} Selecting the number of shared factors ($k$) and the number of study-specific factors ($j_s, s=1, \dots, S$) is crucial in multi-study factor analysis. However, this is not the primary goal of the present work, as our main objective is to estimate the shared and study-specific covariance and precision matrix and the corresponding shared and condition-specific networks. To assess the impact of selecting the correct number of factors on networks estimation, simulation results are presented using both the true, known, number of factors and the number of factors estimated using the multiple regression-based screeplot analysis approach of \cite{zoski1993using}. This comparison allows us to decouple the issue of selecting the number of factors from assessing the accuracy of shared and study-specific network structures, while also demonstrating how MSFA-X would perform in practice, where the true number of factors is unknown. Further detail is provided in the Supplement.

\section{Simulation Studies}

To assess the robustness of MSFA-X, we conducted extensive simulation across ten different settings, varying the number of factors, sample size, and number of studies (Table \ref{tab:simDesign}). For each setting, we generate a collection of 100 different datasets $\mathbf{x}_{is}$ sampled from the multivariate distributions $ \mathbf{x}_{is} \sim \mathcal{N}(\mathbf{0}, \bsPH \bsPH^\top + \mathbf{\Lambda}_s \mathbf{\Lambda}_s^\top + \bsG + \mathbf{H}_s)$. The entries of the shared and study-specific factors $\Phi$ and $\mathbf{\Lambda}_s, s \in 1, \dots, S$, are set equal to $0, -1,$ or $1$, ensuring that each loading had either the maximum possible magnitude (1 or -1) or no contribution (0). The elements of the residual covariance matrices $\bsG = diag(\gamma_1, \dots, \gamma_p)$ and $\mathbf{H}_s = diag(\eta_{1s}, \dots, \eta_{ps})$, are drawn randomly from uniform distributions (Table \ref{tab:simDesign}). Factor scores are sampled from $\mathbf{f}_{is} \sim N_k(\bm{0},\bm{I}_k)$ and $\mathbf{l}_{is} \sim N_{j_s}(\bm{0},\bm{I}_{j_s})$, while the residual noise are samples as $\mathbf{e}_{is} = \mathbf{g}_{is} + \mathbf{h}_{is} \sim N_p(\bm{0},\bsG + \mathbf{H}_s)$. We then generate simulated multivariate normal data with all these components, consistent with the underlying factor structure specified in Equation \ref{eq:MSFAX}.

To evaluate the performance of MSFA-X, we explored two benchmark methods. The first approach is based on the application of the graphical lasso without the multi-study factor model. This method involved three steps:  (i) estimate separate GGMs for each study using the graphical lasso, (ii) estimate a a shared GGM by pooling the data from all $S$ studies and applying the graphical lasso, and (iii) define the study-specific GGMs as the graph obtained by subtracting the adjacency matrix of the shared GGM estimated among the pooled data in (ii) from the adjacency matrix of each within-study GGM estimated in (i). 

The second benchmark method we explored applied the joint graphical lasso (JGL) to estimate the study-specific networks in step (i). JGL requires selecting two tuning parameters, $\lambda_1 \geq 0$, which controls the networks' sparsity, and $\lambda_2 \geq 0$, which controls the similarity between groups. We used a BIC-type criterion to select the optimal tuning parameters over a coarse grid of options ($\lambda_1 \in (0,0.001, 0.01, 0.1,1), \lambda_2 \in (0, 0.001, 0.01, 0.1, 1)$). However, preliminary JGL runs across the ten simulation settings described in Table \ref{tab:simDesign} consistently selected $\lambda_2=0$, reducing JGL to the standard graphical lasso. This makes our second benchmark equivalent to the first approach described above. Thus, we report only results from the graphical lasso approach.

\begin{table}[h!]
    \centering
    \caption{Simulation study design.}
    \begin{tabular}{p{1cm}p{2.25cm}p{1.5cm}p{1cm}p{1cm}p{1cm}p{1.5cm}p{1.5cm}p{1cm}p{1.5cm}}
    Setting & Description & \# Studies ($S$) &	$n_s$& $p$ & $k$ &	$j_s$ & $(\Gamma, \mathbf{H}_s)$ & Exact Zeros\footnote{``Zero entries" were set to either exactly zero or to a small number below a Bonferroni-corrected significance threshold for the Fisher-transformed partial correlation. See Supplement for details.)} \\
    \hline 
1 & Baseline &	2&	1600, 1600	& 60 &	2 &	2,2	& equal & Yes \\
2 & Change \# of predictors &	2 &	1600, 1600	& 12 &	2 &	2,2	& equal & Yes \\
3 &Change \# of studies	& 4	& 1600, 1600, 1600, 1600 &	60	& 2	& 2,2,2,2	& equal & Yes \\
4 &Change \# of factors	& 2	& 1600, 1600 &	60	& 4	& 3,5 & equal & Yes \\
5 &Small sample size &	2& 250, 250	& 60 & 2 & 2,2 & equal & Yes \\
6 &Unequal sample size &	2 &	1600, 250 &	60	& 2	& 2,2 & equal & Yes \\
7 & Unequal noise &	2 &	1600, 1600 &	60 &	2 &	2,2	& $\Gamma > \mathbf{H}_s$ & Yes \\
8 & Unequal noise & 2 &	1600, 1600 &	60 &	2 &	2,2	& $\mathbf{H}_s > \Gamma$ & Yes\\
9 & Not true zeros &	2&	1600, 1600	& 60 &	2 &	2,2	& equal & No \\			
10 & Mimic HAPO application &	2 &	2887, 576 &	60 &	2 & 2,2	& equal & No \\
    \end{tabular}
    \label{tab:simDesign}
\end{table}

To assess MSFA-X's performance, we use the modified RV coefficient \cite{smilde2009matrix} to assess similarities between the estimated and true adjacency matrices. The modified-RV coefficient is an extension of the RV coefficient of \cite{robert1976unifying} developed for high-dimensional data. Additionally, we examined Euclidean distance and a cosine similarity metrics to further assess performance (Supplementary Section 1.2). 

\begin{figure}
    \centering
    \begin{subfigure}[b]{0.4\textwidth}
    \centering
    \includegraphics[width=\textwidth]{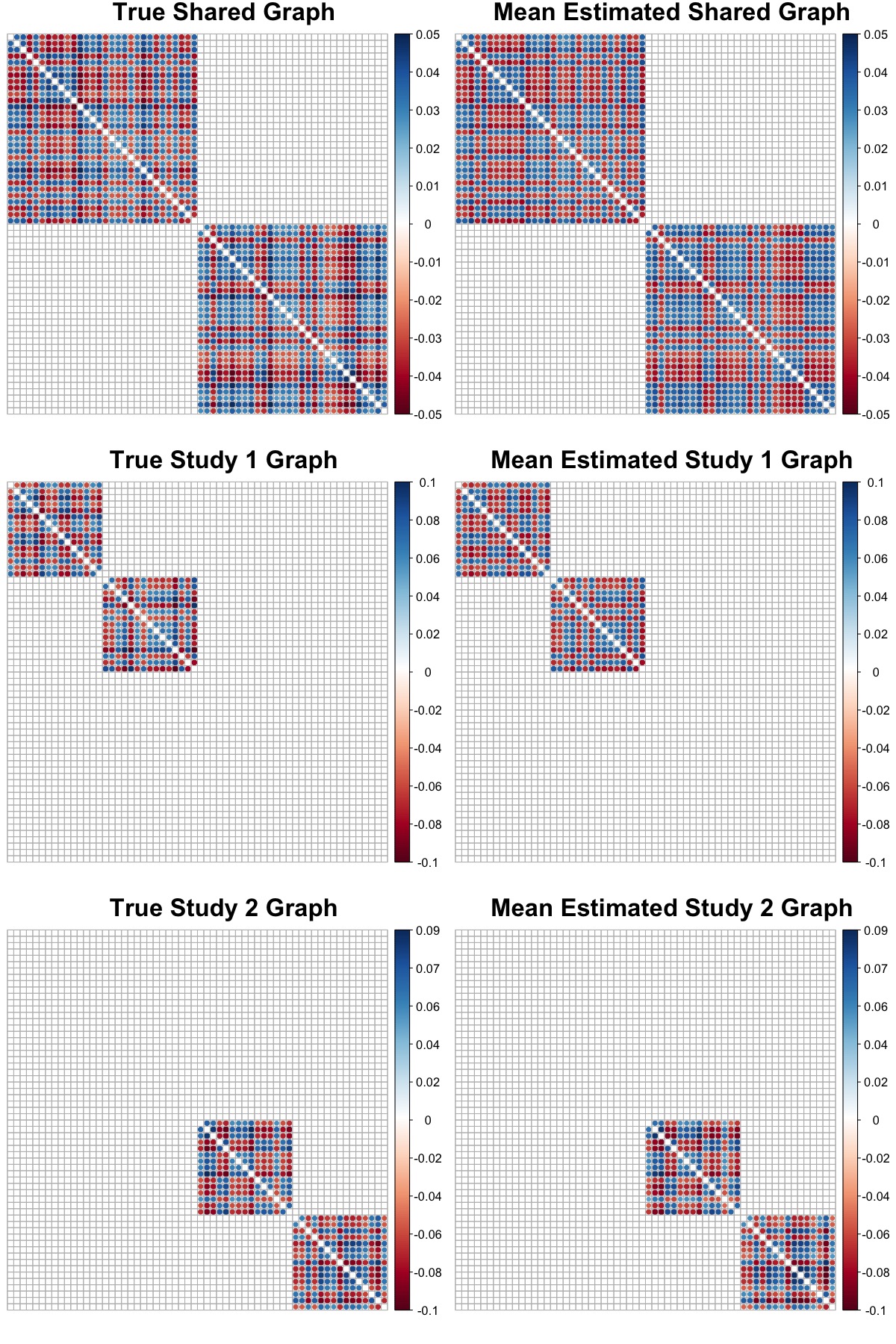}
    \caption{Setting 1}
    \end{subfigure}
    \begin{subfigure}[b]{0.4\textwidth}
    \centering
    \includegraphics[width=\textwidth]{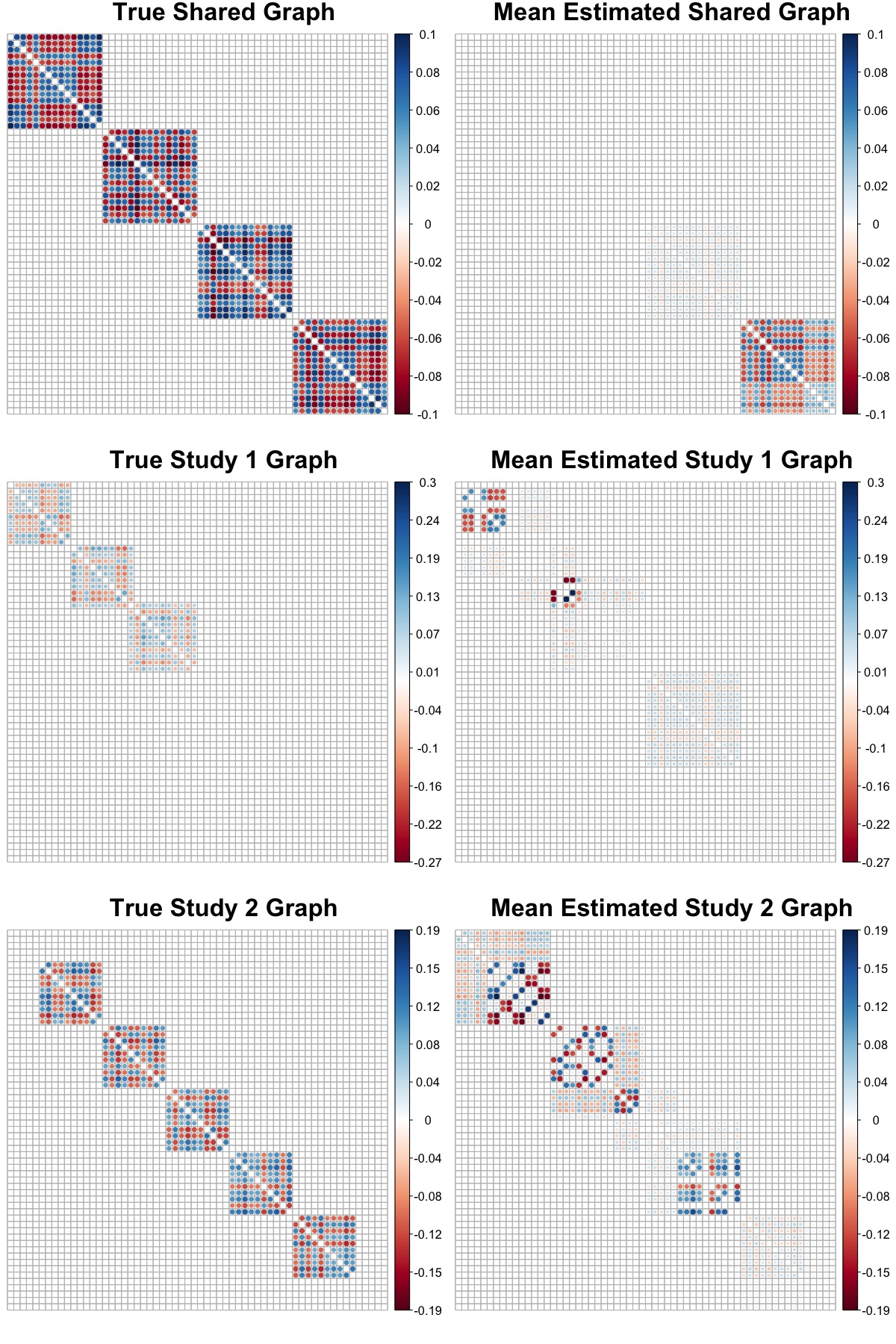}
    \caption{Setting 4}
    \end{subfigure}
    \caption{Mean estimated GGM adjacency matrices across 100 simulated datasets for simulation setting 1 (baseline) and setting 4 (increased factor count).}
    \label{fig:visualsimresults}
\end{figure}

Table \ref{tab:mrv} reports the median and the corresponding 95\% quantile interval of the matrix RV coefficient between the estimated and true network in simulation settings 1,2,4,7, and 10. Results for additional settings and error metrics are available in the Supplement (Supplementary Tables 5-7; Supplementary Figures 1-3). A visual inspection of adjacency matrices estimated by MSFA-X shows that the general structure of GGMs was well recovered  (Figure \ref{fig:visualsimresults}; Supplementary Figures 4-11). These results demonstrate that MSFA-X successfully recovers graphical structures induced by latent variables and outperforms standard graphical modeling approaches, despite identifiability limitations.

\begin{table}
\begin{longtable}{llllll}
\caption{Matrix RV coefficients for simulation settings 1,2,4,7, and 10. A higher value of the matrix RV coefficient corresponds to better estimation of the gold standard matrix in the simulation. The maximum value for the RV coefficient is 1. Est. Fac. refers to a simulation in which we estimated the number of factors using our mreg-based method. True Fac. refers to a simulation in which we used the true number of factors from the known data generating process.} \\
Method&Setting&Study&Median&2.5th percentile&97.5th percentile \\
\hline
\hline
\endhead
MSFA-X: Est. Fac.&Setting 1&Shared&0.9946&0.9938&0.9954 \\
MSFA-X: True Fac.&Setting 1&Shared&0.9946&0.9938&0.9954 \\
glasso&Setting 1&Shared&0.5412&0.5394&0.5428 \\
MSFA-X: Est. Fac.&Setting 1&Study 1&0.9925&0.989&0.9947 \\
MSFA-X: True Fac.&Setting 1&Study 1&0.9925&0.989&0.9947\\
glasso&Setting 1&Study 1&0.0135&0.0123&0.0146\\
MSFA-X: Est. Fac.&Setting 1&Study 2&0.9948&0.9898&0.9968\\
MSFA-X: True Fac.&Setting 1&Study 2&0.9948&0.9898&0.9968\\
glasso&Setting 1&Study 2&0.015&0.0143&0.0155\\
\hline
MSFA-X: Est. Fac.&Setting 2&Shared&0.7672&0.389&0.8522\\
MSFA-X: True Fac.&Setting 2&Shared&0.8313&0.3601&0.9952\\
glasso&Setting 2&Shared&0.6599&0.6518&0.6676\\
MSFA-X: Est. Fac.&Setting 2&Study 1&0.7935&0.1576&0.9185\\
MSFA-X: True Fac.&Setting 2&Study 1&0.8026&0.2127&0.9893\\
glasso&Setting 2&Study 1&0.0489&0.0354&0.0591\\
MSFA-X: Est. Fac.&Setting 2&Study 2&0.799&0.6682&0.8581\\
MSFA-X: True Fac.&Setting 2&Study 2&0.9308&0.3932&0.9945\\
glasso&Setting 2&Study 2&0.0634&0.0531&0.0742\\
\hline
MSFA-X: Est. Fac.&Setting 4&Shared&0.6262&0.5975&0.6291 \\
MSFA-X: True Fac.&Setting 4&Shared&0.9926&0.9913&0.9936 \\ 
glasso&Setting 4&Shared&0.512&0.5088&0.5159 \\
MSFA-X: Est. Fac.&Setting 4&Study 1&0.493&0.4509&0.5059\\
MSFA-X: True Fac.&Setting 4&Study 1&0.9873&0.9783&0.9911\\
glasso&Setting 4&Study 1&0.1008&0.0952&0.1051\\
MSFA-X: Est. Fac.&Setting 4&Study 2&0.6248&0.527&0.7668\\
MSFA-X: True Fac.&Setting 4&Study 2&0.9914&0.9883&0.9935\\
glasso&Setting 4&Study 2&0.0795&0.0758&0.0841\\
\hline
MSFA-X: Est. Fac.&Setting 7&Shared&0.9989&0.9986&0.9992\\
MSFA-X: True Fac.&Setting 7&Shared&0.9989&0.9986&0.9992\\
glasso&Setting 7&Shared&0.5426&0.5413&0.5441\\
MSFA-X: Est. Fac.&Setting 7&Study 1&0.9855&0.9812&0.9881\\
MSFA-X: True Fac.&Setting 7&Study 1&0.9855&0.9812&0.9881\\
glasso&Setting 7&Study 1&0.015&0.0143&0.0158\\
MSFA-X: Est. Fac.&Setting 7&Study 2&0.991&0.9864&0.9947\\
MSFA-X: True Fac.&Setting 7&Study 2&0.991&0.9864&0.9947\\
glasso&Setting 7&Study 2&0.0153&0.0146&0.0159\\
\hline
MSFA-X: Est. Fac.&Setting 10&Shared&0.8475&0.737&0.9248\\
MSFA-X: True Fac.&Setting 10&Shared&0.9952&0.994&0.9961\\
glasso&Setting 10&Shared&0.5607&0.5569&0.5646\\
MSFA-X: Est. Fac.&Setting 10&Study 1&0.9921&0.7799&0.9964\\
MSFA-X: True Fac.&Setting 10&Study 1&0.9946&0.9892&0.9964\\
glasso&Setting 10&Study 1&0.0274&0.0207&0.0361\\
MSFA-X: Est. Fac.&Setting 10&Study 2&0.7932&0.5376&0.9072\\
MSFA-X: True Fac.&Setting 10&Study 2&0.9983&0.997&0.999\\
glasso&Setting 10&Study 2&0.0634&0.0622&0.0644\\
\label{tab:mrv}
\end{longtable}
\end{table}

\clearpage

\section{Application: Metabolomic Networks of Gestational Diabetes in the HAPO Study}

\begin{figure}[h!]
    \centering
    \includegraphics[height=0.6\textheight]{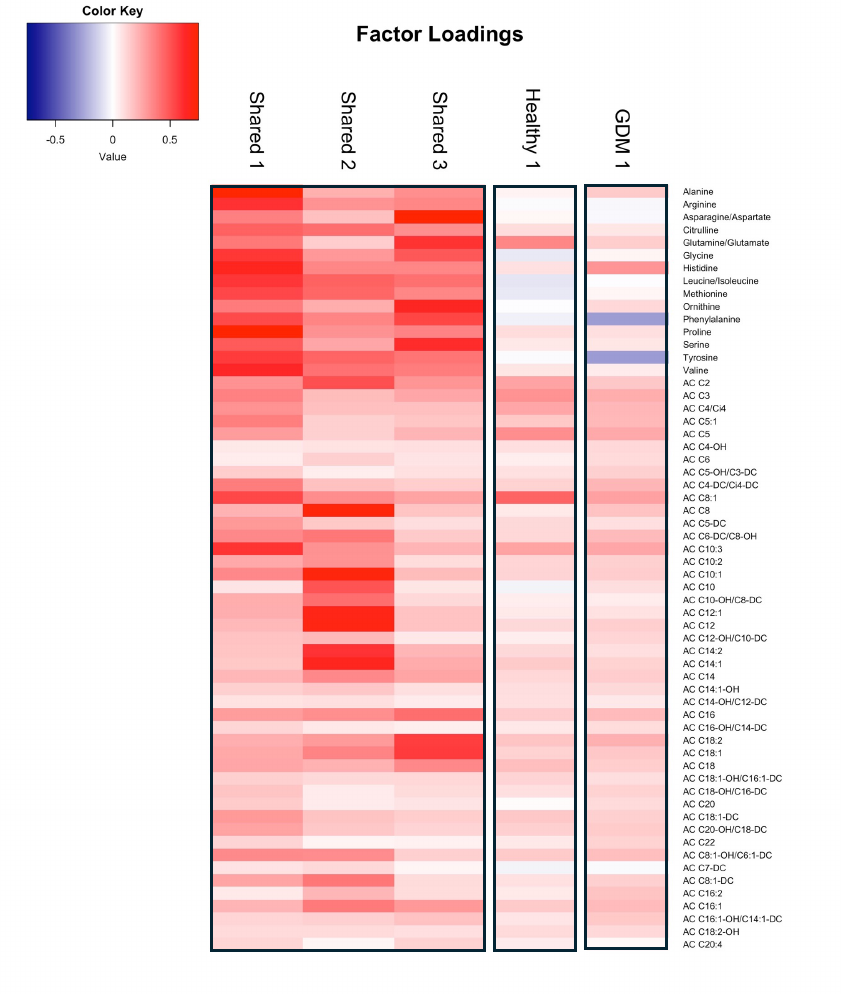}
    \caption{MSFA-estimated factor loadings for the shared factors and factors specific to women without and with GDM in the HAPO metabolomics dataset. The varimax transformation has been applied to the matrix of shared factors. For all factors, factor signs have been assigned such that the loading entries of largest magnitude are positive.}
    \label{fig:hapoheatmap}
\end{figure}

To apply MSFA-X to the HAPO data, we first selected the number of shared and condition-specific factors with the multiple regression approach described in the Supplement. This approach identified 3 shared factors, one study-specific factor for women without GDM, and one study-specific factors for women with GDM (Figure \ref{fig:hapoheatmap}).

The three shared components confirm coordination among amino acids (AAs) and medium- to long-chain acylcarnitines (AC C8, AC C10, AC C12, AC C14) in metabolic networks that have been well-documented in this HAPO population \cite{scholtens2016metabolic}. The first shared component loads heavily on the six essential AAs assayed in this set of metabolites (histidine, leucine/isoleucine, methionine, phenylalanine, and valine) while the third shared component loads more heavily on non-essential or conditionally essential AAs (asparagine/aspartate, glutamine/glutamate, serine) as well as phenylalanine and ornithine, a non-proteogenic AA that has been associated with GDM \cite{cetin2005maternal}. The study specific-factor for women without GDM suggests coordination among glutamine/glutamate, the short-chain acylcarnitines AC C2, AC C3, AC C4/Ci4, AC C5:1, and AC C5 as well as AC C8:1. The study-specific factor for women with GDM demonstrates a uniquely strong dependency between three aromatic amino acides: histidine,phenylalanine, and tyrosine. Notably, phenylalanine and tyrosine are related through direct conversion through a single hydroxyl group, and both have documented associations with GDM \cite{wu2024association}.

\begin{figure}
    \centering
    \begin{subfigure}[b]{\textwidth}
    \centering
    \includegraphics[width=0.315\textwidth]{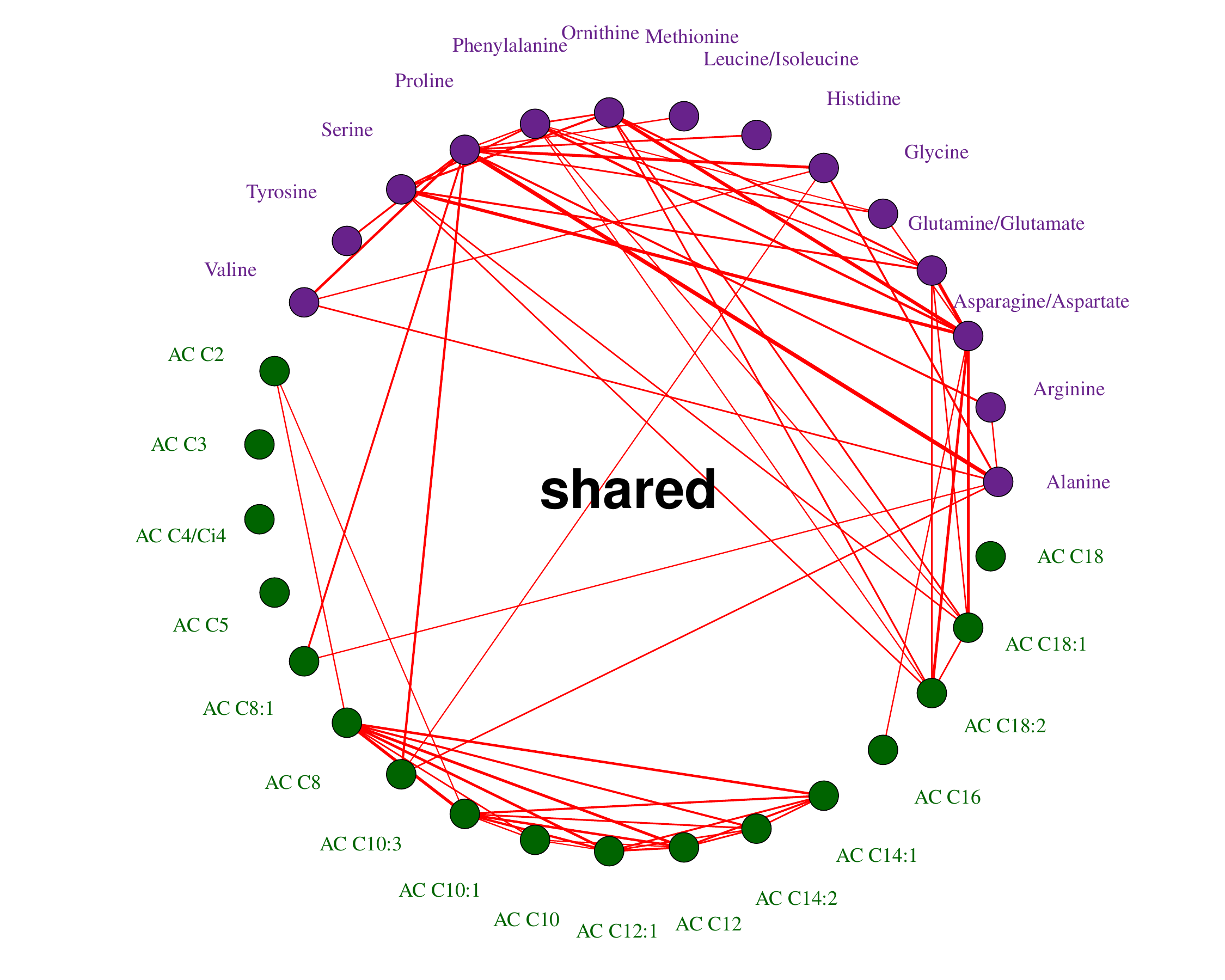}
    \includegraphics[width=0.53\textwidth]{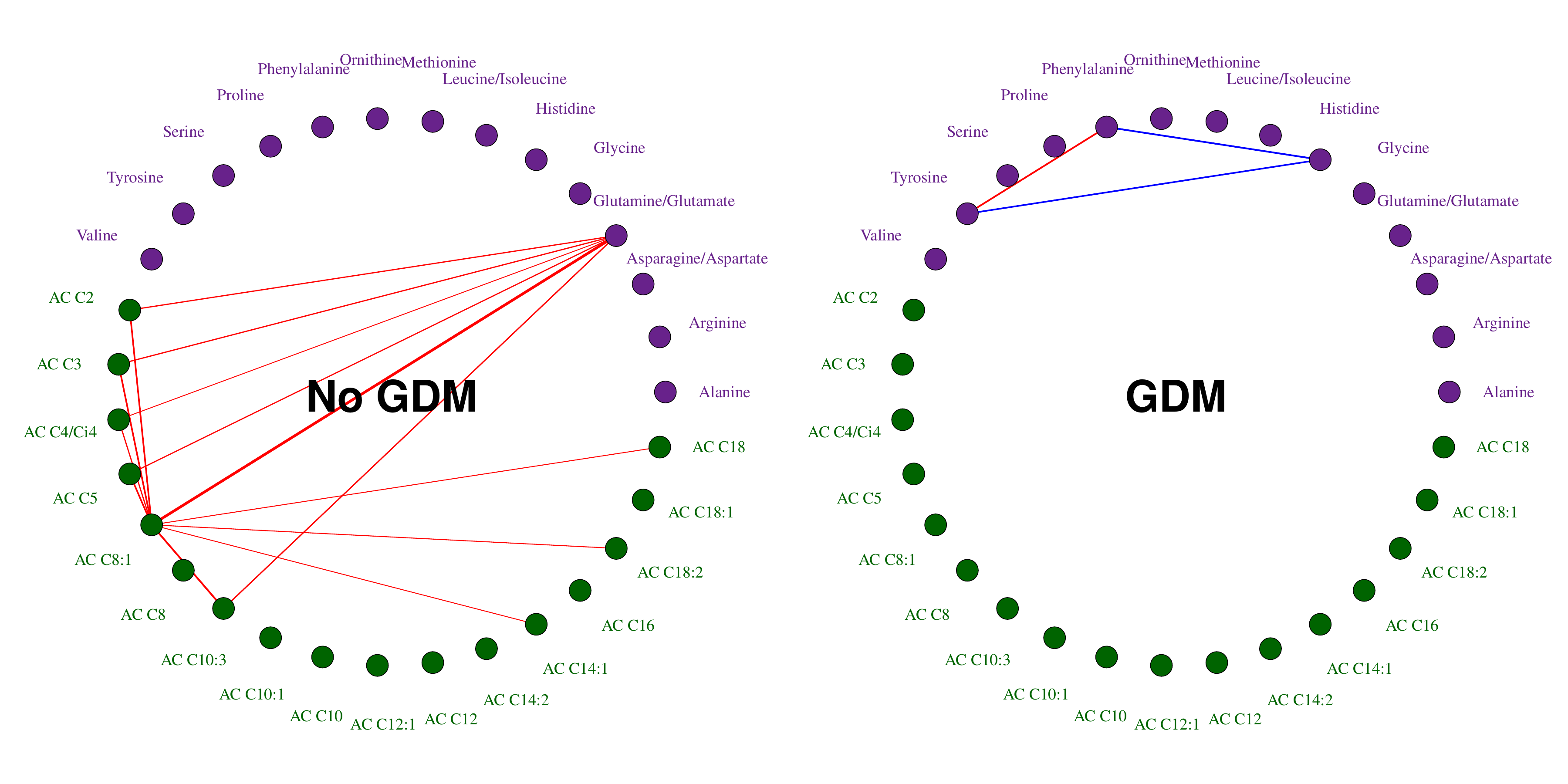}
    \caption{}
    \label{fig:gdm_results}
    \end{subfigure}
    \\ 
    \begin{subfigure}[t]{0.4\textwidth}
    \centering
    \includegraphics[width=\textwidth]{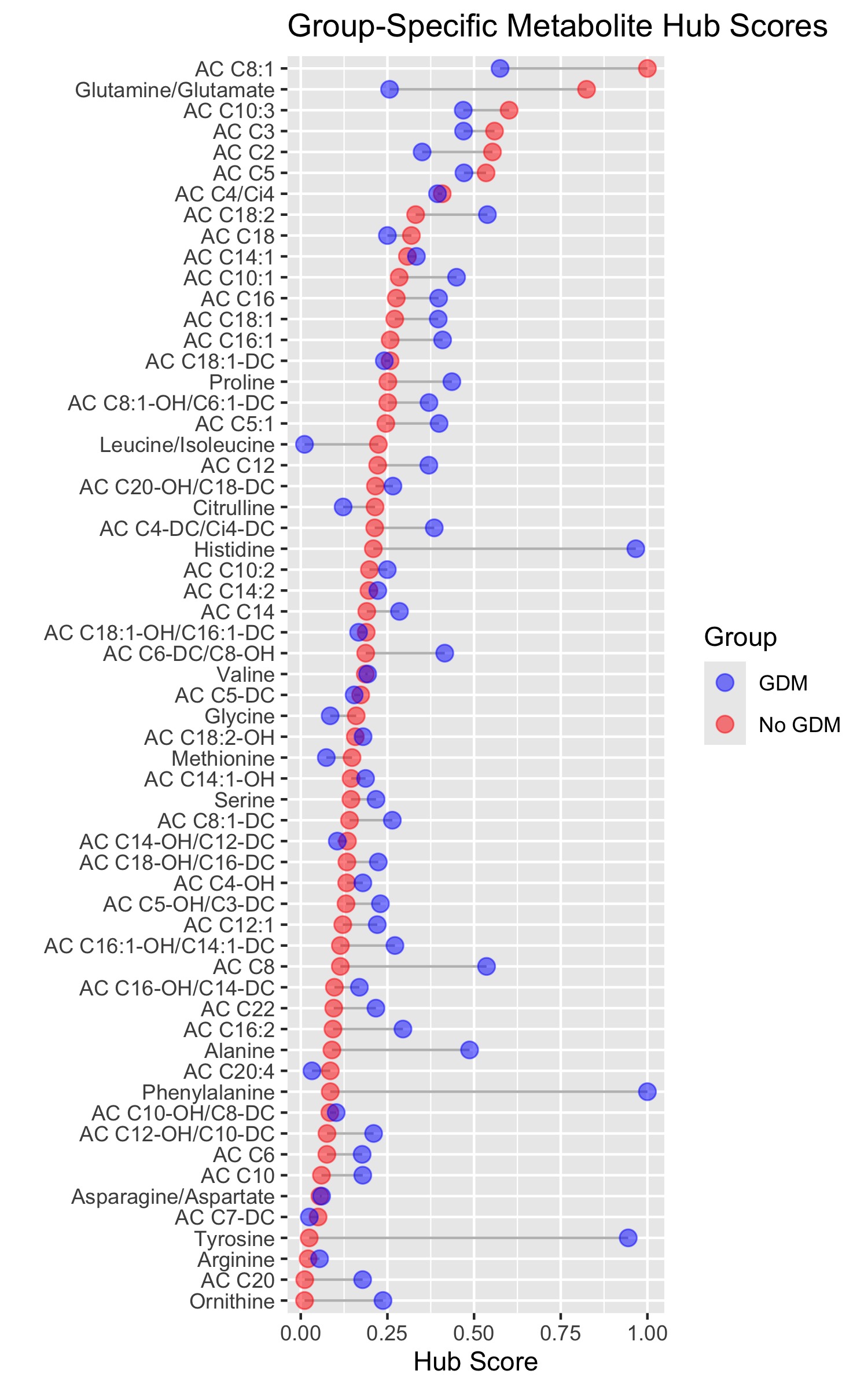}
    \caption{}
    \label{fig:lollipop}
    \end{subfigure}
    \begin{subfigure}[t]{0.4\textwidth}
    \centering
    \includegraphics[width=\textwidth]{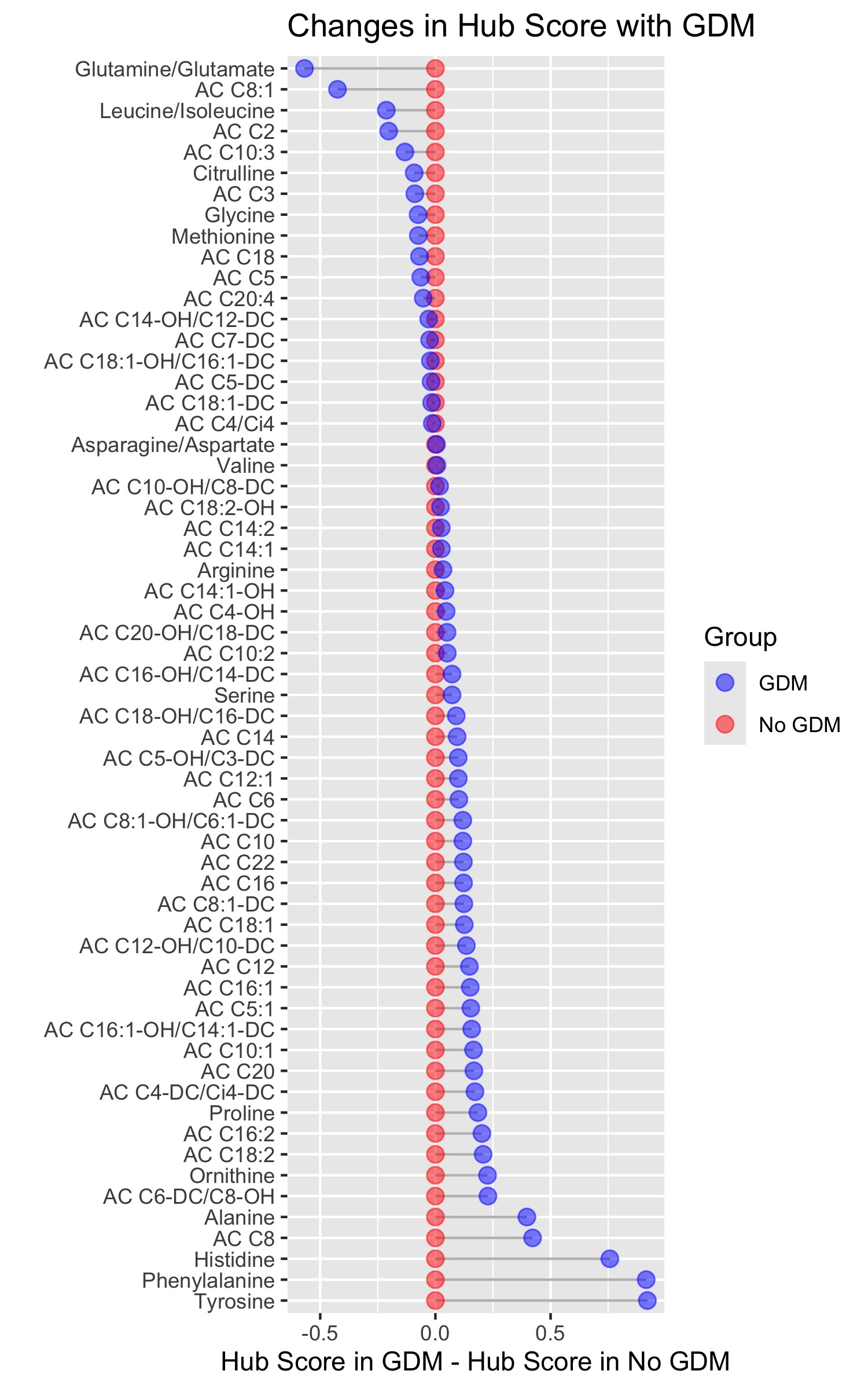}
    \caption{}
    \label{fig:lollipop2}
    \end{subfigure}
    \caption{(a) MSFA-X estimated networks of conditional dependencies between metabolite changes in response to the oral glucose tolerance test. Green (purple) nodes indicate amino acids (acylcarnitines). Red (blue) edges indicate positive (negative) partial correlations. Edge width is proportional to the magnitude of the partial correlation. (b-c) Hub score of metabolites in the networks of women with GDM (blue) and without GDM (red). In (b), metabolites are ordered by hub score in the network of women without GDM. In (c), metabolites are ordered by difference in hub score relative to the hub score among women without GDM.}
\end{figure}

The thresholded shared and study-specific GGMs are shown in Figure \ref{fig:gdm_results}, with thresholding based on empirical p-values estimated using the approach in Section 3.2.  The shared network reveals two distinct highly-connected subnetworks. One subnetwork consists of the amino acids (AAs) along with four long-chain acylcarnitines (ACs; AC C16, AC C18, AC C18:1, and AC C18:2).  The second subnetwork consists of several short- and medium-chain ACs, with the majority of edges connecting the medium-chain ACs.

The study-specific networks reveal clear differences in metabolism after glucose uptake between women with and without GDM. Among women without GDM, glutamine/glutamate is a highly connected node serving as a connector between the AAs and the short- and medium-chain ACs. Additionally, AC C8:1 exhibits conditional associations with several short-chain ACs (AC C2, AC C3, AC C4/Ci4, AC C5). In contrast, leucine/isoleucine and glutamine/glutamate have much smaller hub score in the GDM network, and glutamine/glutamate does not function as a connector between AAs and ACs. Other AA patterns also differ, with histidine, phenylalanine, and tyrosine exhibiting strong conditional associations that are specific to women with GDM. Conducting the same analysis with the graphical lasso benchmark reveals that the benchmark was unable to disentangle these metabolic patterns between women with and without GDM (Supplementary Section 3.3).

To further assess these differences quantitatively, we also analyzed the hub score, an eigenvector-based measure of node centrality that indicates a node's overall influence in the network \cite{kleinberg1998authoritative}. Using the \texttt{hub\_score} function of the \texttt{igraph} R package, we compared hub scores of each metabolite between women with and without GDM (Figure \ref{fig:lollipop}, Figure \ref{fig:lollipop2}). This analysis further corroborates the lower involvement glutamine/glutamate and the higher involvement of histidine, phenylalanine, and tyrosine in the network of women with GDM compared to women without GDM. Additionally, leucine/isoleucine and AC C8:1 showed lower connectivity in women with GDM than those without.

The sample sizes between the two groups in this application are substantially different (N=576 women with GDM vs. N=2887 women without GDM), leading to higher power for detecting network edges in the shared network and the non-GDM network compared to the GDM network. To ensure that the observed differences between women with and without GDM were not due to sample size discrepancies, we conducted a sensitivity analysis in which we downsampled the non-GDM group to balance the sample size. To account for the stochasticity of the downsampling, we ran MSFA-X on 100 different such subsamples and took the mean estimates of the model parameters over the 100 iterations. We observed differences in the top edges of the study-specific networks (particularly the non-GDM network), likely due to the reduced power to estimate the study-specific signals; however, key network motifs were preserved in the shared and GDM-specific networks (Supplementary Figure 14) and the comparison of hub scores between the non-GDM and GDM networks yielded similar results to the original analysis (Supplementary Figure 15).  These results support the robustness of our analyses, i.e., the metabolic differences estimated by MSFA-X were not artifacts of unequal sample sizes  (Supplementary Section 3.4). 

In factor analysis, it is crucial to validate the estimated factors by associating each factor with an outcome of interest. In the HAPO Study, one such outcome is newborn adiposity, as it is a critical risk factor associated with childhood overweight \cite{winter2010newborn}. Thus, we validated our latent factors by estimating their association with newborn adiposity. 

For infants born to women participating in the HAPO Study, newborn adiposity was assessed using the sum of three skinfold measurements ("newborn sum of skinfolds") as previously described by \cite{catalano2012hyperglycemia}. This prior study found that GDM was positively associated with the newborn sum of skinfolds in the HAPO Study. In this spirit, we proceeded to validate our latent factors by modeling the association between the newborn sum of skinfolds and maternal factor scores for each of the five factors identified by MSFA-X (Figure \ref{fig:hapoheatmap}). The factor score for the $i^{th}$ individual is computed by projecting their vector of metabolite ratios $X^{i} =(X^{(i)}_{1}, \dots,X^{(i)}_{60})$  onto each factor. Given the factor loadings $F = (F_1, \dots, F_{60})$, this projection is calculated as:

\begin{equation}
proj_{F}(X^{i}) = \frac{\sum_{j=1}^{60} F_jX^{(i)}_j}{\sqrt{\sum_{j=1}^{60}F_j^2}}
\end{equation}

Larger values of $proj_{F}(X^{i})$ indicate a higher score of participant $i$'s metabolite profile with factor $F$.  For women without GDM, projections are calculated for the two shared factors $\Phi_1$ and $\Phi_2$ and the two non-GDM-specific factors $\Lambda_{11}$ and $\Lambda_{12}$. For women with GDM, projections are calculated for the two shared factors $\Phi_1$ and $\Phi_2$ and the two GDM-specific factors $\Lambda_{21}$ and $\Lambda_{22}$. 

\begin{figure}
\centering
\begin{subfigure}[b]{\textwidth}
\centering
\includegraphics[width=0.73\textwidth]{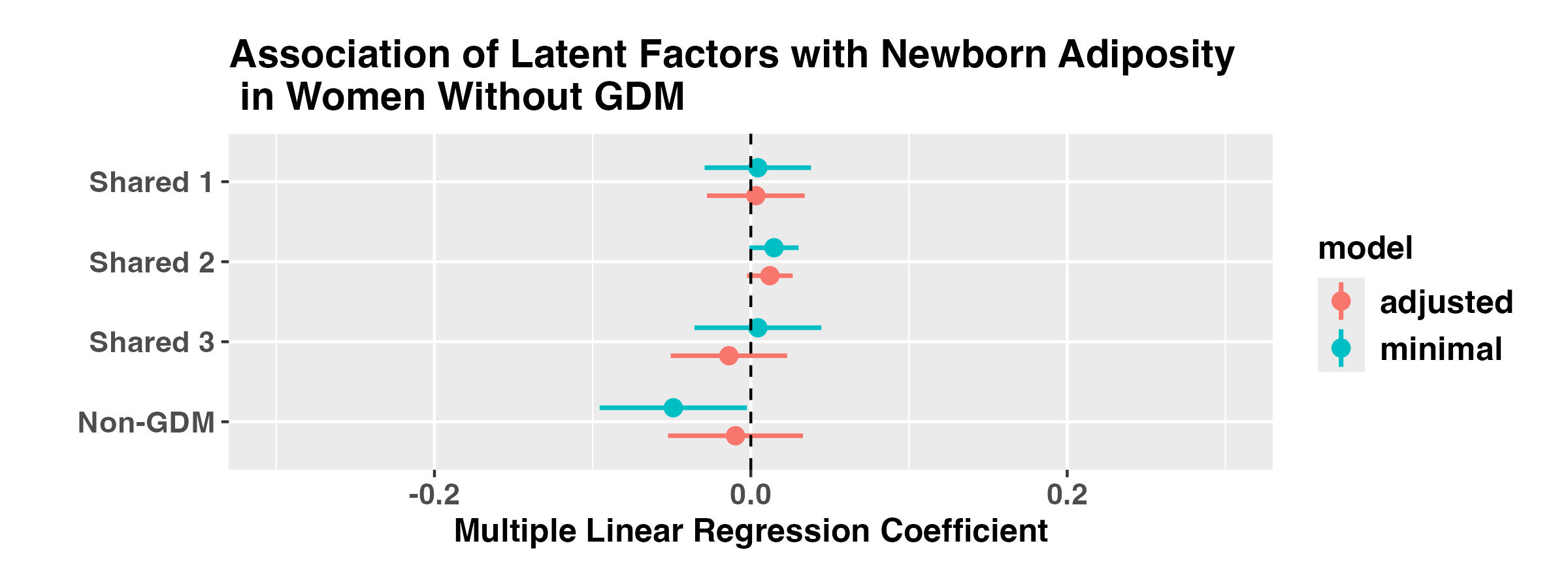}
\end{subfigure}
\\
\begin{subfigure}[b]{\textwidth}
\centering
\includegraphics[width=0.7\textwidth]{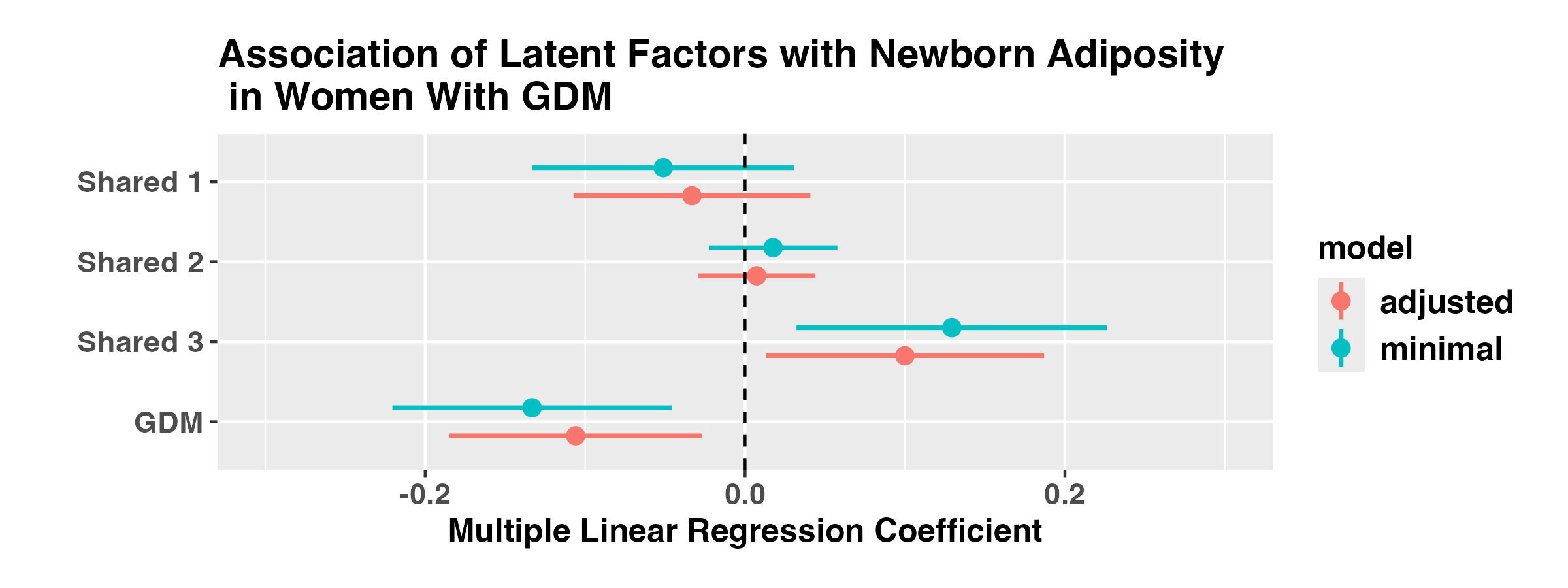}
\end{subfigure}
\caption{Latent variables learned by MSFA are associated with newborn adiposity in infants born to women with gestational diabetes.} 
\label{fig:forest}
\end{figure}

Within the groups of women with and without GDM, we separately fit multiple linear regression models to evaluate the associations between the log-transformed sum of skinfolds and each factor score. We fit both a minimal model and a full model adjusting for maternal and neonatal covariates including mean arterial pressure, BMI, age, and maternal height at the time of the OGTT, neonatal sex, gestational age at delivery, fasting plasma glucose, ancestry group, and storage time of the sample (Figure \ref{fig:forest}). In the minimal model among women without GDM, only the study-specific factor was associated with newborn adiposity ($\Lambda_{noGDM}$: effect = $-0.054$, $p = 0.024$, 95\% CI: $[-0.101, -0.007]$), and this association was attenuated after adjusting for confounders $(p=0.624)$. In contrast, in women with GDM, shared factors 2 and 3 as well as the GDM-specific factor were associated with newborn adiposity in a minimal model ($\Phi_2$: effect = $0.088$, $p=0.026$, 95\% CI: $[0.011, 0.166]$; $\Phi_3$: effect = $-0.110$, $p=0.026$, 95\% CI: $[-0.206, -0.014]$; $\Lambda_{GDM}$: effect = $-0.148$, $p=0.002$, 95\% CI: $[-0.241, -0.054]$;). The association of $\Lambda_{GDM}$ with the newborn sum of skinfolds remained significant after adjusting for covariates ($\Lambda_{GDM}$: effect = $-0.106$, $p = 0.014$, 95\% CI: $[-0.191, -0.021]$), while the association of $\Phi_3$ remained marginally significant (effect = $-0.085$, $p = 0.053$, 95 \% CI: $[-0.172,0.001]$).

\clearpage

\section{Discussion}

Here, we have introduced MSFA-X, a method leveraging multi-study factor analysis together with network modeling to facilitate the identification of shared and divergent network patterns across multiple conditions. By applying our method to metabolomic profiles of oral glucose tolerance testing during gestation in women with and without GDM in the HAPO study, we identified key latent metabolomic factors involved in normal and dysregulated glucose metabolism and associated with newborn adiposity. Network models estimated with MSFA-X showed that certain amino acids show markedly different network involvement depending on GDM status. Glutamine/glutamate is a key network hub in healthy glucose metabolism during pregnancy, highly connected to the short and medium-chain acylcarnitines, particularly AC C8:1. In contrast, glutamine/glutamate is unconnected to the rest of the network in GDM; several aromatic amino acids (histidine, tyrosine, phenylalanine) become network hubs instead. The absence of edges connecting glutamine/glutamate to the rest of the network largely decouples the acylcarnitines from the amino acids, suggesting systemic differences in glucose metabolism may be effectuated by disruption of key network nodes.  

We validated the biological relevance of the latent factors  underlying the MSFA networks by assessing their association with newborn adiposity, an important neonatal outcome that has been shown to be associated with maternal GDM \cite{logan2016development} and predictive of childhood overweight \cite{winter2010newborn}. Shared factor 3, which loads heavily on several non-essential amino acids (asparagine/aspartate, glutamine/glutamate, serine) as well as phenylalanine, ornithine, and acylcarnitines C18, C18:1, and C18:2, is of particular interest in that it is associated with newborn adiposity in women with GDM, but not in women without GDM. This suggests that in the presence of GDM, alterations in these maternal metabolites confer additional risk of newborn adiposity to the infant; however, in women without GDM, the infant remains resilient to such alterations in the maternal metabolome. The GDM-specific factor identifies a clear signature  related to the aromatic AAs (histidine, phenylalanine, and tyrosine), and this signature is strongly associated with newborn adiposity, remaining significant after adjusting for confounders. The direction of the factor loadings suggests that lower 1hr post-OGTT/fasting ratios of histidine and/or higher ratios of phenylalanine and tyrosine could reflect GDM-specific metabolic dysregulation that confers increased risk of newborn adiposity to infants born to women with GDM. These findings provide strong evidence that the latent factors identified by MSFA-X capture biologically meaningful variation in metabolomic profiles that relates to newborn adiposity.

MSFA-X is inspired by metabolomics data: functional mechanisms involving groups of metabolites are a natural context for the application of latent variable analysis, and the informative correlation structure of metabolomics data is well-modeled by network analysis. However, MSFA-X is a general tool that can be applied to any type of multivariate normal data, including data transformed to approximate normality.  

The problem of comparing networks across multiple studies or conditions is often considered under the umbrella term of differential network analysis. Typical differential network analysis methods for GGMs do not focus on deconvoluting the shared and study-specific signal, but either on (i) estimating a GGM for each study which contains both signals or (ii) directly estimating the differential network between studies \cite{shojaie2021differential}. While these methods can produce accurate maps of conditional dependencies present in each study, they are not designed to allow the user to separate the shared and study-specific contributions to these dependencies.   In contrast, MSFA-X accomplishes exactly this goal, providing a distinctive and interpretable perspective.  MSFA-X is able to recover shared and study-specific conditional independence patterns (GGMs) and outperform a naive benchmark based on the graphical lasso. 

MSFA-X has important limitations. First, factor analysis can be a restrictive method due to the numerous assumptions made regarding normality and independence of factors and residual noise, and MSFA-X is no exception. Copula methods such as the nonparanormal transformation have been widely used to facilitate the application of GGMs to non-normally distributed data, and we propose such approaches could also be applied to MSFA-X \cite{liu2009nonparanormal}.  Future work could model correlated residuals by allowing different matrix structures for $\bsG$ and $\mathbf{H}_s$. For example, van Kesteren and Kievit (2021) present an open-source framework for modeling correlated residuals in the context of exploratory factor analysis in neuroscience \cite{van2021exploratory}. A similar approach could be applied to MSFA-X \cite{van2021exploratory}. Second, MSFA-X is limited in that it is a likelihood-based method with a non-identifiable likelihood. Our approach determines a region of non-identifiability and selects an estimate at the center of this region, meaning the error in these parameter estimates is bounded: we cannot be further off than the longest distance from the center of this region to its border. However, $\bsG$ and $\mathbf{H}_s$ are not the final estimands of interest in this problem; rather, we are trying to estimate the shared precision matrix $\hat\Theta_{MSFA-X}^{shared} = (\Phi\Phi^T + \Gamma)^{-1}$ and the study-specific precision matrices $\hat\Theta_{MSFA-X}^{study;s} = (\Lambda_s\Lambda_s^T + \mathbf{H}_s)^{-1}$. Future work could explore how the bounded error in the estimation of $\bsG$ and $\mathbf{H}_s$ propagates to these latter estimands. Closed-form expressions for inverses of sums of matrices may prove useful; several previous works specifically address matrix inversion error \cite{miller1981inverse,lefebvre2000propagation}. In our simulation studies, the relative error of MSFA-X was similar across a broad range of simulation studies, suggesting that such a bound does exist and could potentially be identified in closed form. A third limitation of MSFA-X is that, in contrast to the graphical lasso, it does not produce precision matrix estimates with exact zeros. This means that thresholding is required in order to produce a GGM that is not a completely connected graph. It would be interesting to explore regularization methods with structural specifications that could produce exact zeros in the factor loadings matrices and, consequently, sparse GGMs that do not require thresholding. Finally, MSFA-X, like all factor analysis methods, requires estimation of the number of factors. Our simulation study demonstrates that the accurate estimation of this quantity is an important part of accurate GGM estimation itself. We correctly identified the number of factors in many cases for small numbers of factors using our CNG-based approach, but our method does not work as well for settings with $>2$ factors per study. Future work is needed here, especially for high-dimensional biological data which are likely to have many factors.

MSFA-X is a tool for researchers to disentangle shared and study-specific network signals when working with data across multiple studies or conditions. This is a unique perspective compared to many existing methods for differential network analysis that do not deconvolute shared and study-specific signals. MSFA-X will assist researchers in gaining new insights into the direct dependencies underlying multi-study data, leading to improved understanding of complex biological function. The MSFA-X method is available as an extension of the R package \texttt{MSFA} \cite{de2019multi} at \url{https://github.com/katehoffshutta/MSFA}.

\section{Acknowledgements}
The authors wish to acknowledge Nathan P. Gill for his valuable input to early conversations on the method. We gratefully acknowledge the participants of the HAPO Study.

\section{Funding}
Research reported in this paper was supported by the National Heart, Lung, and Blood Institute of the National Institutes of Health under award number R01LM013444 and by the National Institute of Diabetes and Digestive and Kidney Diseases under award numbers R01DK095963 and R01DK117491, and by the National Institute of Child Health and Human Development under award numbers R01HD34242 and R01-HD34243. RDV was supported the National Institutes of Health under grant number NIGMS/NIH COBRE CBHD P20GM109035 and ``Programma per Giovani Ricercatori Rita Levi Montalcini'' granted by the Italian Ministry of Education, University, and Research. KHS was supported in part by the National Heart, Lung, and Blood Institute of the National Institutes of Health under awards number T32HL007427 and K25HL175222. The content is solely the responsibility of the authors and does not necessarily represent the official views of the National Institutes of Health.



\newpage

\putbib
\end{bibunit}

\begin{bibunit}





\begin{center}
{\huge{SUPPLEMENTARY MATERIALS}}
\end{center}

\setcounter{section}{0}
\setcounter{figure}{0} 
\setcounter{table}{0} 

\renewcommand{\figurename}{Supplementary Figure}
\renewcommand{\tablename}{Supplementary Table}








\section{Supplement to Simulation Studies}

\subsection{Estimating the Number of Factors}

There are several methods for determining the number of factors in standard factor analysis \cite{raiche2013non}, each of which could serve as a foundation for determining the number of factors in multi-study factor analysis. Here, we adopt a two-step process. First, we use the Zoski and Jurs multiple regression (``mreg") scree test \cite{zoski1993using} as implemented in the R package \texttt{nFactors} \cite{raiche2020package}. We apply mreg on each study separately in order to determine the total number of factors $T_s, s=1, \dots, S$, where $T_s = K + J_s$ for all $s$. This step provides an upper bound on the value of $K$: we must have $K < min_s (T_s)$ in order to satisfy $T_s = K + J_s$ and $J_s > 0$ for all $s$. Let $K^* = min_s (T_s) - 1$; we begin by running MSFA with $K = K^*$ and $J_s = T_s - K^*, s= 1, \dots, S$. This method works well in many but not in all of our simulation scenarios (Supplementary Table \ref{tab:nfac_correct_prop}). 

\begin{table}[h]
\centering
\begin{tabular}{ccc}
Setting & Proportion $K$ Correct & Proportion $J_s$ Correct \\
\hline
Setting 1&1&1;1 \\
Setting 2&0.03&0.03;0.03 \\
Setting 3&0.78&0.78;0.78;0.78;0.78 \\
Setting 4&0&1;0.72 \\
Setting 5&1&1;1 \\
Setting 6&0.96&0.96;0.96 \\
Setting 7&1&1;1 \\
Setting 8&1&1;1 \\
Setting 9&1&1;1 \\
Setting 10&0&0.88;0.02 \\
\end{tabular}
\caption{Proportion of times out of 100 simulations in which our mreg-based method for estimating the number of factors correctly estimated the true number of factors.}
\label{tab:nfac_correct_prop}
\end{table}

We further investigated the distributions of factors in settings in which our method performed poorly (Settings 2,3,4, and 10; Supplementary Tables 2, 3, 4, and 5). These results indicate an overall trend that our method tends to underselect the number of shared factors and overselect the number of study-specific factors, an area of focus for future work.

\begin{table}
\caption{Distribution of estimated factor counts in Setting 2. The correct number of factors was $K=J_1=J_2=2$.}
\centering
\begin{tabular}{c|c}
 K  &  Frequency\\
 \hline
 1  & 97 \\
 2 & 3 \\
\end{tabular}

\begin{tabular}{c|c}
$J_1$  &  Frequency\\
 \hline
 2  & 3 \\
 3 & 97 
\end{tabular}

\begin{tabular}{c|c}
 $J_2$  &  Frequency\\
 \hline
 2  & 3 \\
 3 & 97 
\end{tabular}
\end{table}

\begin{table}
\caption{Distribution of estimated factor counts in Setting 3. The correct number of factors was $K=J_1=J_2=J_3=J_4=2$.}
\centering
\begin{tabular}{c|c}
 K  &  Frequency\\
 \hline
 1  & 22 \\
 2 & 88 \\
\end{tabular}

\begin{tabular}{c|c}
$J_1$  &  Frequency\\
 \hline
 2  & 78 \\
 3 & 22 
\end{tabular}

\begin{tabular}{c|c}
 $J_2$  &  Frequency\\
 \hline
 2  & 78 \\
 3 & 22 
\end{tabular}

\begin{tabular}{c|c}
 $J_3$  &  Frequency\\
 \hline
 2  & 78 \\
 3 & 22 
\end{tabular}

\begin{tabular}{c|c}
 $J_4$  &  Frequency\\
 \hline
 2  & 78 \\
 3 & 22 
\end{tabular}

\end{table}

\begin{table}[h]
\caption{Distribution of estimated factor counts in Setting 4. The correct number of factors was $K=4,J_1=3$, and $J_2=5$.}
\centering
\begin{tabular}{c|c}
 K  &  Frequency\\
 \hline
 1 & 100
\end{tabular}

\begin{tabular}{c|c}
 $J_1$  &  Frequency\\
 \hline
 3  &  100
\end{tabular}

\begin{tabular}{c|c}
 $J_2$  &  Frequency\\
 \hline
 3  & 28 \\
 5 & 72 \\
\end{tabular}
\end{table}

\begin{table}[h]
\caption{Distribution of estimated factor counts in Setting 10. The correct number of factors was $K=J_1=J_2=2$.}
\centering
\begin{tabular}{c|c}
 K  &  Frequency\\
\hline
 4 & 12 \\
 5 & 88 
\end{tabular}

\begin{tabular}{c|c}
$J_1$ &  Frequency \\
\hline
 1  & 88 \\
 2 & 12
\end{tabular}

\begin{tabular}{c|c}
 $J_2$  &  Frequency\\
 \hline
 1  & 2 \\
 2 & 86 \\
 3 & 12 \\
\end{tabular}
\end{table}
 
\clearpage

\subsection{Numerical Results}

Performance of the estimator was assessed using the modified-RV coefficient as described in the main manuscript along with the two additional measures described below: relative Euclidean distance and cosine similarity. Here, we use the notation $\mathcal{G}$ to represent the adjacency matrix of a GGM and $\mathcal{G}_{ij}$ to represent its $(i,j)^{th}$ entry, i.e., the partial correlation between the $i^{th}$ and $j^{th}$ predictors in the model. Similarly, we use $\hat{\mathcal{G}}$ and $\hat{\mathcal{G}}_{ij}$ to refer to an estimated adjacency matrix.

We first calculate a correlation measure between the estimated and true adjacency matrices using the modified-RV coefficient as described by \cite{smilde2009matrix}. The modified-RV coefficient is an extension of the RV coefficient of Robert and Escoufier \cite{robert1976unifying} developed for high-dimensional data.  We next investigate the distance between the estimated and true adjacency matrices by defining a quantity we call the relative Euclidean distance (RE):
\begin{equation*}
RE = \frac{\sqrt{\sum_{i<j} (\hat{\mathcal{G}}_{ij} - \mathcal{G}_{ij})^2}}{\sqrt{\sum_{i<j} \mathcal{G}_{ij}^2}}
\end{equation*}

The numerator of the RE is the Euclidean distance between the estimated and true lower triangular entries of the adjacency matrix; we specify this subset of entries due to the symmetry of the estimator and the fact that the diagonal of $\mathcal{G}$ is 0 since we do not consider self-edges in a GGM. To facilitate comparison across adjacency matrices with different structures, we scale by the size of $\mathcal{G}$ in the denominator.Finally, we calculate the cosine similarity between the lower triangular entries of $\mathcal{G}$ and $\mathcal{\hat{G}}$:
\begin{equation}
Cosine(\mathcal{G},\mathcal{\hat{G}}) = \frac{\sum_{i<j} \mathcal{G}_{ij}\mathcal{\hat{G}}_{ij}}{\sum_{i<j} \mathcal{G}_{ij}^2 \sum_{i<j} \mathcal{\hat{G}}_{ij}^2}
\end{equation}

\begin{footnotesize}
\begin{longtable}{cccccccccccc}
\caption{Matrix RV coefficients for all ten simulation settings. A higher value of the matrix RV coefficient corresponds to better estimation of the gold standard matrix in the simulation. The maximum value for the matrix RV coefficient is 1.} \\
Method&Setting&Study&Median&2.5th percentile&97.5th percentile \\
\hline
\hline
\endhead
MSFA-X: Est. Fac.&Setting 1&Shared&0.9946&0.9938&0.9954 \\
MSFA-X: True Fac.&Setting 1&Shared&0.9946&0.9938&0.9954 \\
glasso&Setting 1&Shared&0.5412&0.5394&0.5428 \\
MSFA-X: Est. Fac.&Setting 1&Study 1&0.9925&0.989&0.9947 \\
MSFA-X: True Fac.&Setting 1&Study 1&0.9925&0.989&0.9947 \\
glasso&Setting 1&Study 1&0.0135&0.0123&0.0146 \\
MSFA-X: Est. Fac.&Setting 1&Study 2&0.9948&0.9898&0.9968 \\
MSFA-X: True Fac.&Setting 1&Study 2&0.9948&0.9898&0.9968 \\
glasso&Setting 1&Study 2&0.015&0.0143&0.0155 \\
\hline
MSFA-X: Est. Fac.&Setting 2&Shared&0.7672&0.389&0.8522 \\
MSFA-X: True Fac.&Setting 2&Shared&0.8313&0.3601&0.9952 \\
glasso&Setting 2&Shared&0.6599&0.6518&0.6676 \\
MSFA-X: Est. Fac.&Setting 2&Study 1&0.7935&0.1576&0.9185 \\
MSFA-X: True Fac.&Setting 2&Study 1&0.8026&0.2127&0.9893 \\
glasso&Setting 2&Study 1&0.0489&0.0354&0.0591 \\
MSFA-X: Est. Fac.&Setting 2&Study 2&0.799&0.6682&0.8581 \\
MSFA-X: True Fac.&Setting 2&Study 2&0.9308&0.3932&0.9945 \\
glasso&Setting 2&Study 2&0.0634&0.0531&0.0742 \\
\hline
MSFA-X: Est. Fac.&Setting 3&Shared&0.9952&0.8257&0.9958 \\
MSFA-X: True Fac.&Setting 3&Shared&0.9952&0.9946&0.9958 \\
glasso&Setting 3&Shared&0.4283&0.4262&0.4302 \\
MSFA-X: Est. Fac.&Setting 3&Study 1&0.9926&0.8785&0.9949 \\
MSFA-X: True Fac.&Setting 3&Study 1&0.9931&0.9902&0.9952 \\
glasso&Setting 3&Study 1&0.0299&0.0202&0.0385 \\
MSFA-X: Est. Fac.&Setting 3&Study 2&0.9945&0.8515&0.9968 \\
MSFA-X: True Fac.&Setting 3&Study 2&0.9949&0.9907&0.9968 \\
glasso&Setting 3&Study 2&0.0278&0.02&0.0359 \\
MSFA-X: Est. Fac.&Setting 3&Study 3&0.9948&0.8562&0.9971 \\
MSFA-X: True Fac.&Setting 3&Study 3&0.9955&0.9912&0.9971 \\
glasso&Setting 3&Study 3&0.0408&0.0344&0.0485 \\
MSFA-X: Est. Fac.&Setting 3&Study 4&0.9948&0.8618&0.9974 \\
MSFA-X: True Fac.&Setting 3&Study 4&0.9953&0.9936&0.9975 \\
glasso&Setting 3&Study 4&0.0417&0.0349&0.0499 \\
\hline
MSFA-X: Est. Fac.&Setting 4&Shared&0.6262&0.5975&0.6291 \\
MSFA-X: True Fac.&Setting 4&Shared&0.9926&0.9913&0.9936 \\
glasso&Setting 4&Shared&0.512&0.5088&0.5159 \\
MSFA-X: Est. Fac.&Setting 4&Study 1&0.493&0.4509&0.5059 \\
MSFA-X: True Fac.&Setting 4&Study 1&0.9873&0.9783&0.9911 \\
glasso&Setting 4&Study 1&0.1008&0.0952&0.1051 \\
MSFA-X: Est. Fac.&Setting 4&Study 2&0.6248&0.527&0.7668 \\
MSFA-X: True Fac.&Setting 4&Study 2&0.9914&0.9883&0.9935 \\
glasso&Setting 4&Study 2&0.0795&0.0758&0.0841 \\
\hline
MSFA-X: Est. Fac.&Setting 5&Shared&0.9917&0.9891&0.9935 \\
MSFA-X: True Fac.&Setting 5&Shared&0.9917&0.9891&0.9935 \\
glasso&Setting 5&Shared&0.5066&0.5007&0.5128 \\
MSFA-X: Est. Fac.&Setting 5&Study 1&0.9786&0.9546&0.9891 \\
MSFA-X: True Fac.&Setting 5&Study 1&0.9786&0.9546&0.9891 \\
glasso&Setting 5&Study 1&0.0246&0.0219&0.0289 \\
MSFA-X: Est. Fac.&Setting 5&Study 2&0.9742&0.9441&0.985 \\
MSFA-X: True Fac.&Setting 5&Study 2&0.9742&0.9441&0.985 \\
glasso&Setting 5&Study 2&0.0224&0.0207&0.0241 \\
\hline
MSFA-X: Est. Fac.&Setting 6&Shared&0.9934&0.6455&0.9952 \\
MSFA-X: True Fac.&Setting 6&Shared&0.9934&0.6455&0.9952 \\
glasso&Setting 6&Shared&0.5554&0.5505&0.5593 \\
MSFA-X: Est. Fac.&Setting 6&Study 1&0.9853&0.6674&0.9942 \\
MSFA-X: True Fac.&Setting 6&Study 1&0.9853&0.6674&0.9942 \\
glasso&Setting 6&Study 1&0.018&0.0172&0.0187 \\
MSFA-X: Est. Fac.&Setting 6&Study 2&0.9705&0.6277&0.9865 \\
MSFA-X: True Fac.&Setting 6&Study 2&0.9705&0.6277&0.9865 \\
glasso&Setting 6&Study 2&-0.0175&-0.0221&-0.0097 \\
\hline
MSFA-X: Est. Fac.&Setting 7&Shared&0.9989&0.9986&0.9992 \\
MSFA-X: True Fac.&Setting 7&Shared&0.9989&0.9986&0.9992 \\
glasso&Setting 7&Shared&0.5426&0.5413&0.5441 \\
MSFA-X: Est. Fac.&Setting 7&Study 1&0.9855&0.9812&0.9881 \\
MSFA-X: True Fac.&Setting 7&Study 1&0.9855&0.9812&0.9881 \\
glasso&Setting 7&Study 1&0.015&0.0143&0.0158 \\
MSFA-X: Est. Fac.&Setting 7&Study 2&0.991&0.9864&0.9947 \\
MSFA-X: True Fac.&Setting 7&Study 2&0.991&0.9864&0.9947 \\
glasso&Setting 7&Study 2&0.0153&0.0146&0.0159 \\
\hline
MSFA-X: Est. Fac.&Setting 8&Shared&0.9861&0.9848&0.9873 \\
MSFA-X: True Fac.&Setting 8&Shared&0.9861&0.9848&0.9873 \\
glasso&Setting 8&Shared&0.5399&0.5378&0.5419 \\
MSFA-X: Est. Fac.&Setting 8&Study 1&0.9968&0.9934&0.998 \\
MSFA-X: True Fac.&Setting 8&Study 1&0.9968&0.9934&0.998 \\
glasso&Setting 8&Study 1&0.0109&0.0097&0.0124 \\
MSFA-X: Est. Fac.&Setting 8&Study 2&0.9943&0.9892&0.9964 \\
MSFA-X: True Fac.&Setting 8&Study 2&0.9943&0.9892&0.9964 \\
glasso&Setting 8&Study 2&0.0145&0.0137&0.015 \\
\hline
MSFA-X: Est. Fac.&Setting 9&Shared&0.9949&0.9942&0.9956 \\
MSFA-X: True Fac.&Setting 9&Shared&0.9949&0.9942&0.9956 \\
glasso&Setting 9&Shared&0.5466&0.5438&0.5492 \\
MSFA-X: Est. Fac.&Setting 9&Study 1&0.9925&0.989&0.9946 \\
MSFA-X: True Fac.&Setting 9&Study 1&0.9925&0.989&0.9946 \\
glasso&Setting 9&Study 1&0.0139&0.0129&0.0151 \\
MSFA-X: Est. Fac.&Setting 9&Study 2&0.9939&0.989&0.9965 \\
MSFA-X: True Fac.&Setting 9&Study 2&0.9939&0.989&0.9965 \\
glasso&Setting 9&Study 2&0.0175&0.0165&0.019 \\
\hline
MSFA-X: Est. Fac.&Setting 10&Shared&0.8475&0.737&0.9248 \\
MSFA-X: True Fac.&Setting 10&Shared&0.9952&0.994&0.9961 \\
glasso&Setting 10&Shared&0.5607&0.5569&0.5646 \\
MSFA-X: Est. Fac.&Setting 10&Study 1&0.9921&0.7799&0.9964 \\
MSFA-X: True Fac.&Setting 10&Study 1&0.9946&0.9892&0.9964 \\
glasso&Setting 10&Study 1&0.0274&0.0207&0.0361 \\
MSFA-X: Est. Fac.&Setting 10&Study 2&0.7932&0.5376&0.9072 \\
MSFA-X: True Fac.&Setting 10&Study 2&0.9983&0.997&0.999 \\
glasso&Setting 10&Study 2&0.0634&0.0622&0.0644 \\
\end{longtable}
\end{footnotesize}

\clearpage

\begin{footnotesize}
\begin{longtable}{cccccccccccc}
\caption{Full summary results for all 10 simulation settings assessed by cosine similarity. A larger value of cosine similarity indicates better agreement of the estimated matrix with the target gold standard matrix in the simulation. The maximum value of cosine similarity is 1.} \\
\label{tab:suppSimResultsCosSim}
Method&Setting&Study&Median&2.5th percentile&97.5th percentile \\
\hline
\hline
\endhead
MSFA-X: Est. Fac.&Setting 1&Shared&0.9929&0.992&0.9939 \\
MSFA-X: True Fac.&Setting 1&Shared&0.9929&0.992&0.9939 \\
glasso&Setting 1&Shared&0.5387&0.5361&0.5408 \\
MSFA-X: Est. Fac.&Setting 1&Study 1&0.9909&0.9858&0.9937 \\
MSFA-X: True Fac.&Setting 1&Study 1&0.9909&0.9858&0.9937 \\
glasso&Setting 1&Study 1&0.0795&0.0743&0.0846 \\
MSFA-X: Est. Fac.&Setting 1&Study 2&0.9933&0.9871&0.9959 \\
MSFA-X: True Fac.&Setting 1&Study 2&0.9933&0.9871&0.9959 \\
glasso&Setting 1&Study 2&0.0634&0.0598&0.0681 \\
\hline
MSFA-X: Est. Fac.&Setting 2&Shared&0.6328&0.3106&0.7027 \\
MSFA-X: True Fac.&Setting 2&Shared&0.8231&0.453&0.9963 \\
glasso&Setting 2&Shared&0.7273&0.7197&0.7368 \\
MSFA-X: Est. Fac.&Setting 2&Study 1&0.7743&0.1514&0.8245 \\
MSFA-X: True Fac.&Setting 2&Study 1&0.8955&0.1238&0.9898 \\
glasso&Setting 2&Study 1&0.273&0.2509&0.2984 \\
MSFA-X: Est. Fac.&Setting 2&Study 2&0.7694&0.6087&0.8088 \\
MSFA-X: True Fac.&Setting 2&Study 2&0.9472&0.3297&0.9942 \\
glasso&Setting 2&Study 2&0.1562&0.1299&0.1817 \\
\hline
MSFA-X: Est. Fac.&Setting 3&Shared&0.9938&0.7027&0.9946 \\
MSFA-X: True Fac.&Setting 3&Shared&0.9939&0.9931&0.9946 \\
glasso&Setting 3&Shared&0.4803&0.4762&0.4835 \\
MSFA-X: Est. Fac.&Setting 3&Study 1&0.9907&0.8124&0.9939 \\
MSFA-X: True Fac.&Setting 3&Study 1&0.9915&0.9863&0.9943 \\
glasso&Setting 3&Study 1&0.1789&0.1702&0.19 \\
MSFA-X: Est. Fac.&Setting 3&Study 2&0.9929&0.7917&0.9959 \\
MSFA-X: True Fac.&Setting 3&Study 2&0.9936&0.9879&0.9959 \\
glasso&Setting 3&Study 2&0.1702&0.1626&0.1781 \\
MSFA-X: Est. Fac.&Setting 3&Study 3&0.993&0.8005&0.9962 \\
MSFA-X: True Fac.&Setting 3&Study 3&0.994&0.9883&0.9962 \\
glasso&Setting 3&Study 3&0.1659&0.1569&0.174 \\
MSFA-X: Est. Fac.&Setting 3&Study 4&0.9929&0.798&0.9964 \\
MSFA-X: True Fac.&Setting 3&Study 4&0.9935&0.9911&0.9965 \\
glasso&Setting 3&Study 4&0.1554&0.1458&0.1637 \\
\hline
MSFA-X: Est. Fac.&Setting 4&Shared&0.4878&0.4781&0.4915 \\
MSFA-X: True Fac.&Setting 4&Shared&0.9926&0.9914&0.9935 \\
glasso&Setting 4&Shared&0.5913&0.5874&0.5956 \\
MSFA-X: Est. Fac.&Setting 4&Study 1&0.3347&0.2894&0.3541 \\
MSFA-X: True Fac.&Setting 4&Study 1&0.9871&0.9787&0.9909 \\
glasso&Setting 4&Study 1&0.1305&0.1255&0.1364 \\
MSFA-X: Est. Fac.&Setting 4&Study 2&0.5667&0.4351&0.7192 \\
MSFA-X: True Fac.&Setting 4&Study 2&0.9917&0.9884&0.9934 \\
glasso&Setting 4&Study 2&0.1372&0.1325&0.1425 \\
\hline
MSFA-X: Est. Fac.&Setting 5&Shared&0.9891&0.9858&0.9912 \\
MSFA-X: True Fac.&Setting 5&Shared&0.9891&0.9858&0.9912 \\
glasso&Setting 5&Shared&0.5261&0.521&0.533 \\
MSFA-X: Est. Fac.&Setting 5&Study 1&0.9712&0.9352&0.9854 \\
MSFA-X: True Fac.&Setting 5&Study 1&0.9712&0.9352&0.9854 \\
glasso&Setting 5&Study 1&0.0679&0.0581&0.0777 \\
MSFA-X: Est. Fac.&Setting 5&Study 2&0.966&0.9283&0.98 \\
MSFA-X: True Fac.&Setting 5&Study 2&0.966&0.9283&0.98 \\
glasso&Setting 5&Study 2&0.054&0.0479&0.0615 \\
\hline
MSFA-X: Est. Fac.&Setting 6&Shared&0.9913&0.4915&0.9936 \\
MSFA-X: True Fac.&Setting 6&Shared&0.9913&0.491&0.9936 \\
glasso&Setting 6&Shared&0.5683&0.5626&0.5735 \\
MSFA-X: Est. Fac.&Setting 6&Study 1&0.9785&0.4979&0.993 \\
MSFA-X: True Fac.&Setting 6&Study 1&0.9785&0.4979&0.993 \\
glasso&Setting 6&Study 1&0.0172&0.0151&0.0188 \\
MSFA-X: Est. Fac.&Setting 6&Study 2&0.9588&0.4674&0.9817 \\
MSFA-X: True Fac.&Setting 6&Study 2&0.9588&0.4674&0.9817 \\
glasso&Setting 6&Study 2&0.1861&0.1734&0.1956 \\
\hline
MSFA-X: Est. Fac.&Setting 7&Shared&0.9986&0.9982&0.9989 \\
MSFA-X: True Fac.&Setting 7&Shared&0.9986&0.9982&0.9989 \\
glasso&Setting 7&Shared&0.5354&0.5329&0.538 \\
MSFA-X: Est. Fac.&Setting 7&Study 1&0.9829&0.977&0.9862 \\
MSFA-X: True Fac.&Setting 7&Study 1&0.9829&0.977&0.9862 \\
glasso&Setting 7&Study 1&0.067&0.0618&0.0716 \\
MSFA-X: Est. Fac.&Setting 7&Study 2&0.9893&0.9835&0.9937 \\
MSFA-X: True Fac.&Setting 7&Study 2&0.9893&0.9835&0.9937 \\
glasso&Setting 7&Study 2&0.062&0.0582&0.0664 \\
\hline
MSFA-X: Est. Fac.&Setting 8&Shared&0.9818&0.9803&0.9835 \\
MSFA-X: True Fac.&Setting 8&Shared&0.9818&0.9803&0.9835 \\
glasso&Setting 8&Shared&0.5411&0.5382&0.5439 \\
MSFA-X: Est. Fac.&Setting 8&Study 1&0.9956&0.9903&0.9973 \\
MSFA-X: True Fac.&Setting 8&Study 1&0.9956&0.9903&0.9973 \\
glasso&Setting 8&Study 1&0.0929&0.0866&0.0984 \\
MSFA-X: Est. Fac.&Setting 8&Study 2&0.9928&0.9862&0.9954 \\
MSFA-X: True Fac.&Setting 8&Study 2&0.9928&0.9862&0.9954 \\
glasso&Setting 8&Study 2&0.0652&0.0618&0.0703 \\
\hline
MSFA-X: Est. Fac.&Setting 9&Shared&0.9934&0.9926&0.9943 \\
MSFA-X: True Fac.&Setting 9&Shared&0.9934&0.9926&0.9943 \\
glasso&Setting 9&Shared&0.546&0.543&0.5486 \\
MSFA-X: Est. Fac.&Setting 9&Study 1&0.9911&0.9861&0.9936 \\
MSFA-X: True Fac.&Setting 9&Study 1&0.9911&0.9861&0.9936 \\
glasso&Setting 9&Study 1&0.0759&0.072&0.0797 \\
MSFA-X: Est. Fac.&Setting 9&Study 2&0.9919&0.9846&0.9956 \\
MSFA-X: True Fac.&Setting 9&Study 2&0.9919&0.9846&0.9956 \\
glasso&Setting 9&Study 2&0.0676&0.0642&0.0721 \\
\hline
MSFA-X: Est. Fac.&Setting 10&Shared&0.9175&0.828&0.9672 \\
MSFA-X: True Fac.&Setting 10&Shared&0.9946&0.9933&0.9956 \\
glasso&Setting 10&Shared&0.6443&0.6398&0.6482 \\
MSFA-X: Est. Fac.&Setting 10&Study 1&0.9858&0.7077&0.9934 \\
MSFA-X: True Fac.&Setting 10&Study 1&0.9901&0.98&0.9935 \\
glasso&Setting 10&Study 1&0.1274&0.1122&0.1388 \\
MSFA-X: Est. Fac.&Setting 10&Study 2&0.7595&0.6012&0.9265 \\
MSFA-X: True Fac.&Setting 10&Study 2&0.9969&0.9946&0.9982 \\
glasso&Setting 10&Study 2&0.0076&0.007&0.0082 \\
\end{longtable}
\end{footnotesize}

\clearpage

\begin{footnotesize}
\begin{longtable}{cccccccccccc}
\caption{Relative Euclidean distance between estimated and gold standard matrices. A lower value corresponds to better estimation of the gold standard matrix in the simulation.} \\
Method&Setting&Study&Median&2.5th percentile&97.5th percentile \\
\hline
\hline
\endhead
MSFA-X: Est. Fac.&Setting 1&Shared&0.119&0.1268 \\
MSFA-X: True Fac.&Setting 1&Shared&0.119&0.1268 \\
glasso&Setting 1&Shared&1.5103&1.5198 \\
MSFA-X: Est. Fac.&Setting 1&Study 1&0.1348&0.1683 \\
MSFA-X: True Fac.&Setting 1&Study 1&0.1348&0.1683 \\
glasso&Setting 1&Study 1&1.4288&1.4462 \\
MSFA-X: Est. Fac.&Setting 1&Study 2&0.1156&0.1606 \\
MSFA-X: True Fac.&Setting 1&Study 2&0.1156&0.1606 \\
glasso&Setting 1&Study 2&1.4417&1.4698 \\
\hline
MSFA-X: Est. Fac.&Setting 2&Shared&0.7934&1.1239 \\
MSFA-X: True Fac.&Setting 2&Shared&0.6279&1.1694 \\
glasso&Setting 2&Shared&0.7721&0.7875 \\
MSFA-X: Est. Fac.&Setting 2&Study 1&0.7901&1.5054 \\
MSFA-X: True Fac.&Setting 2&Study 1&0.4737&1.2934 \\
glasso&Setting 2&Study 1&0.9929&1.0072 \\
MSFA-X: Est. Fac.&Setting 2&Study 2&0.7103&1.057 \\
MSFA-X: True Fac.&Setting 2&Study 2&0.3289&1.0569 \\
glasso&Setting 2&Study 2&1.0733&1.0918 \\
\hline
MSFA-X: Est. Fac.&Setting 3&Shared&0.1115&0.7115 \\
MSFA-X: True Fac.&Setting 3&Shared&0.1105&0.1172 \\
glasso&Setting 3&Shared&1.7358&1.7529 \\
MSFA-X: Est. Fac.&Setting 3&Study 1&0.1366&0.7111 \\
MSFA-X: True Fac.&Setting 3&Study 1&0.1305&0.1654 \\
glasso&Setting 3&Study 1&1.7449&1.7698 \\
MSFA-X: Est. Fac.&Setting 3&Study 2&0.1196&0.7658 \\
MSFA-X: True Fac.&Setting 3&Study 2&0.1134&0.1555 \\
glasso&Setting 3&Study 2&1.7676&1.7972 \\
MSFA-X: Est. Fac.&Setting 3&Study 3&0.1188&0.7445 \\
MSFA-X: True Fac.&Setting 3&Study 3&0.1095&0.1533 \\
glasso&Setting 3&Study 3&1.7401&1.7796 \\
MSFA-X: Est. Fac.&Setting 3&Study 4&0.1191&0.7501 \\
MSFA-X: True Fac.&Setting 3&Study 4&0.1139&0.1335 \\
glasso&Setting 3&Study 4&1.7677&1.8061 \\
\hline
MSFA-X: Est. Fac.&Setting 4&Shared&0.8731&0.8785 \\
MSFA-X: True Fac.&Setting 4&Shared&0.1214&0.1311 \\
glasso&Setting 4&Shared&1.2613&1.2726 \\
MSFA-X: Est. Fac.&Setting 4&Study 1&1.1642&1.1967 \\
MSFA-X: True Fac.&Setting 4&Study 1&0.1607&0.2063 \\
glasso&Setting 4&Study 1&1.5127&1.5314 \\
MSFA-X: Est. Fac.&Setting 4&Study 2&0.9348&0.9716 \\
MSFA-X: True Fac.&Setting 4&Study 2&0.1291&0.1524 \\
glasso&Setting 4&Study 2&1.2132&1.2244 \\
\hline
MSFA-X: Est. Fac.&Setting 5&Shared&0.1476&0.1684 \\
MSFA-X: True Fac.&Setting 5&Shared&0.1476&0.1684 \\
glasso&Setting 5&Shared&1.5576&1.5766 \\
MSFA-X: Est. Fac.&Setting 5&Study 1&0.2399&0.3598 \\
MSFA-X: True Fac.&Setting 5&Study 1&0.2399&0.3598 \\
glasso&Setting 5&Study 1&1.5478&1.5945 \\
MSFA-X: Est. Fac.&Setting 5&Study 2&0.2614&0.3803 \\
MSFA-X: True Fac.&Setting 5&Study 2&0.2614&0.3803 \\
glasso&Setting 5&Study 2&1.5406&1.5919 \\
\hline
MSFA-X: Est. Fac.&Setting 6&Shared&0.1317&1.0193 \\
MSFA-X: True Fac.&Setting 6&Shared&0.1317&1.0199 \\
glasso&Setting 6&Shared&1.4006&1.4207 \\
MSFA-X: Est. Fac.&Setting 6&Study 1&0.2073&0.9988 \\
MSFA-X: True Fac.&Setting 6&Study 1&0.2073&0.9988 \\
glasso&Setting 6&Study 1&1.2633&1.2865 \\
MSFA-X: Est. Fac.&Setting 6&Study 2&0.2878&1.0463 \\
MSFA-X: True Fac.&Setting 6&Study 2&0.2878&1.047 \\
glasso&Setting 6&Study 2&1.5897&1.6346 \\
\hline
MSFA-X: Est. Fac.&Setting 7&Shared&0.0535&0.0611 \\
MSFA-X: True Fac.&Setting 7&Shared&0.0535&0.0611 \\
glasso&Setting 7&Shared&1.5403&1.5499 \\
MSFA-X: Est. Fac.&Setting 7&Study 1&0.1842&0.2137 \\
MSFA-X: True Fac.&Setting 7&Study 1&0.1842&0.2137 \\
glasso&Setting 7&Study 1&1.4332&1.4574 \\
MSFA-X: Est. Fac.&Setting 7&Study 2&0.1461&0.1809 \\
MSFA-X: True Fac.&Setting 7&Study 2&0.1461&0.1809 \\
glasso&Setting 7&Study 2&1.438&1.4716 \\
\hline
MSFA-X: Est. Fac.&Setting 8&Shared&0.1901&0.1982 \\
MSFA-X: True Fac.&Setting 8&Shared&0.1901&0.1982 \\
glasso&Setting 8&Shared&1.4819&1.4928 \\
MSFA-X: Est. Fac.&Setting 8&Study 1&0.0953&0.14 \\
MSFA-X: True Fac.&Setting 8&Study 1&0.0953&0.14 \\
glasso&Setting 8&Study 1&1.4194&1.4408 \\
MSFA-X: Est. Fac.&Setting 8&Study 2&0.1214&0.1673 \\
MSFA-X: True Fac.&Setting 8&Study 2&0.1214&0.1673 \\
glasso&Setting 8&Study 2&1.446&1.4752 \\
\hline
MSFA-X: Est. Fac.&Setting 9&Shared&0.1145&0.1219 \\
MSFA-X: True Fac.&Setting 9&Shared&0.1145&0.1219 \\
glasso&Setting 9&Shared&1.4747&1.4868 \\
MSFA-X: Est. Fac.&Setting 9&Study 1&0.1336&0.1667 \\
MSFA-X: True Fac.&Setting 9&Study 1&0.1336&0.1667 \\
glasso&Setting 9&Study 1&1.4416&1.4614 \\
MSFA-X: Est. Fac.&Setting 9&Study 2&0.1271&0.1757 \\
MSFA-X: True Fac.&Setting 9&Study 2&0.1271&0.1757 \\
glasso&Setting 9&Study 2&1.4063&1.4288 \\
\hline
MSFA-X: Est. Fac.&Setting 10&Shared&0.4314&0.6733 \\
MSFA-X: True Fac.&Setting 10&Shared&0.1038&0.1154 \\
glasso&Setting 10&Shared&1.14&1.1528 \\
MSFA-X: Est. Fac.&Setting 10&Study 1&0.1686&0.988 \\
MSFA-X: True Fac.&Setting 10&Study 1&0.1404&0.2001 \\
glasso&Setting 10&Study 1&1.9721&2.0069 \\
MSFA-X: Est. Fac.&Setting 10&Study 2&0.8169&1.2543 \\
MSFA-X: True Fac.&Setting 10&Study 2&0.0793&0.1043 \\
glasso&Setting 10&Study 2&1.6057&1.6503 \\
\end{longtable}
\end{footnotesize}

\clearpage
\subsection{Visual Results}

\subsubsection{Violin plots of matrix RV, cosine similarity, and relative Euclidean distance} 

\begin{figure}[h!]
    \centering
    \includegraphics[width=0.4\textwidth]{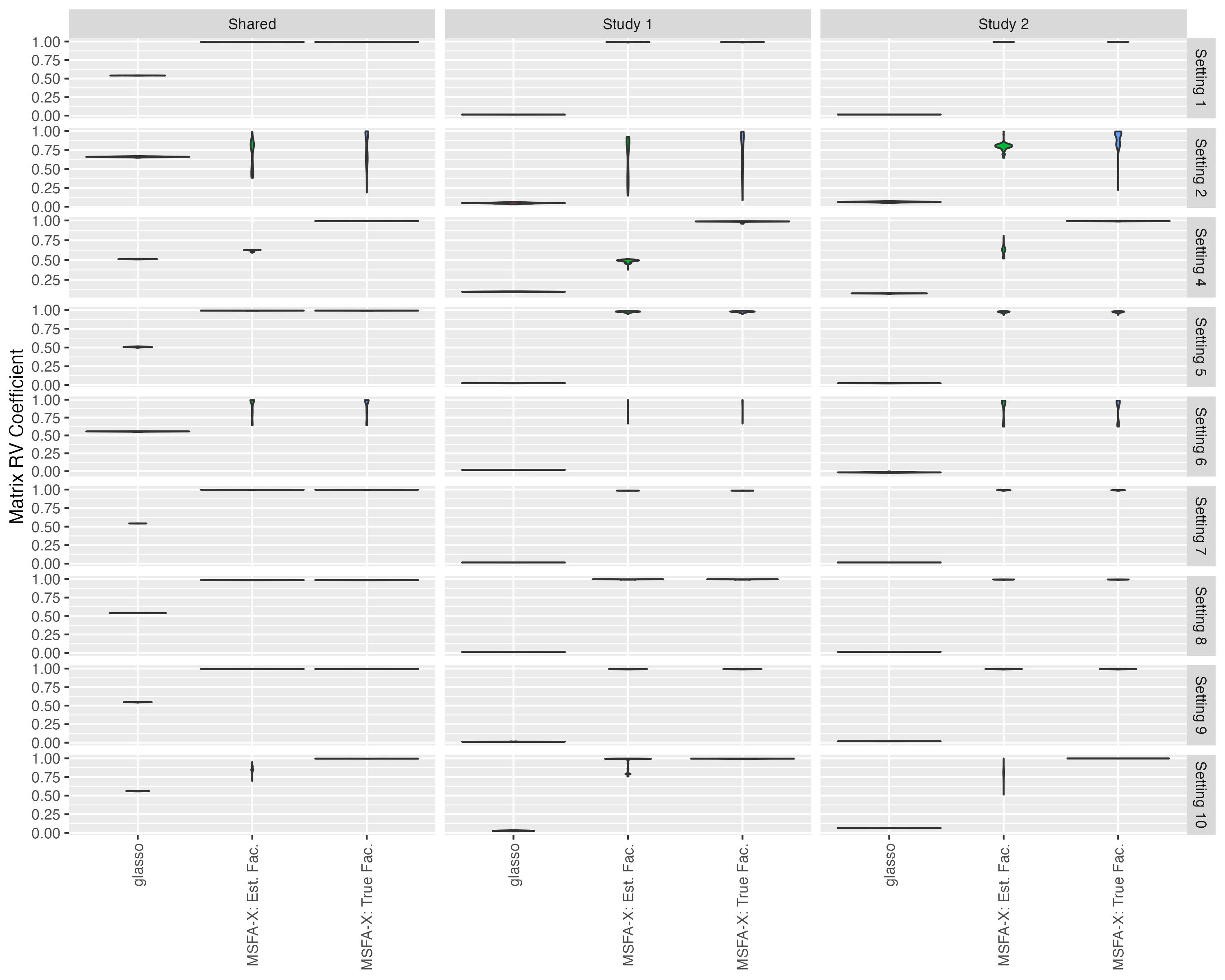}
    \includegraphics[width=0.4\textwidth]{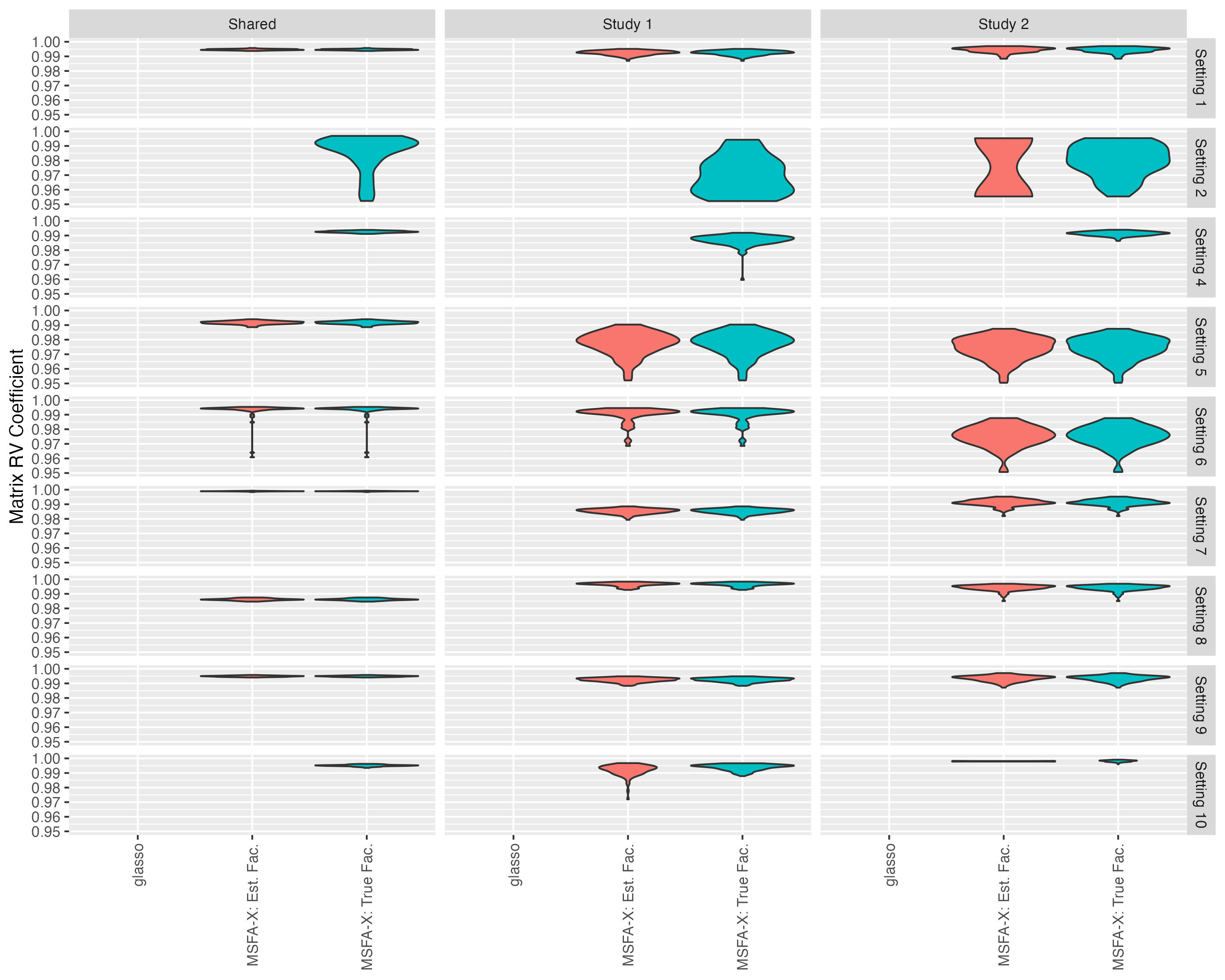}
    \includegraphics[width=0.4\textwidth]{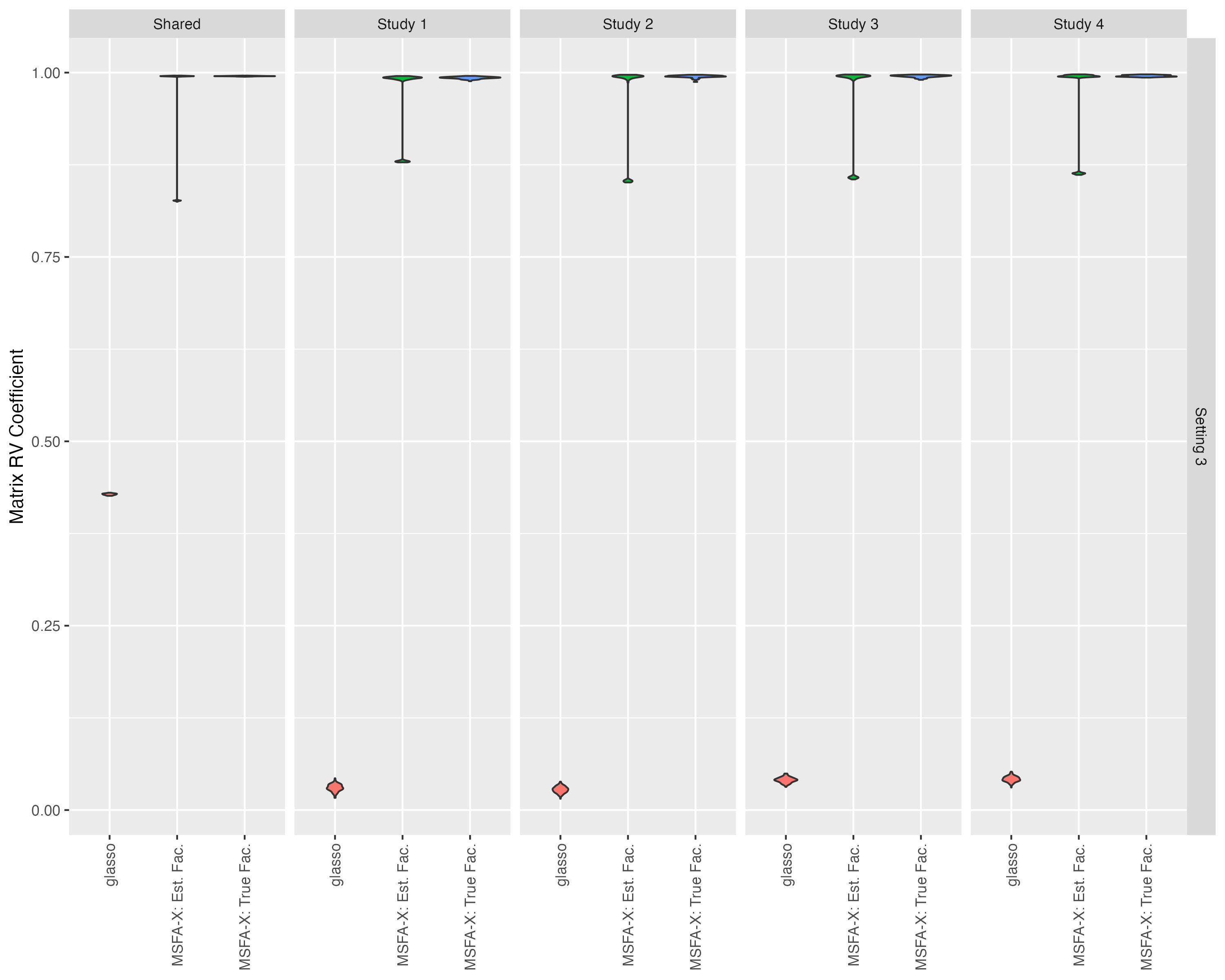}
    \caption{Matrix RV coefficient for the glasso benchmark, MSFA-X with estimated factor count, and MSFA-X with true factor count. Setting 3 is shown in a separate plot because the number of studies is different than the other settings.}
\end{figure}

\begin{figure}
    \centering
    \includegraphics[width=0.4\textwidth]{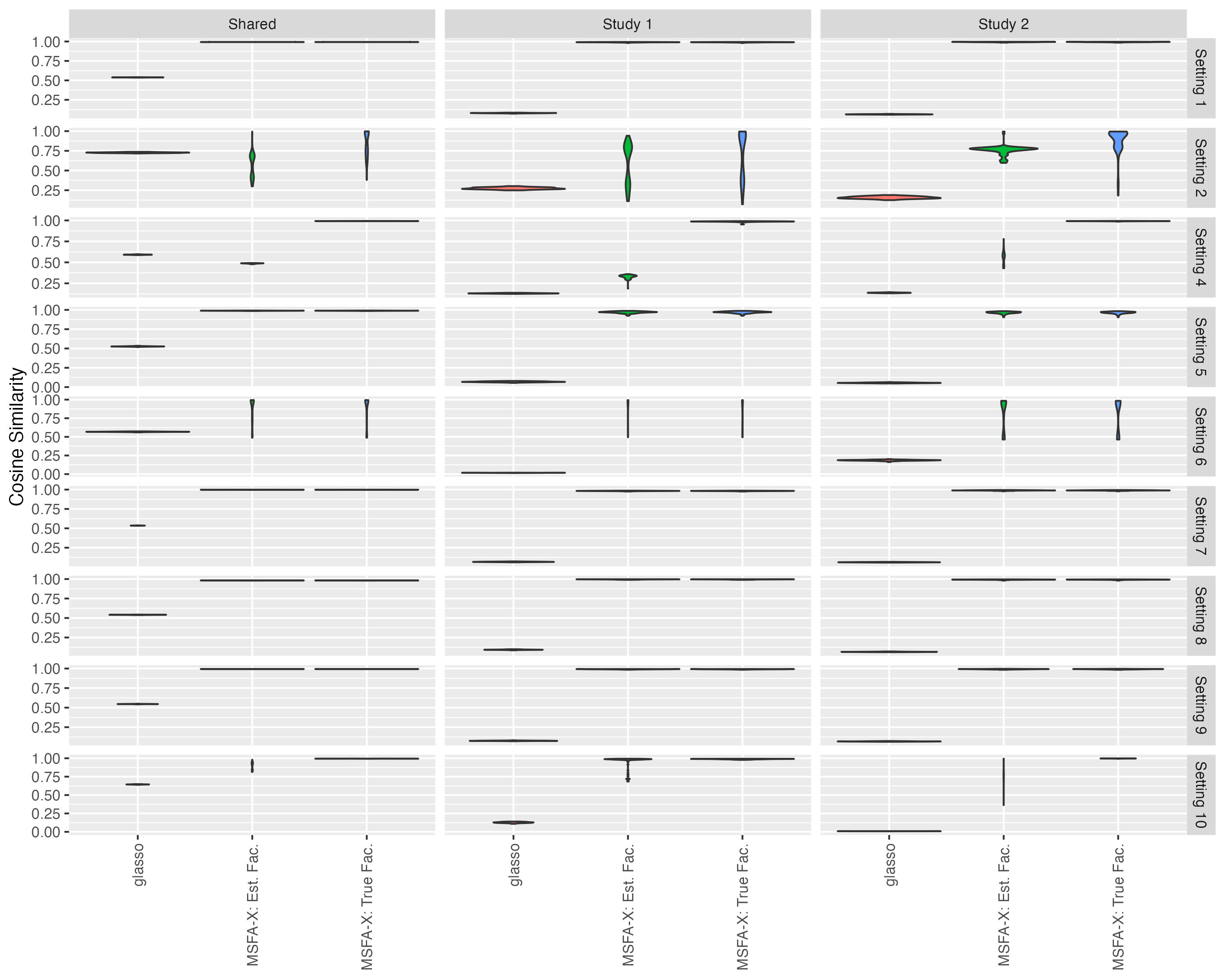}
    \includegraphics[width=0.4\textwidth]{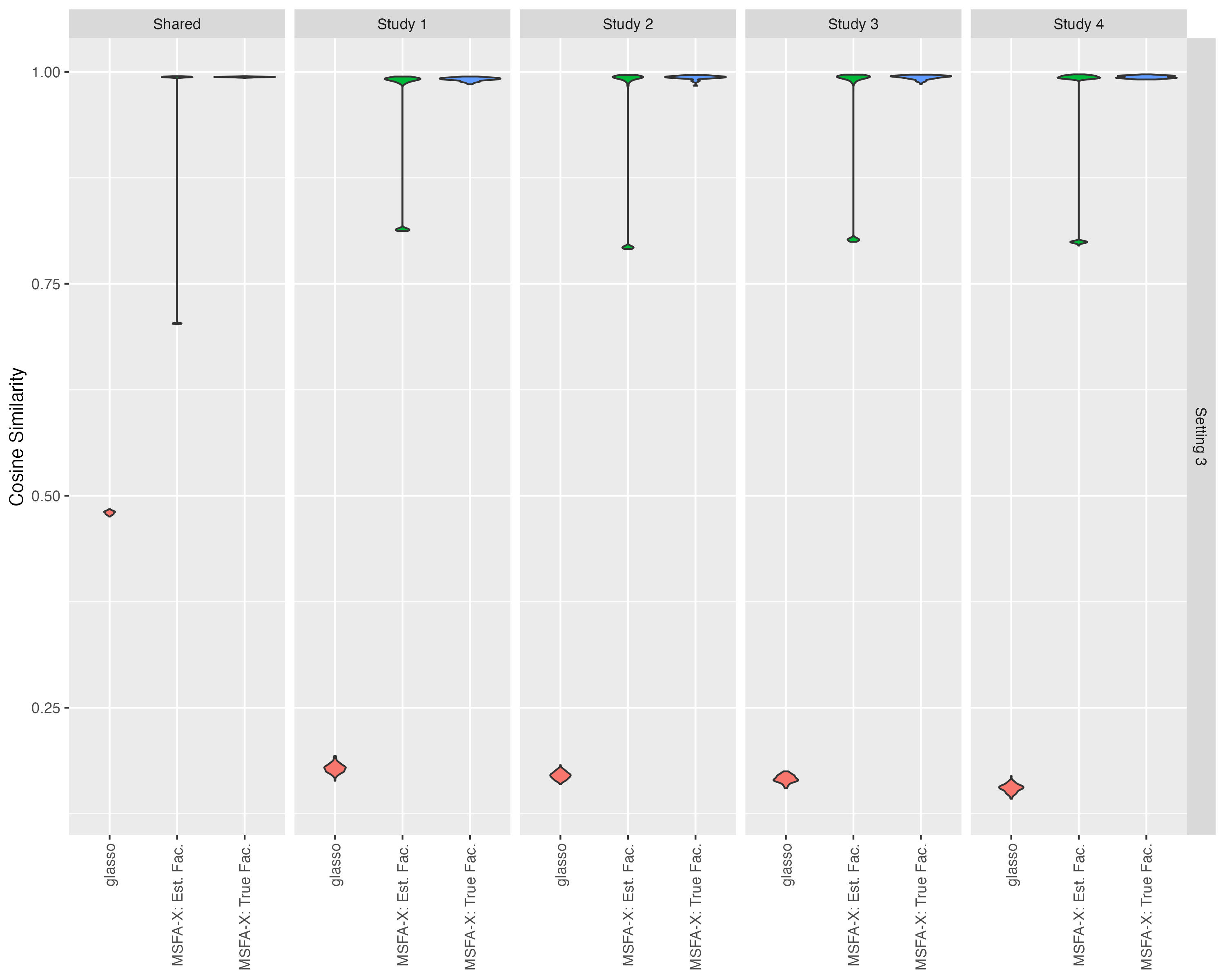}
    \caption{Cosine similarity for the glasso benchmark, MSFA-X with estimated factor count, and MSFA-X with true factor count. Setting 3 is shown in a separate plot because the number of studies is different than the other settings.}
\end{figure}

\begin{figure}
    \centering
    \includegraphics[width=0.4\textwidth]{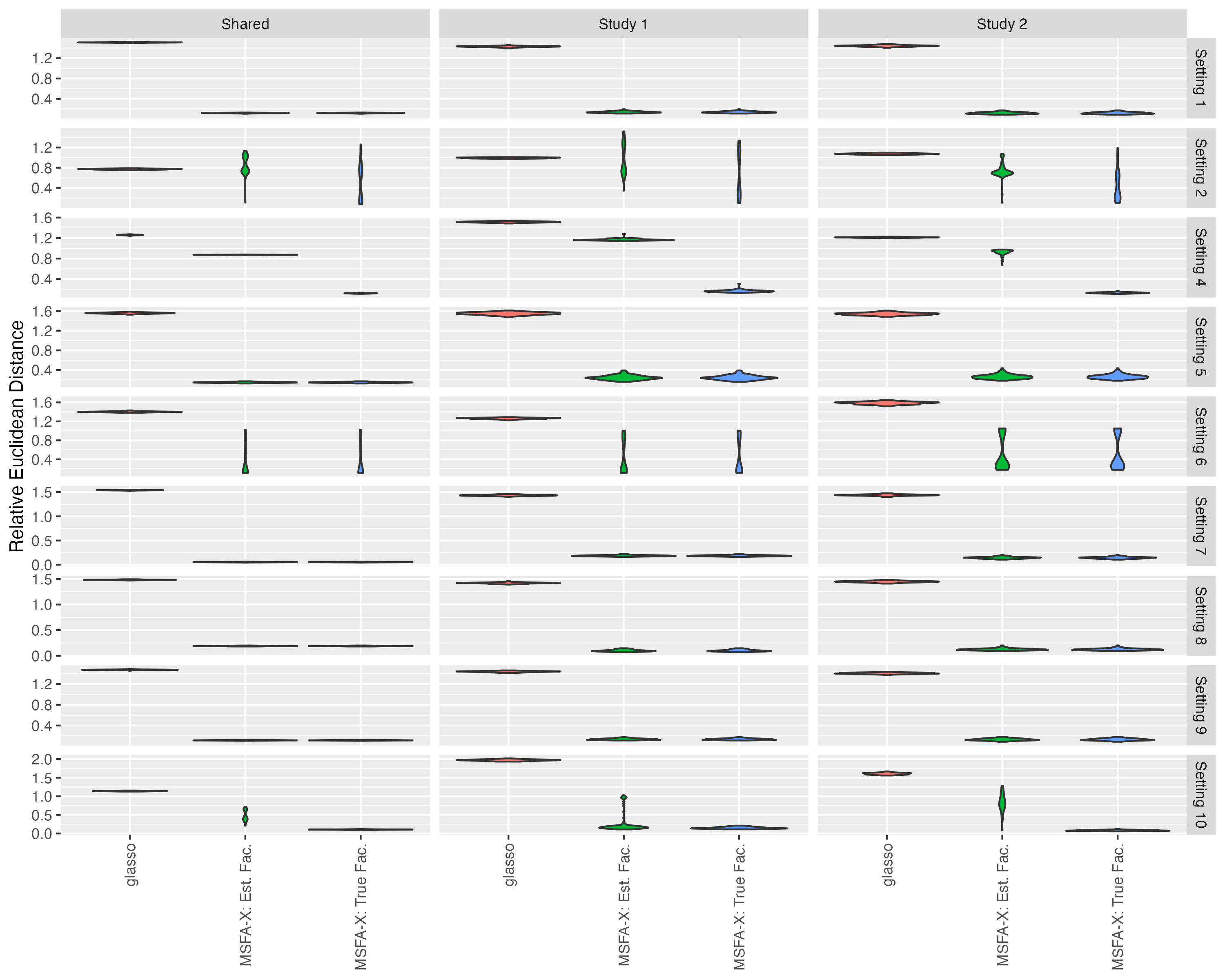}
    \includegraphics[width=0.4\textwidth]{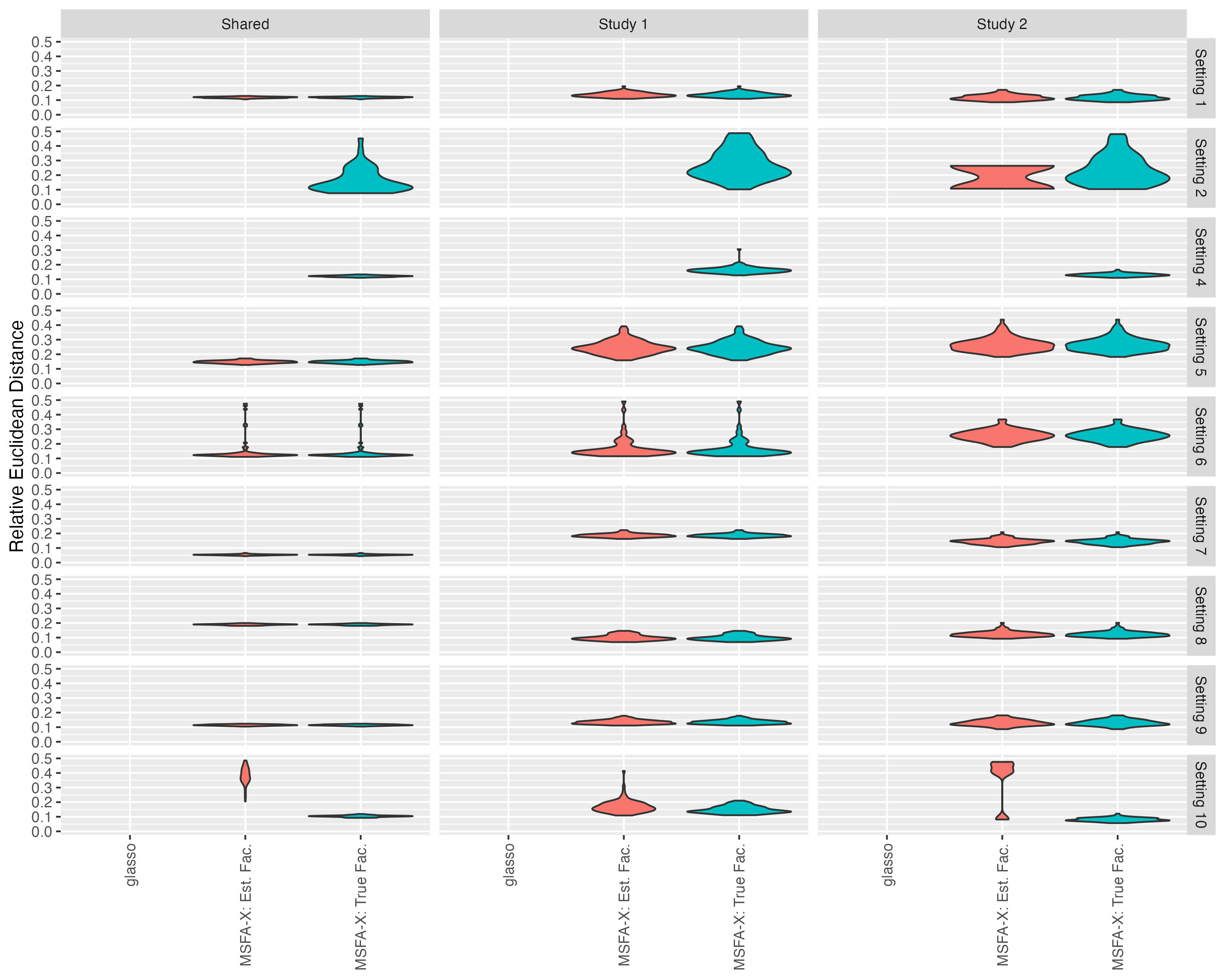}
    \includegraphics[width=0.4\textwidth]{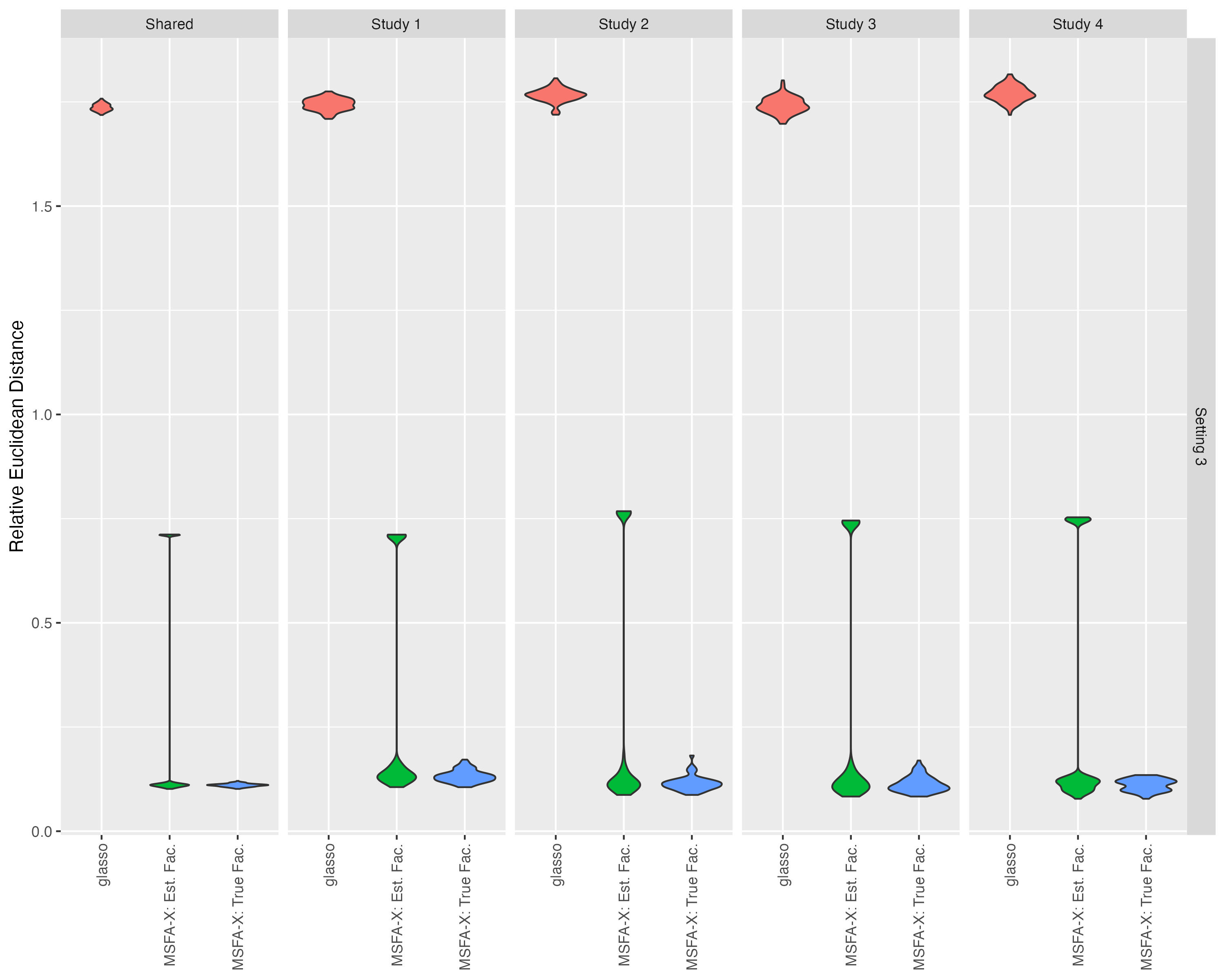}
    \caption{Relative Euclidean distance for the glasso benchmark, MSFA-X with estimated factor count, and MSFA-X with true factor count. Setting 3 is shown in a separate plot because the number of studies is different than the other settings.}
\end{figure}

\clearpage

\subsubsection{Mean estimates} Below are plots of the mean estimated GGM adjacency matrices across 100 simulated datasets for all ten simulation settings, including those not presented in the main manuscript.

\newpage

\begin{figure}
    \centering
    \includegraphics[height=0.8\textheight]{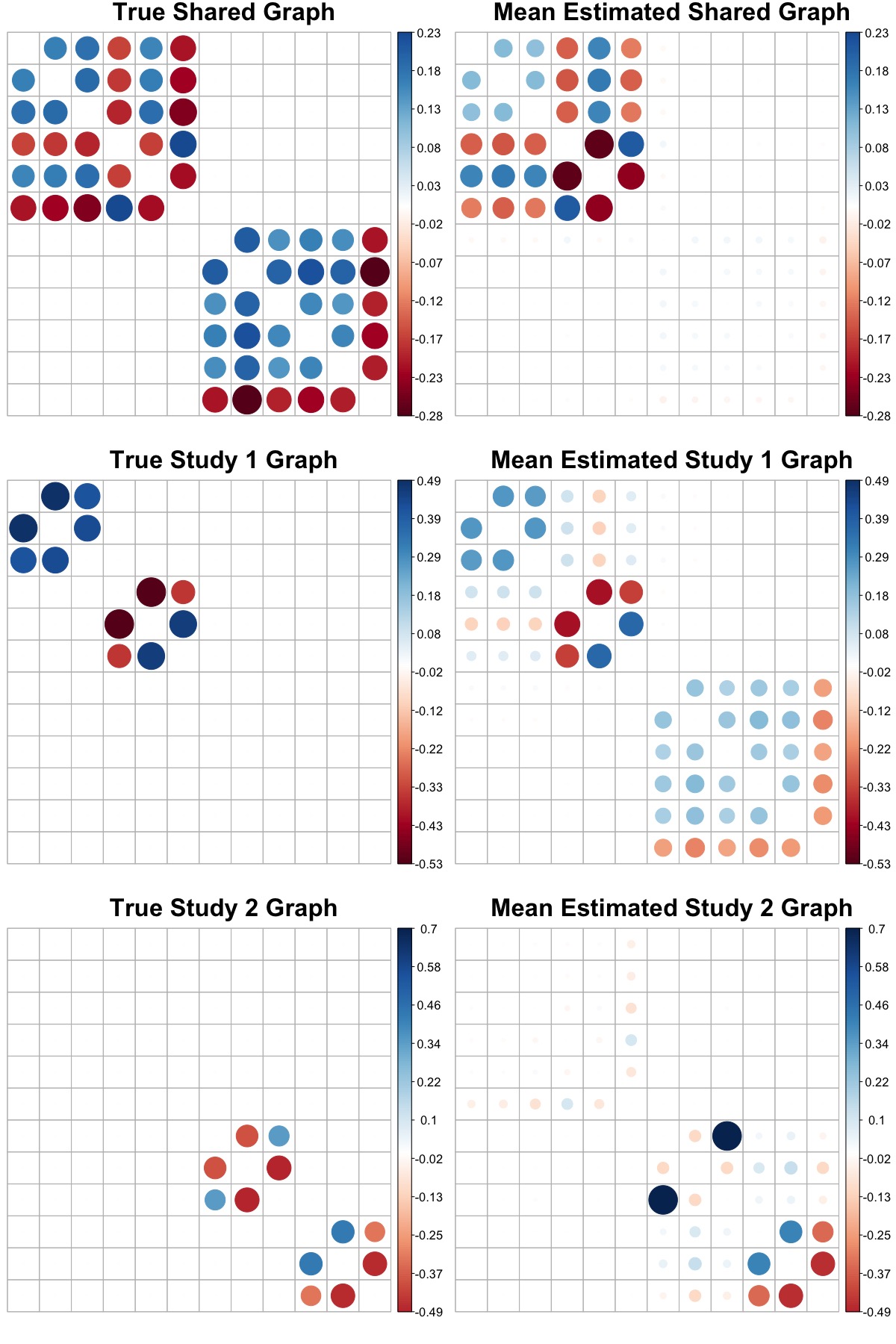}
    \caption{Setting 2}
\end{figure}

\begin{figure}
    \centering
    \includegraphics[height=0.8\textheight]{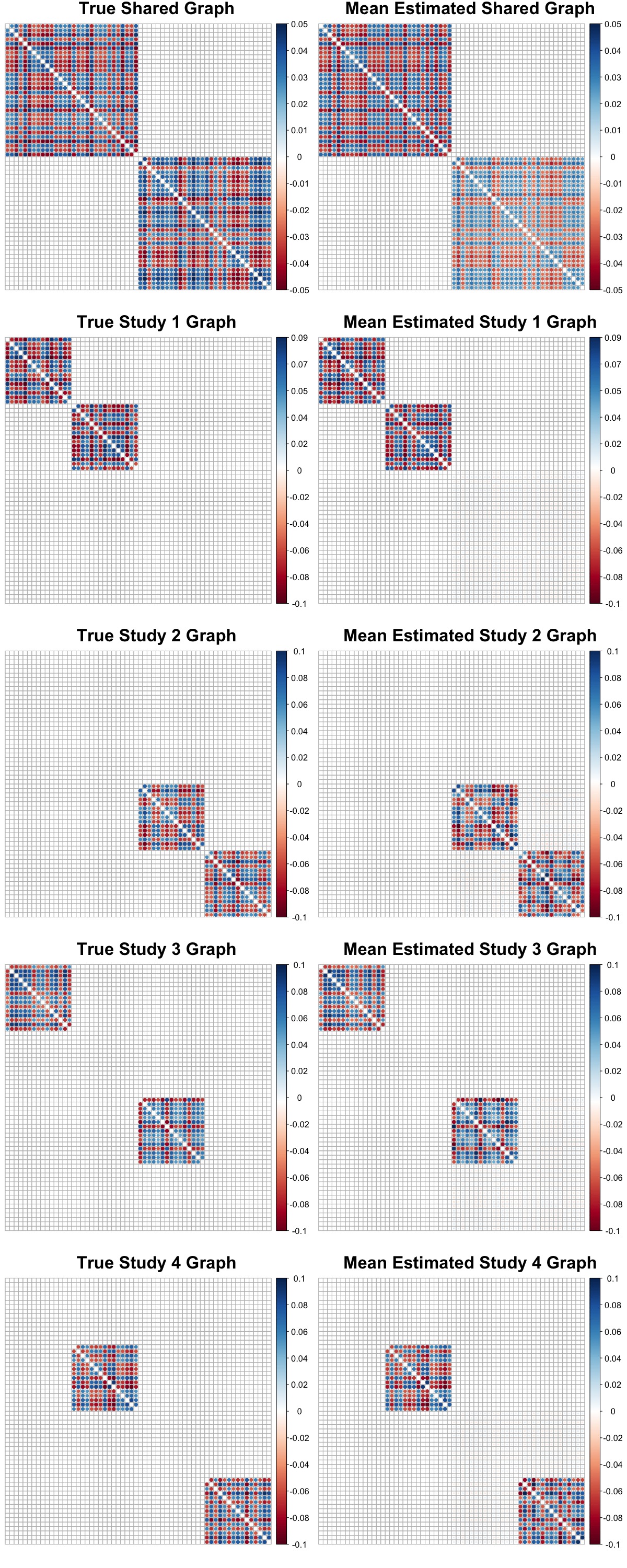}
    \caption{Setting 3}
\end{figure}

\begin{figure}
    \centering
    \includegraphics[height=0.8\textheight]{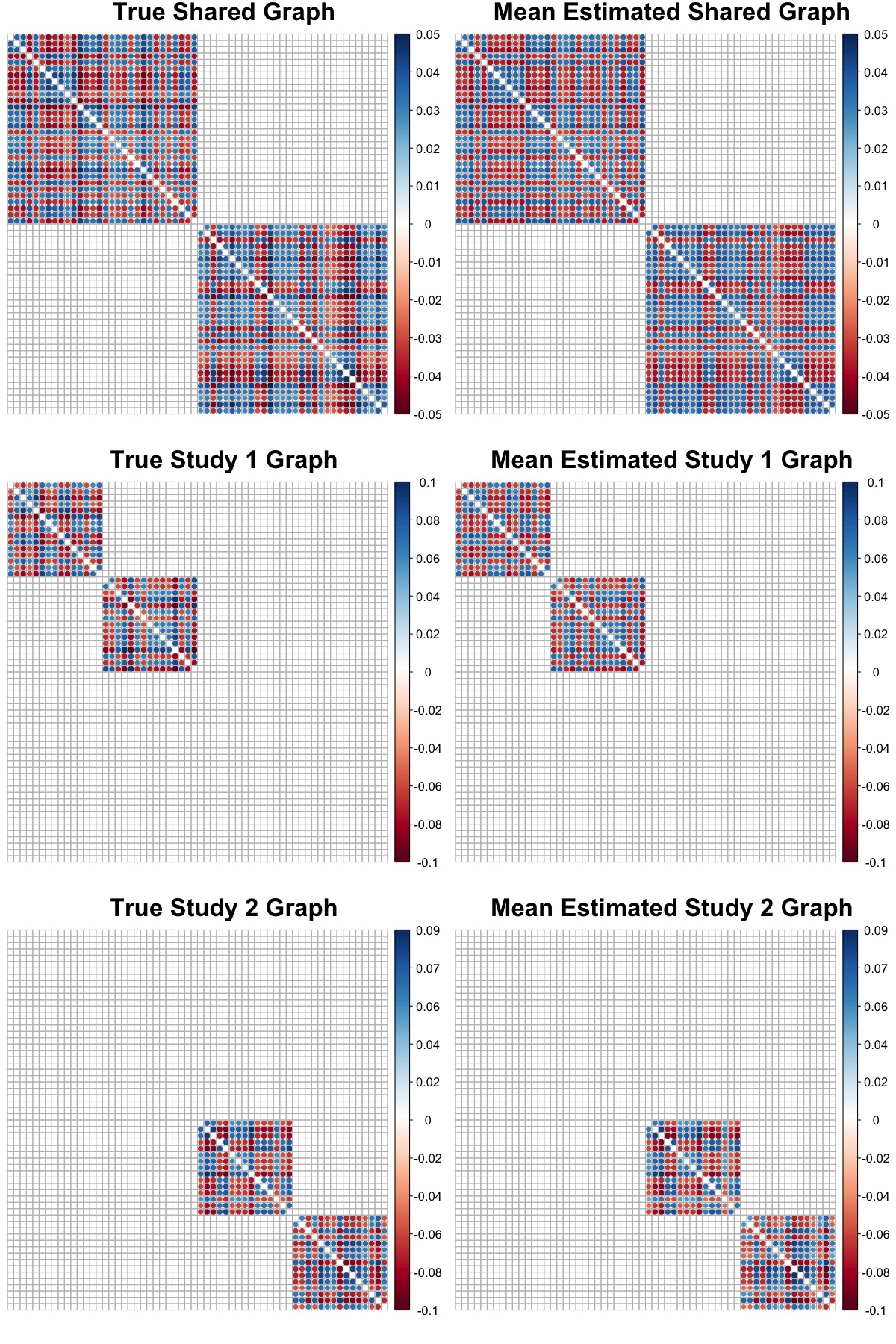}
    \caption{Setting 5}
\end{figure}

\begin{figure}
    \centering
    \includegraphics[height=0.8\textheight]{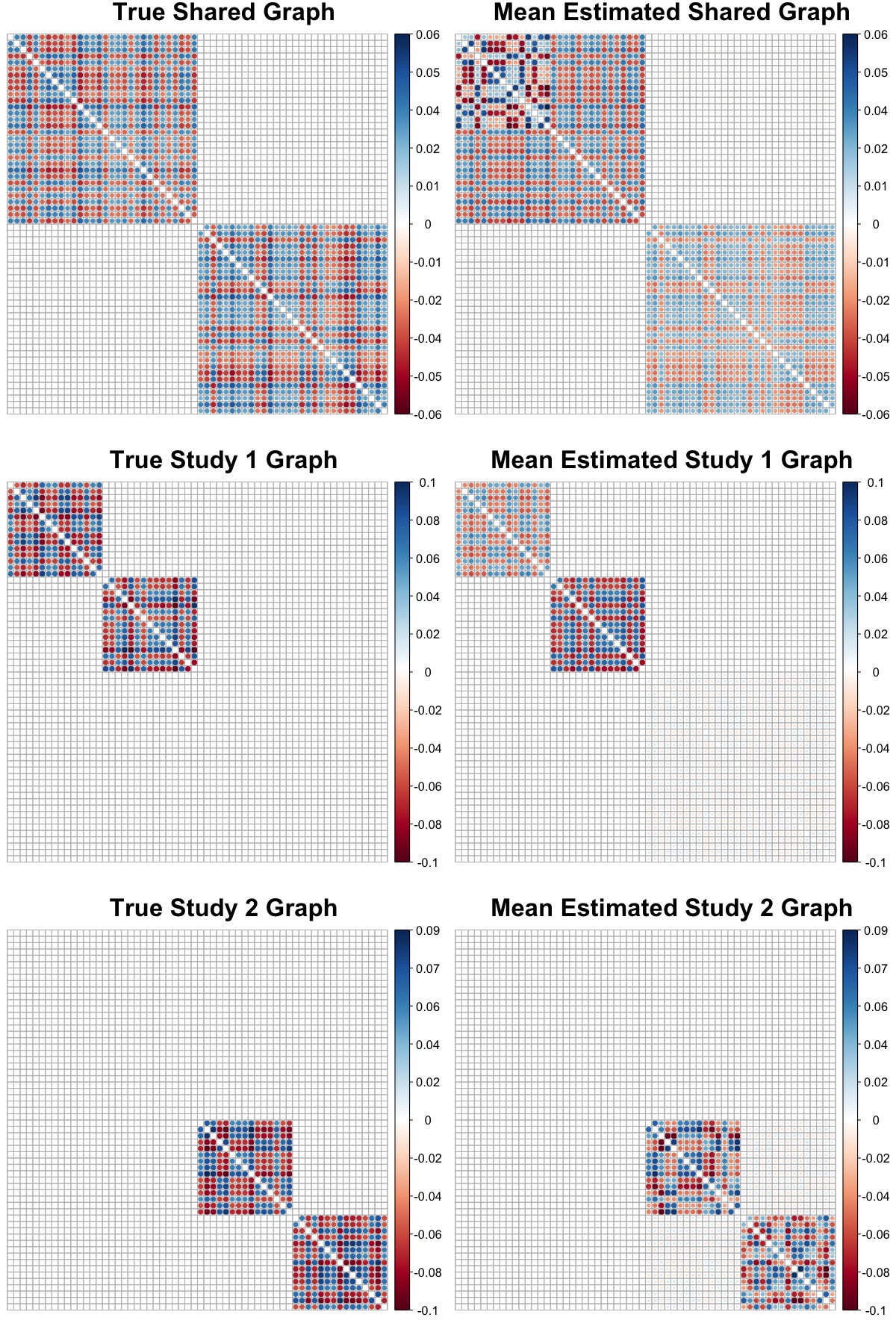}
    \caption{Setting 6}
\end{figure}

\begin{figure}
    \centering
    \includegraphics[height=0.8\textheight]{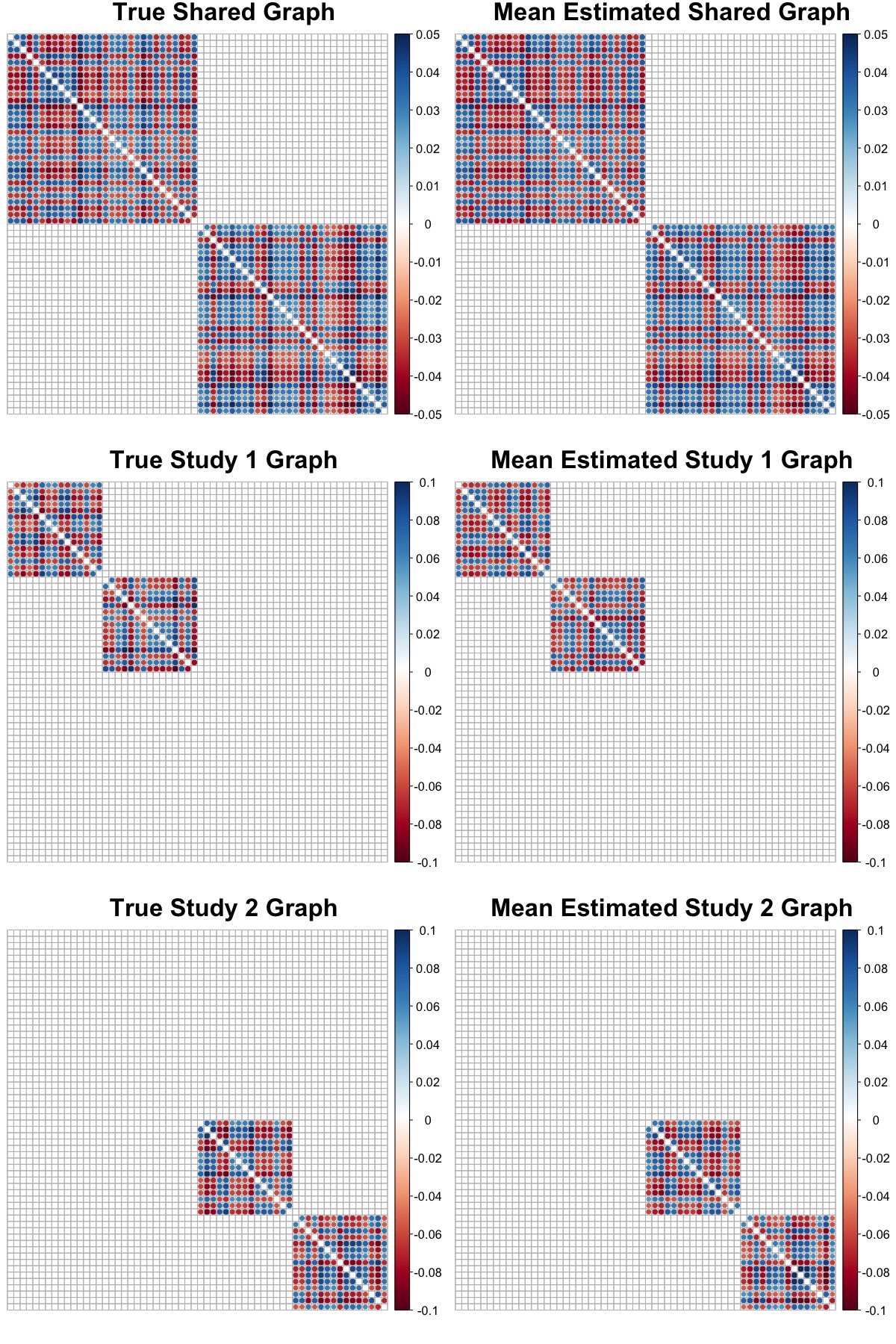}
    \caption{Setting 7}
\end{figure}

\begin{figure}
    \centering
    \includegraphics[height=0.8\textheight]{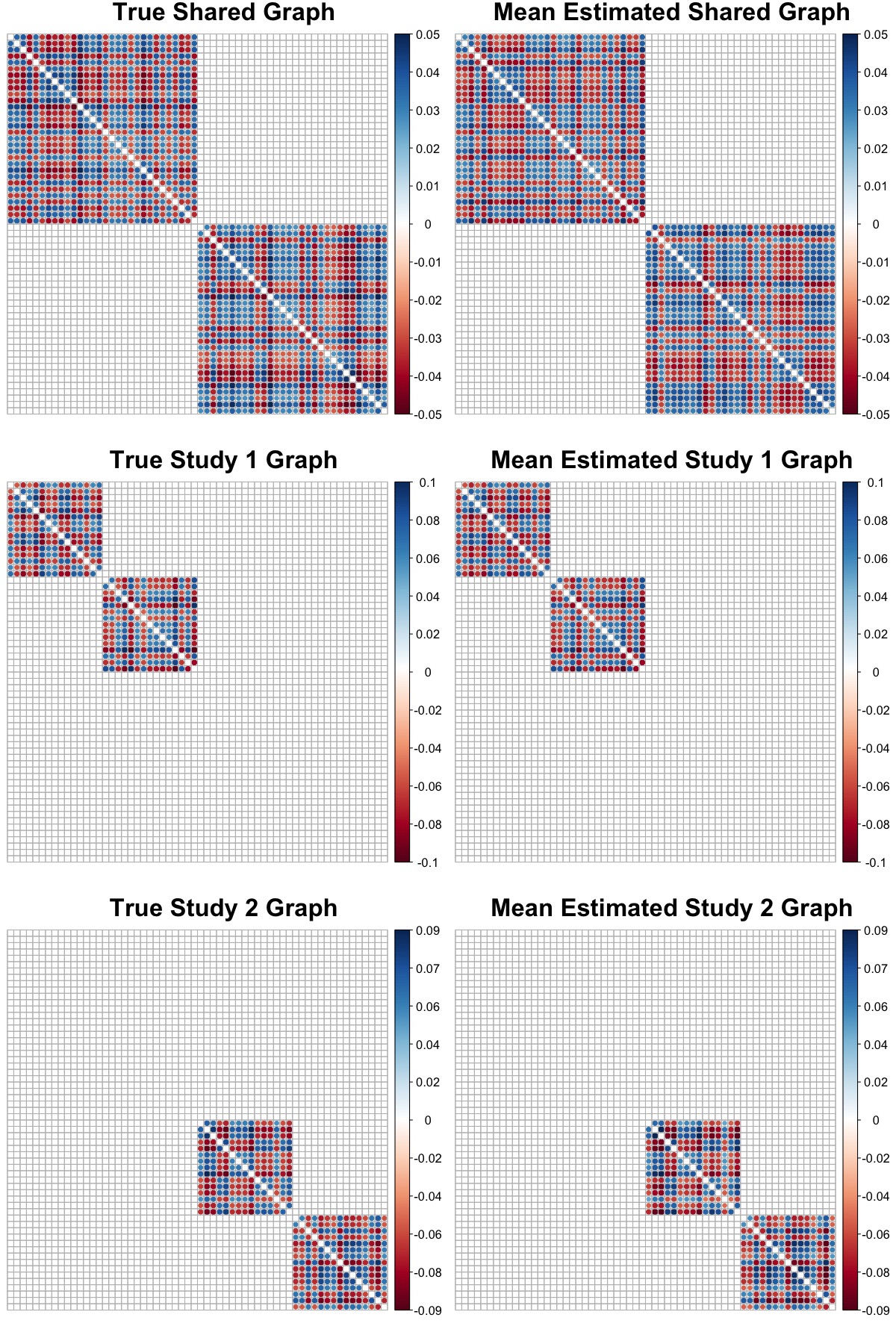}
    \caption{Setting 8}
\end{figure}

\begin{figure}
    \centering
    \includegraphics[height=0.8\textheight]{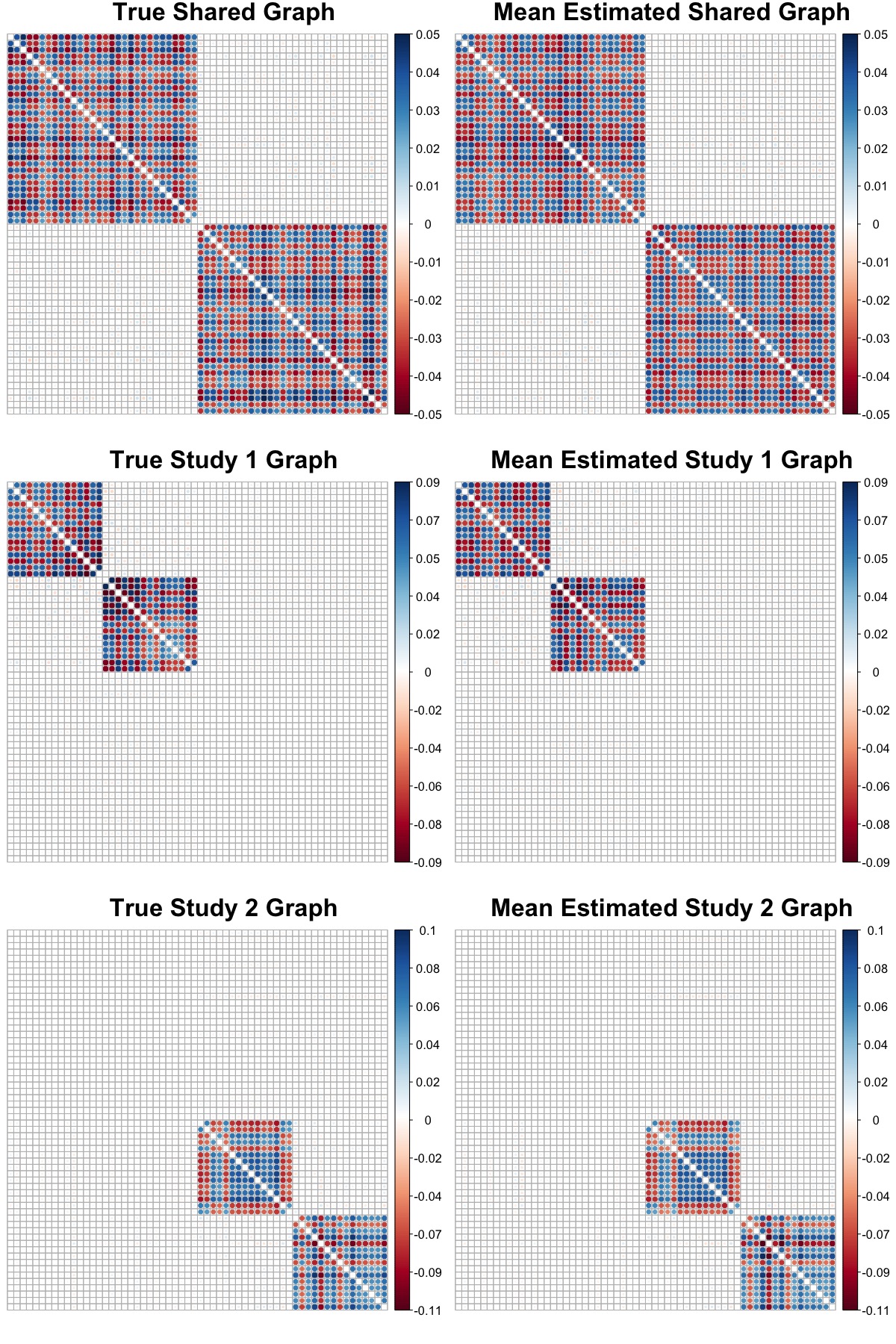}
    \caption{Setting 9}
\end{figure}

\begin{figure}
    \centering
    \includegraphics[height=0.8\textheight]{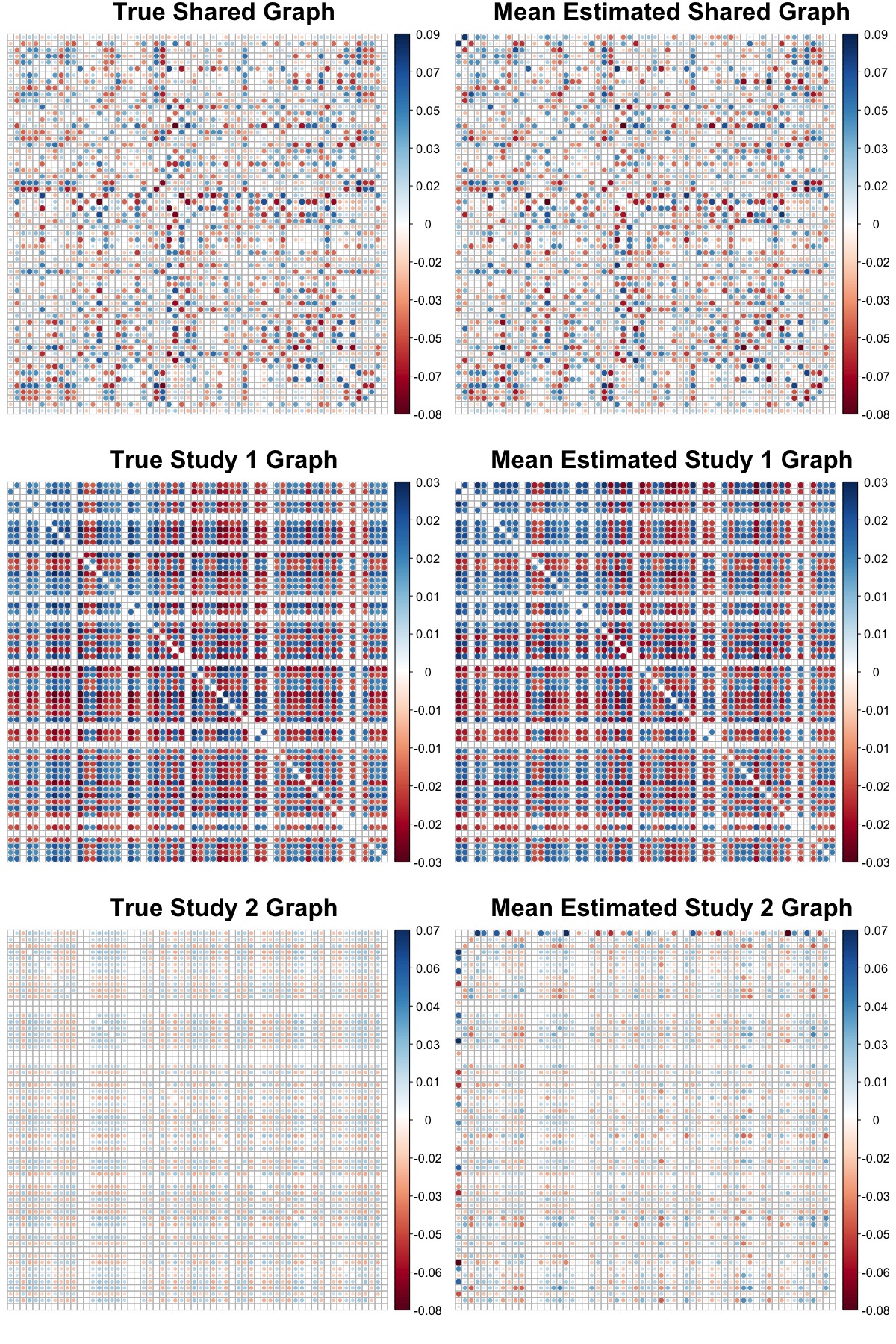}
    \caption{Setting 10}
\end{figure}

\clearpage

\section{Supplement to HAPO Application}

\subsection{Data preprocessing}

We intended to exclude metabolites with $>50$ percent of measurements missing; no metabolites met this threshold and so none were excluded. For metabolites missing $\leq 50$ percent, missing values were assumed to be below the limit of detection and imputed as half of the minimum detected value. After imputation, the log ratios of the changes in metabolite levels from fasting to 1-hr post-glucose intake were calculated as follows:
\begin{equation}
D = \log_2 \frac{\text{Metabolite level at 1 hr post-glucose intake}}{\text{Metabolite level at fasting}}    
\end{equation}

These log ratios were then centered to have mean 0 within each disease group (women without and women with GDM), consistent with the MSFA framework shown in Equation (1) in the main manuscript and the assumption that latent variables have mean zero. We did not standardize the variables to unit variance, as we specifically wish to model the study-specific variances in this case and doing such standardization would remove the signal of interest.

To incorporate the effect of important covariates in the network structure, we began by performing a linear regression of each standardized log metabolite ratio against important confounders, including maternal age at OGTT, mean arterial pressure at OGTT, body mass index (BMI) at OGTT, maternal height at OGTT, fasting plasma glucose, ancestry group, and sample storage time. The residuals from this regression represent the remaining variation in log metabolite ratios after adjusting for these variables. Covariate-adjusted network structures can thus be estimated using these residuals as the input to MSFA-X. Because MSFA-X uses centered predictors, we again centered the residuals to have mean 0 within each group. 

\clearpage

\subsection{Graphical lasso benchmark results}

\begin{figure}[h!]
    \centering
    \caption{Benchmark graphical lasso results for shared and study-specific networks in the HAPO data application. No edges with Bonferroni-significant p-values were detected in the non-GDM network.}
    \includegraphics[width=0.65\textwidth]{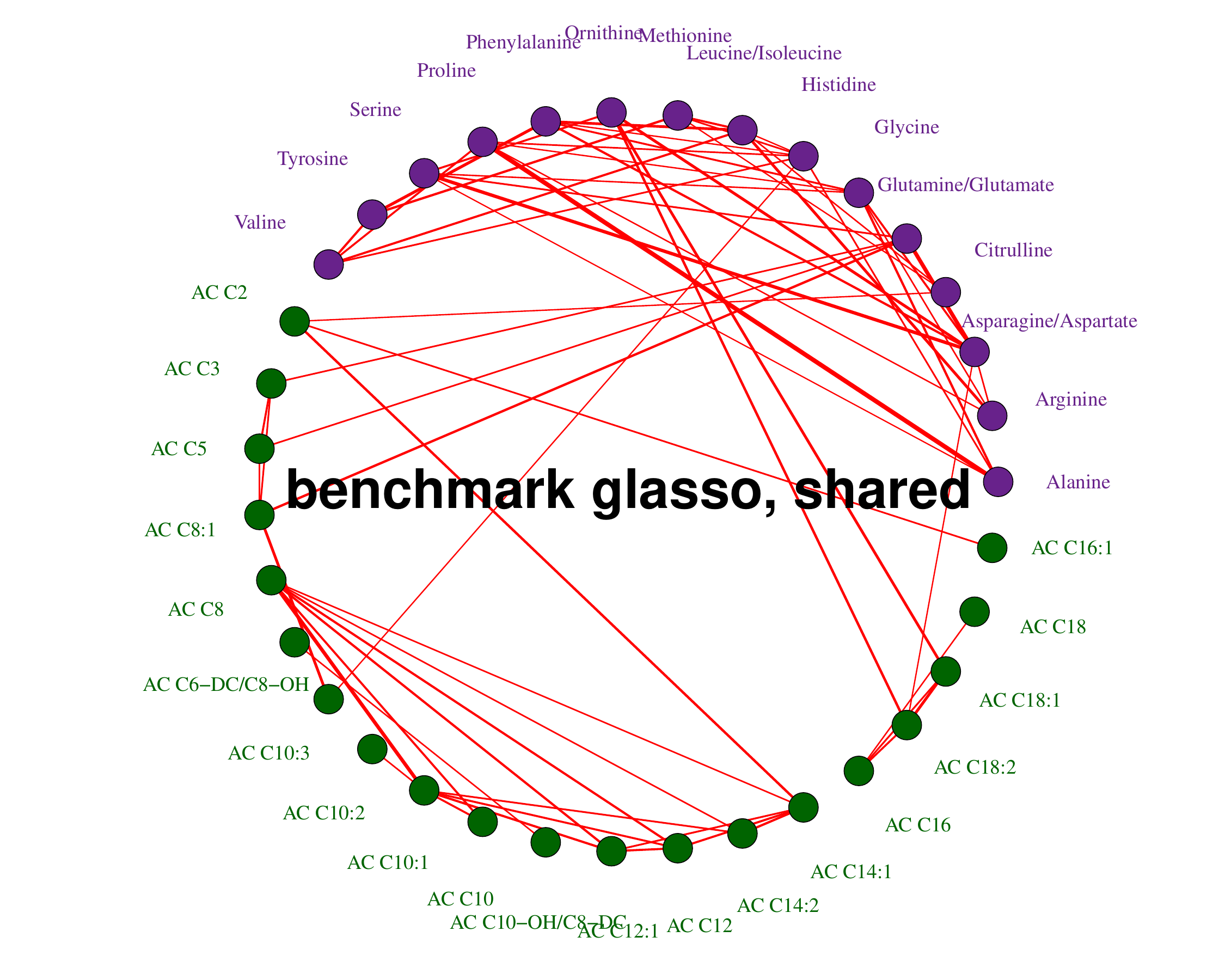}
    \includegraphics[width=0.6\textwidth]{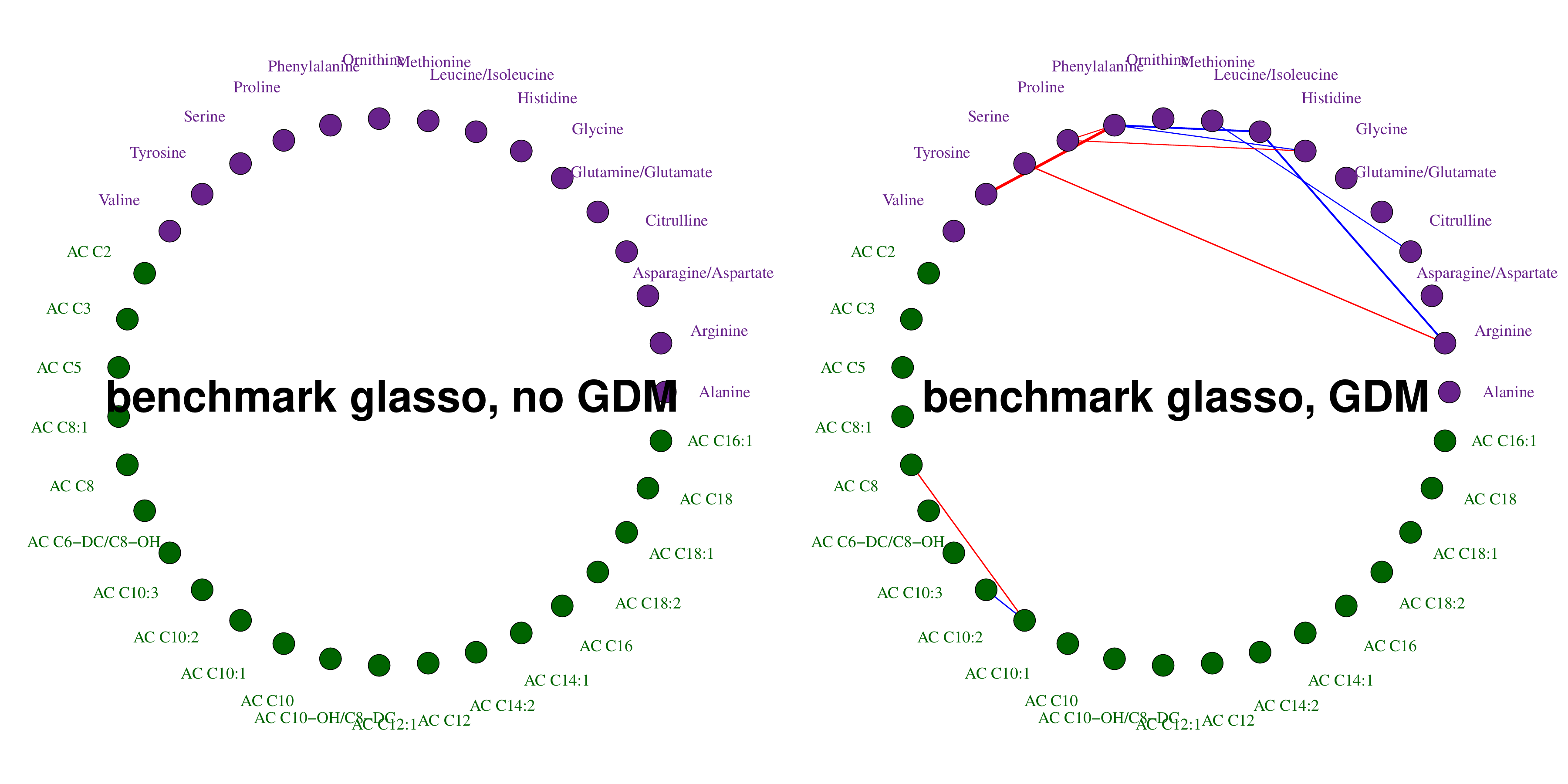}
    \label{fig:benchmark_nets}
\end{figure}

\begin{figure}
    \centering
    \begin{subfigure}[b]{0.45\textwidth}
    \centering
    \includegraphics[width=\textwidth]{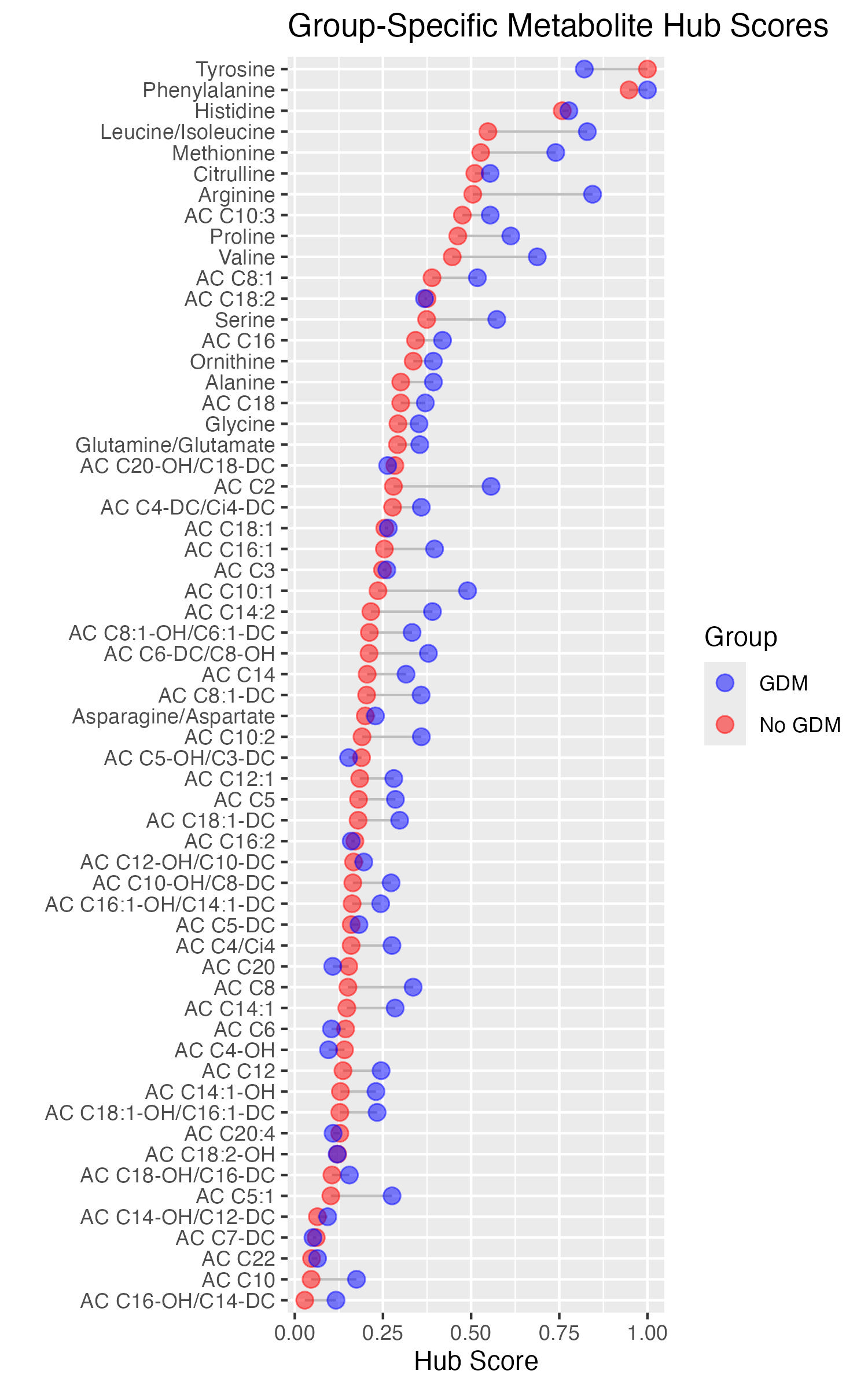}
    \end{subfigure}
    \begin{subfigure}[b]{0.45\textwidth}
    \centering
    \includegraphics[width=\textwidth]{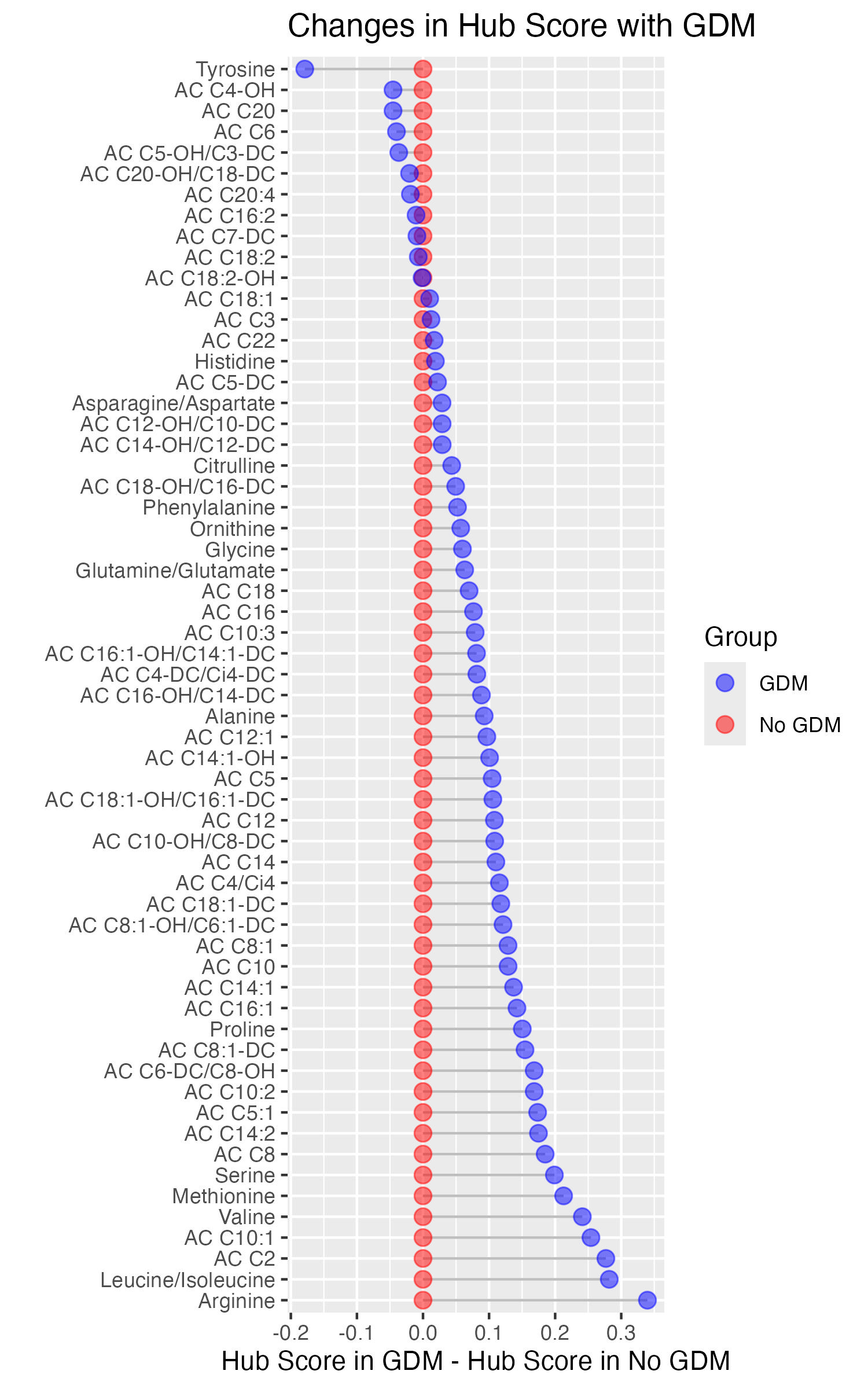}
    \end{subfigure}
    \\
    \begin{subfigure}{0.45\textwidth}
     \centering
     \includegraphics[width=\textwidth]{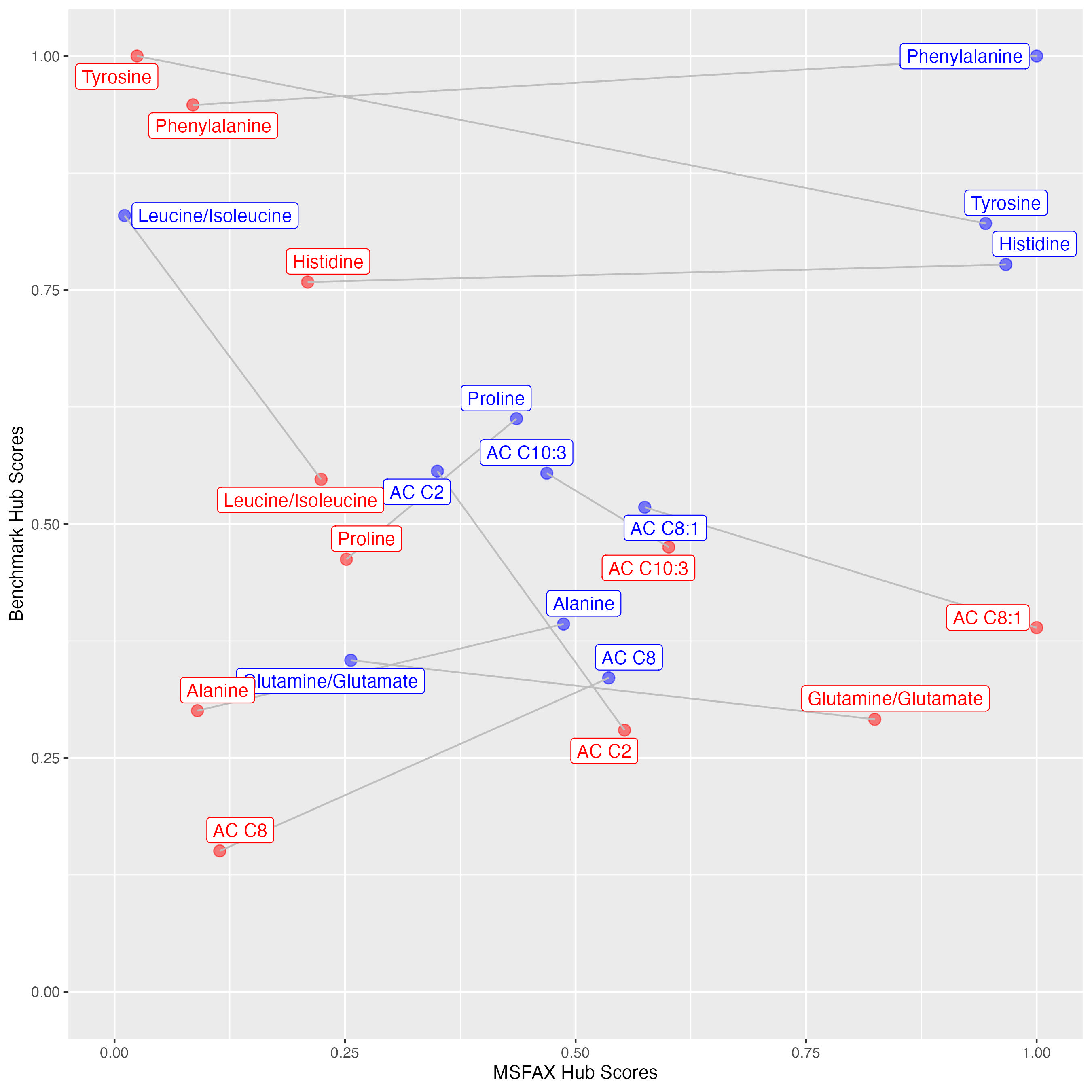}
     \end{subfigure}
    \caption{(a,b) Differences in node importance between women with GDM and women without GDM according to the benchmark graphical lasso model. (c) Biologically relevant differences in leucine/isoleucine and glutamine/glutamate centrality that were observed in the MSFA-X networks are not observed in the benchmark.}
    \label{fig:benchmark_lollipops}
\end{figure}

As with the MSFA-X networks, we analyzed the benchmark graphical lasso networks in two ways: qualitatively, by visualizing the estimated shared and study-specific networks (Supplementary Figure \ref{fig:benchmark_nets}), and quantitatively, by investigating hub scores (\ref{fig:benchmark_lollipops}).

To choose a threshold for visualization in the partial correlation networks calculated with the graphical lasso benchmark, we calculated the critical value for the magnitude of a partial correlation to reject the null hypothesis $H_0: F(\rho_{ij}|X_{-ij})=0$ in favor of the alternative $H_A: F(\rho_{ij}|X_{-ij})\neq0$, where $F(\cdot)$ is the Fisher transformation conventionally used for testing significance of correlations and partial correlations \cite{fisher1,fisher2}. The level of the test was set to $\alpha = 0.05/(60*59)$, corresponding to a Bonferroni-corrected significance level of 0.05 on the collection of tests for each of the 60 choose 2 (1770
) possible network edges. Using the full sample size of 3463, the significance threshold is $t = 0.072$ for the shared network. It is not immediately clear which sample size should be used for this calculation for the study-specific networks, as the full sample is used to estimate them, but the network itself is representative of a smaller subsample. Because the Bonferroni correction is typically quite conservative, we chose to use the more liberal threshold based on the full sample size ($t=0.072$) for the study-specific networks as well.

\newpage

\subsection{Balanced sample sizes sensitivity analysis}

\begin{figure}[h!]
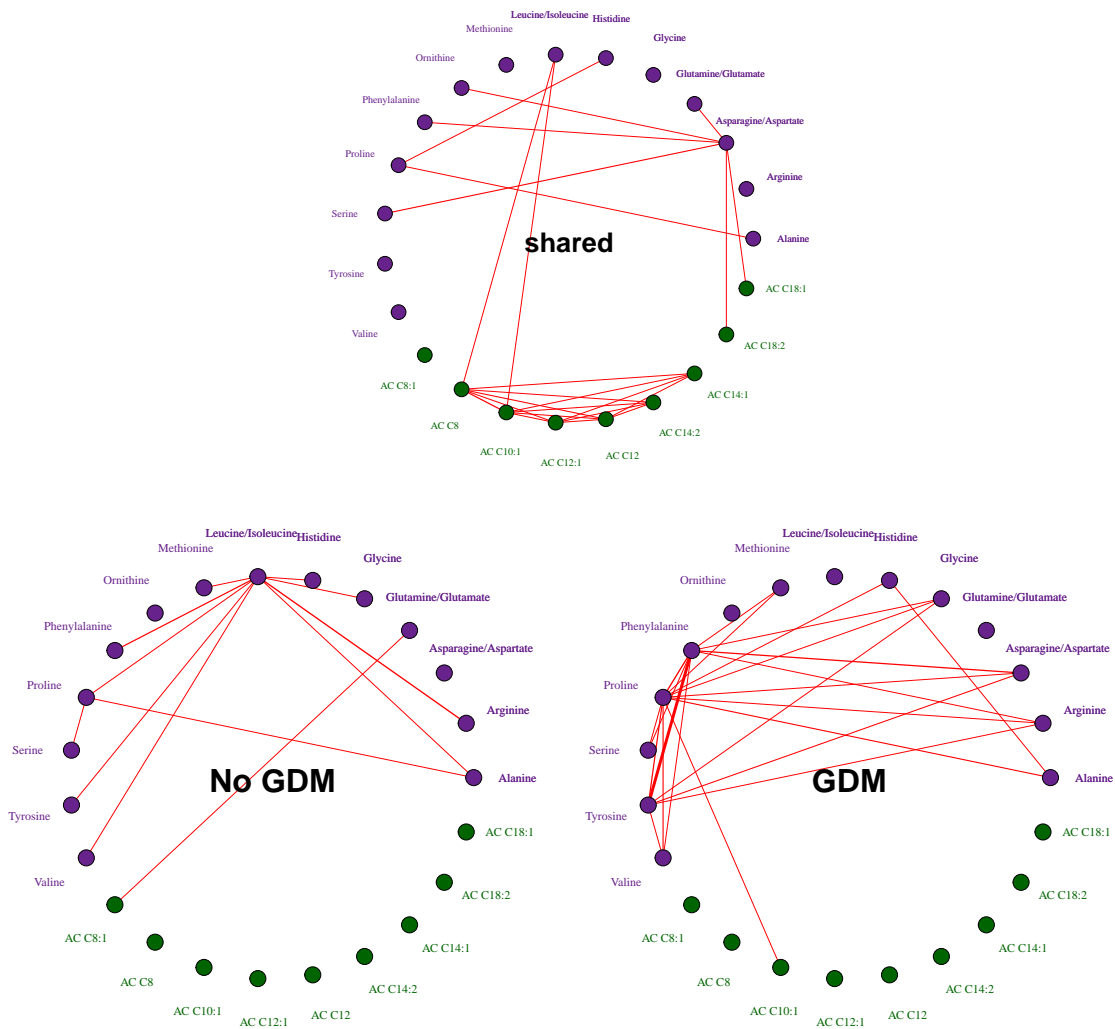

    \centering
    \begin{subfigure}[b]{\textwidth}
    \centering
    \includegraphics[width=0.45\textwidth]{img/gdm/supplementaryFigure14a.pdf}
    \end{subfigure}
    \begin{subfigure}[b]{0.85\textwidth}
    \centering
    \includegraphics[width=\textwidth]{img/gdm/supplementaryFigure14b.pdf}
    \end{subfigure}
    \caption{Shared and study-specific graphs for the balanced case, where the networks shown are the mean of networks estimated from 100 independent samples matching the size of the GDM sample with the size of the healthy control sample (N=576).}
    \label{fig:balanced_sensitivity_networks}
\end{figure}

\begin{figure}[h!]
    \centering
    \begin{subfigure}[b]{0.45\textwidth}
    \centering
    \includegraphics[width=\textwidth]{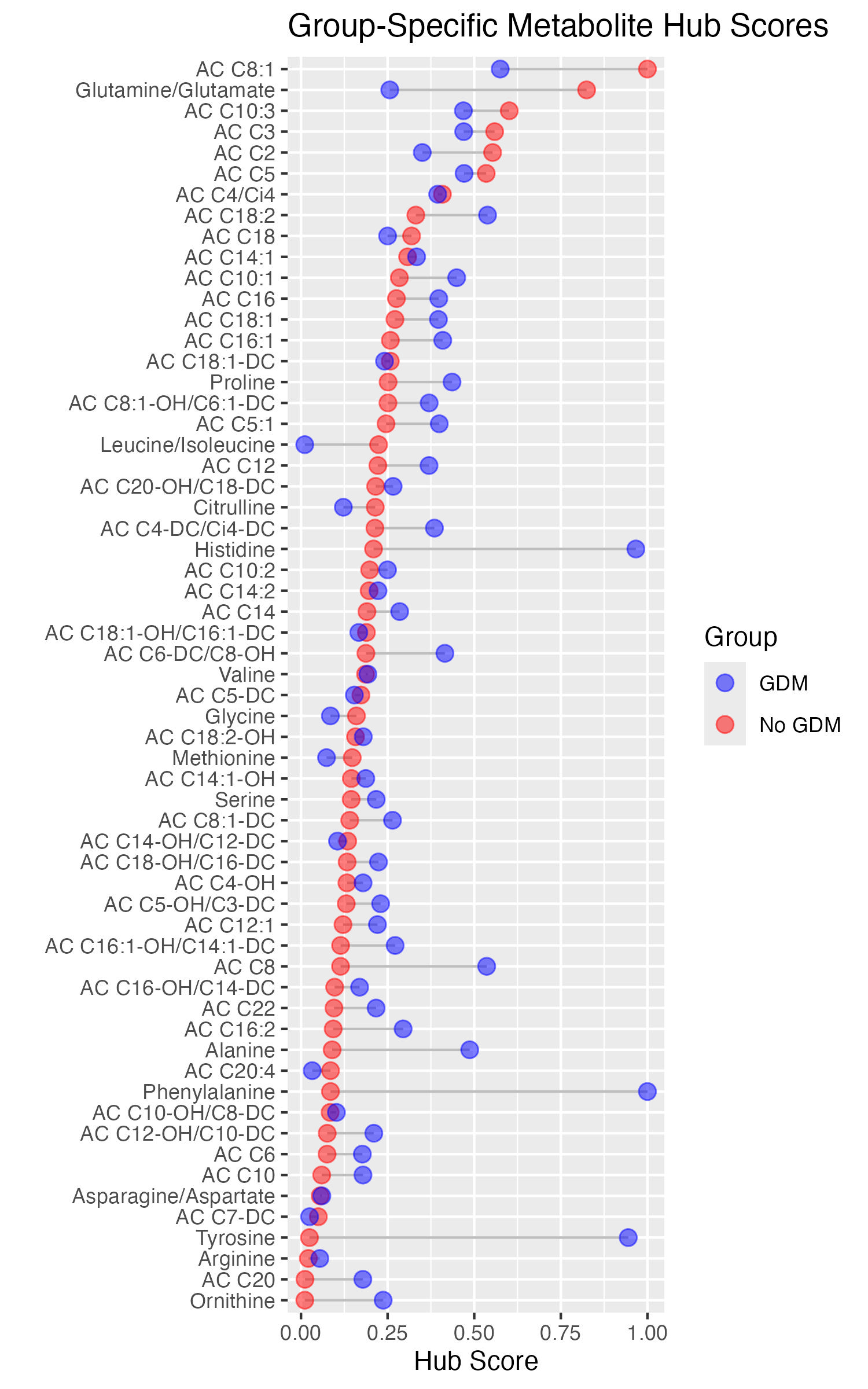}
    \end{subfigure}
    \begin{subfigure}[b]{0.45\textwidth}
    \centering
    \includegraphics[width=\textwidth]{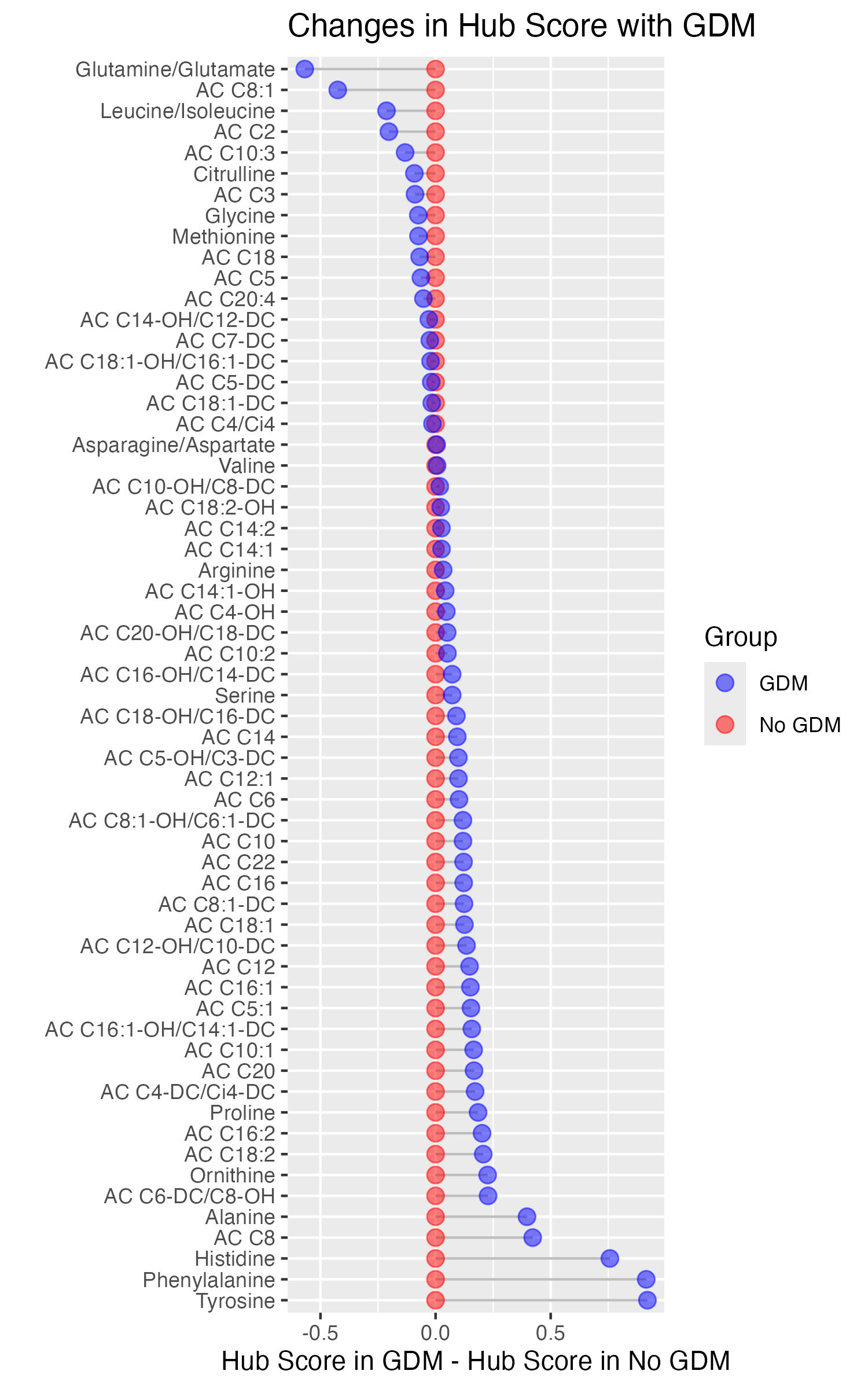}
    \end{subfigure}
    \caption{Differences in node importance between women with GDM and women without GDM according to the balanced sensitivity analysis. Key patterns of the full dataset analysis are reproduced.}
    \label{fig:balanced_sensitivity_lollipops}
\end{figure}

In simulation studies, MSFA-X performed reasonably well in a setting with unequal sample sizes, with slightly worse performance in the smaller study (Simulation 6; Supplementary Table 5,6,7; Supplementary Figure 1). As we wished to gain further confidence that the differences we observed between women with and without GDM were not simply due to the difference in sample size and power, we conducted a sensitivity analysis in which we downsampled the number of women without GDM to balance the samples. Specifically, we obtained 100 random samples of size N=576 by repeatedly sampling  576 women from the 2887 women without GDM in our study population. Sampling was conducted without replacement for each of the 100 samples, but individuals may appear in more than one of the 100 samples overall. We ran MSFA-X to estimate 100 different shared and study-specific networks. We averaged these 100 networks to obtain a single network representative of a balanced version of the full population analysis. The shared and study-specific networks in the balanced analysis both have fewer edges due to the smaller overall sample size resulting from the downsampling, and some changes in top edges are observed, likely due to the reduced power to de-convolute the shared and study-specific signals. In particular, the glutamine/glutamate motif was absent in the shared network and instead, a strong role for isoleucine/leucine as a hub node was apparent (Supplementary Figure \ref{fig:balanced_sensitivity_networks}b). However, key network motifs were preserved in the shared network (e.g., strong alanine-proline connection, asparagine/aspartate as a hub, connections between medium-chain ACs; Supplementary Figure \ref{fig:balanced_sensitivity_networks}a) and in the GDM-specific network (strong tyrosine-phenylalanine connection, although the connection to histidine is notably absent; Supplementary Figure \ref{fig:balanced_sensitivity_networks}c). Moreover, the non-thresholded results based on hub scores show similar results to those in the full analysis, including a decreased role for glutamine/glutamate and leucine/isoleucine in women with GDM vs. without and an increased role of phenylalanine, and tyrosine in women with GDM vs. without (Supplementary Figure \ref{fig:balanced_sensitivity_lollipops}). 

\vspace{1in}

\clearpage

\clearpage
\newpage


\putbib
\end{bibunit}

\end{document}